\documentclass{aa}
\usepackage{graphicx}
\usepackage{txfonts}
\usepackage{lipsum}
\usepackage{subcaption}
\usepackage{lscape}
\usepackage{placeins}
\usepackage{float}
\usepackage{color}
\usepackage{natbib}
\bibpunct{(}{)}{;}{a}{}{,}

\def\vector#1{\mbox{\boldmath $#1$}}

\usepackage[colorlinks=true, 
            linkcolor=blue,
            citecolor=blue,
            urlcolor=blue]{hyperref}
            
\renewcommand{\makeLineNumber}{\relax}

\begin{document}

   \title{Enhancing Compton telescope imaging\\with maximum a posteriori estimation}

   \subtitle{A modified Richardson-Lucy algorithm\\for the Compton Spectrometer and Imager}

   \author{H. Yoneda\inst{1,2},
           T. Siegert\inst{1},  
           I. Martinez-Castellanos\inst{3,4},
           S. Gallego\inst{5},
           C. Karwin\inst{4},
           H. Bates\inst{6},
           S.E. Boggs\inst{7},
           C.Y. Huang\inst{8},\\
           A. Joens\inst{9},
           S. Matsumoto\inst{10},
           S. Mittal\inst{1},
           E. Neights\inst{11,4},
           M. Negro\inst{12},
           U. Oberlack\inst{5},
           K. Okuma\inst{13},
           S. Pike\inst{7},
           J. Roberts\inst{7},\\
           F. Rogers\inst{9},
           Y. Sheng\inst{6},
           T. Takahashi\inst{10},
           A. Valluvan\inst{7},
           Y. Watanabe\inst{10},
           D. Hartmann\inst{6},
           C. Kierans\inst{4},
           J. Tomsick\inst{9},\\
           A. Zoglauer\inst{9}
          }

   \institute{Julius-Maximilians-Universit\"{a}t W\"{u}rzburg, Fakult\"{a}t f\"{u}r Physik und Astronomie, Institut f\"{u}r Theoretische Physik und Astrophysik, Lehrstuhl f\"{u}r Astronomie, Emil-Fischer-Str. 31, D-97074 W\"{u}rzburg, Germany\\
            \email{hiroki.yoneda.phys@gmail.com}
            \and RIKEN Nishina Center, 2-1 Hirosawa, Wako, Saitama 351-0198, Japan
            \and Department of Astronomy, University of Maryland, College Park, MD 20742, USA
            \and NASA Goddard Space Flight Center, 8800 Greenbelt Road, Greenbelt, MD 20771, USA
            \and Institut f\"{u}r Physik \& Exzellenzcluster PRISMA+, Johannes Gutenberg-Universit\"{a}t Mainz, 55099 Mainz, Germany
            \and Department of Physics and Astronomy, Clemson University, Clemson, SC 29634, USA
            \and Department of Astronomy \& Astrophysics, UC San Diego, 9500 Gilman Drive, La Jolla, CA 92093, USA
            \and Institute of Astronomy, National Tsing Hua University, Guangfu Rd., Hsinchu City, 300044, Taiwan
            \and Space Sciences Laboratory, UC Berkeley, 7 Gauss Way, University of California, Berkeley, CA 94720, USA
            \and Kavli Institute for the Physics and Mathematics of the Universe (WPI), The University of Tokyo Institutes for Advanced Study, The University of Tokyo, 5-1-5 Kashiwa-no-ha, Kashiwa, Chiba 277-8583, Japan
            \and The George Washington University, Department of Physics, 725 21st St NW, Washington, DC 20052
            \and Louisiana State University, BatonRouge, LA 70803, USA
            \and Kobayashi-Masukawa Institute, Nagoya University, Furo-cho, Chikusa-ku, Nagoya, Aichi 4648602, Japan}

   \date{Received September 30, 20XX}
   
   \titlerunning{MAP Richardson-Lucy algorithm for COSI}
   \authorrunning{H. Yoneda et al.}

  \abstract{We present a modified Richardson-Lucy (RL) algorithm tailored for image reconstruction in MeV gamma-ray observations, focusing on its application to the upcoming Compton Spectrometer and Imager (COSI) mission. Our method addresses key challenges in MeV gamma-ray astronomy by incorporating Bayesian priors for sparseness and smoothness while optimizing background components simultaneously. We introduce a novel sparsity term suitable for Poisson-sampled data in addition to a smoothness prior, allowing for flexible reconstruction of both point sources and extended emission. The performance of the algorithm is evaluated using simulated three-month COSI observations of gamma-ray lines of $^{44}$Ti (1.157 MeV), $^{26}$Al (1.809 MeV), and positron annihilation (0.511 MeV), respectively, representing various spatial features. Our results demonstrate significant improvements over conventional RL methods, particularly in suppressing artificial structures in point source reconstructions and retaining diffuse spatial structures. This work represents an important step toward establishing a robust data analysis for studying nucleosynthesis, positron annihilation, and other high-energy phenomena in our Galaxy.}


   \maketitle

\section{Introduction}

Observations of gamma-ray lines, such as nuclear lines from radioisotopes and the 0.511 MeV line from positron-electron annihilation, are crucial for understanding the origin of elements in our Universe. Gamma-ray line observations were pioneered by balloon experiments in the 1960s and 1970s \citep[e.g.,][]{Haymes1969}, and our understanding has since deepened through successful satellite missions covering the MeV energy bands, such as CGRO/OSSE \citep{OSSE}, COMPTEL \citep{COMPTEL} and INTEGRAL/SPI \citep{SPI}. These missions have uncovered the distribution of gamma-ray lines across our Galaxy, including the diffuse 0.511 MeV emission from the Galactic disk and bulge \citep[e.g.,][]{Purcell1997,Knoedlseder2005} and the distribution of $^{26}$Al along the Galactic plane \citep{Oberlack1996,Knoedlseder1999,Bouchet2015,Siegert2023_Al26}. Building upon these achievements, the upcoming Compton Spectrometer and Imager (COSI) satellite is scheduled for launch in 2027 \citep{COSIofficial}. 
With a stacked array of germanium cross-strip detectors, COSI is expected to improve gamma-ray line sensitivity by up to an order of magnitude after two years of operation.
The primary science objectives of COSI include mapping the spatial distribution of positron annihilation gamma rays and observing nuclear gamma rays from radioactive nuclei. 

One of the key techniques to achieve the above science goals is reconstructing an all-sky image from Compton scattering events detected by the instrument. COSI employs a Compton telescope \citep{Kamae1987} and measures Compton scattering events with its position sensitive detectors. After event reconstruction \citep[e.g.,][]{Boggs2000,Oberlack2000,Kurfess2000,Zoglauer2007a,Zoglauer2007b,Takashima2022,Yoneda2023}, an event in a Compton telescope is characterized by a Compton scattering angle and a scattered gamma-ray direction. Then, the incoming gamma-ray direction is constrained to a ring in the sky rather than being uniquely determined as a point \citep{vonBallmoos1989}. Consequently, a statistical method is essential to recover the gamma-ray source distribution from the reconstructed Compton scattering events. Such a method must be capable of solving this inverse and ill-posed problem inherent in Compton telescope data analysis.

Traditionally, the Richardson-Lucy (RL) algorithm has been used for image reconstruction in this field \citep{Richardson,Lucy1992,Wilderman1998}. It optimizes the flux on each pixel by iteratively maximizing the surrogate function for the log-likelihood, based on the Expectation–Maximization (EM) algorithm \citep{Bishop2006}. It is widely recognized that the conventional RL algorithm enhances locally bright structures by amplifying statistical fluctuations \citep{Allain2006}. While several modified algorithms, often based on ad-hoc methods, have been proposed to mitigate this issue \citep[e.g.,][]{Green1990,Wang1997,Strong2003,Knoedlseder2005}, recent advancements have shown that incorporating various regularization terms within a Bayesian framework can yield more plausible images, i.e., less artifact-prone \citep{Ikeda2014,Morii2019,FriotGiroux2022,Sakai_2023,Morii2024}. As an example of image reconstruction in astronomy, a Bayesian approach has also succeeded in other fields, e.g., radio imaging of a black hole shadow \citep{Akiyama_2019}.
Another notable example is NIFTY \citep{Selig2013}, which provides a framework for signal inference problems based on information field theory \citep{Ensslin2009}. NIFTY supports Bayesian inference methods and has been successful in the image analysis of various astronomical data, e.g., radio 
\citep{Junklewitz2016}, and GeV gamma-ray observations \citep{Selig2015}.

Despite these advancements, applying such methods to image reconstruction presents unique challenges in MeV gamma-ray astronomy. 
One significant issue is that L1 norm regularization \citep{LASSO} becomes ineffective while it is broadly used for image sparsity in different fields
\citep{Ikeda2014}.
It is the sum of absolute values of model parameters and promotes sparsity in them, namely, tending to make many parameters exactly zero.
Unlike other imaging domains, gamma-ray data follows Poisson statistics, and the reconstructed flux in each pixel is strongly tied to the probability maps of incoming photons. Given that the sum of probability is conserved as 1, the L1 norm becomes almost constant (especially when the exposure throughout the sky is uniform like it will be for COSI), rendering it ineffective as a sparsity term.
Another critical challenge is that instrumental background events dominate signals. Satellites like COSI suffer from significant background originating from activation of the spacecraft material caused by charged particles in orbit \citep[e.g.,][]{Weidenspointner2001,SCHONFELDER2004,SGD2018,Odaka2018,Diehl2018,Cumani2019}. This results in signal-to-background ratios as low as 10\% or even much lower. Thus, it is necessary to carefully consider the background and optimize both signal and background simultaneously.

In this paper, we introduce a modified RL algorithm tailored for COSI, addressing the above challenges. 
The proposed algorithm finds an approximate solution for the maximum a posteriori estimation incorporating regularization terms within the Bayesian framework.
In Section~\ref{sec_COSI}, we briefly overview the COSI mission. Then, in Section~\ref{sec_MAP_RL}, we incorporate a novel sparsity term suitable for Poisson data, as introduced by \cite{Ikeda2014}, while simultaneously optimizing background normalization under a gamma distribution. Our framework allows for the flexible introduction of various regularization terms, including inter-pixel correlations (smoothness), and employs a simple method to compute the RL algorithm. 
To demonstrate the proposed algorithm, we apply it to simulated COSI observation data spanning three months, presenting results for several science cases where sparsity, smoothness, or both are expected to play crucial roles: $^{44}$Ti (only point sources expected at 1.157 MeV), $^{26}$Al (mostly diffuse extended emission with possible point sources, 1.809 MeV), and positron annihilation (so far only diffuse emission detected with possible point sources within the sensitivity range of COSI, 0.511 MeV).
We explain the simulation setup in Section~\ref{sec_simulation}, and show the results in Section~\ref{sec_results}. Through these applications to actual science cases, we illustrate the potential of our method to enhance image reconstruction in MeV gamma-ray astronomy. Finally, we discuss further improvements for the actual data analysis in the future in Section~\ref{sec_discussion}.

\section{The Compton Spectrometer and Imager}
\label{sec_COSI}

\subsection{Overview of the COSI mission}

COSI is an upcoming NASA Small Explorer satellite mission planned for launch in 2027 \citep{COSIofficial}.
Based on successful balloon observations \citep{Lowell2017,Kierans2020,Siegert2020,Beechert2022,Karwin2023,Roberts2024}, 
COSI is designed to survey the entire sky in the energy range of 0.2--5 MeV and study energetic astronomical phenomena through imaging, spectroscopy, and polarimetry.
It has four primary science goals: (1) to uncover the origin of Galactic positrons; (2) to reveal Galactic element formation; (3) to gain insight into extreme environments with polarization; and (4) to probe the physics of multimessenger events.
The first two objectives, in particular, focus on gamma-ray lines to image our Galaxy at specific energies.

\begin{figure}
\centering
    \includegraphics[width = 0.9 \linewidth]{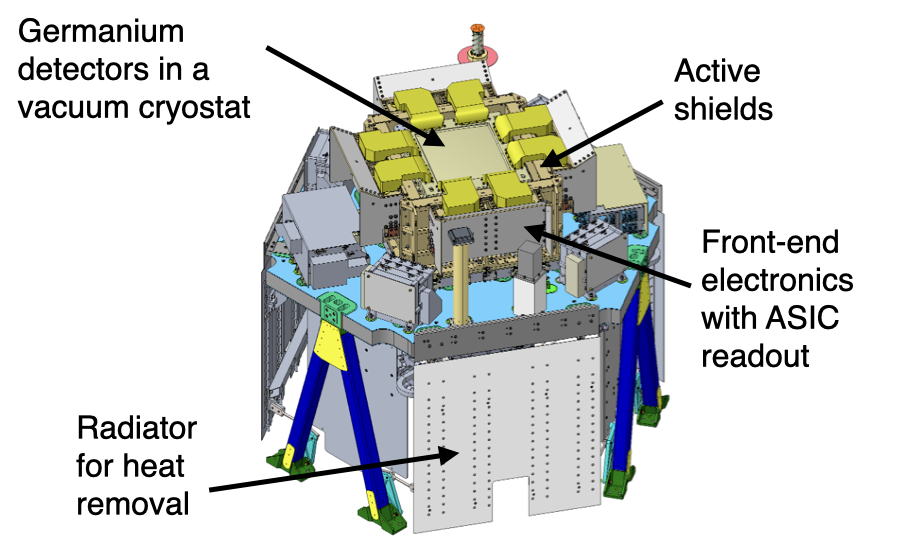}
    \includegraphics[width = 0.9 \linewidth]{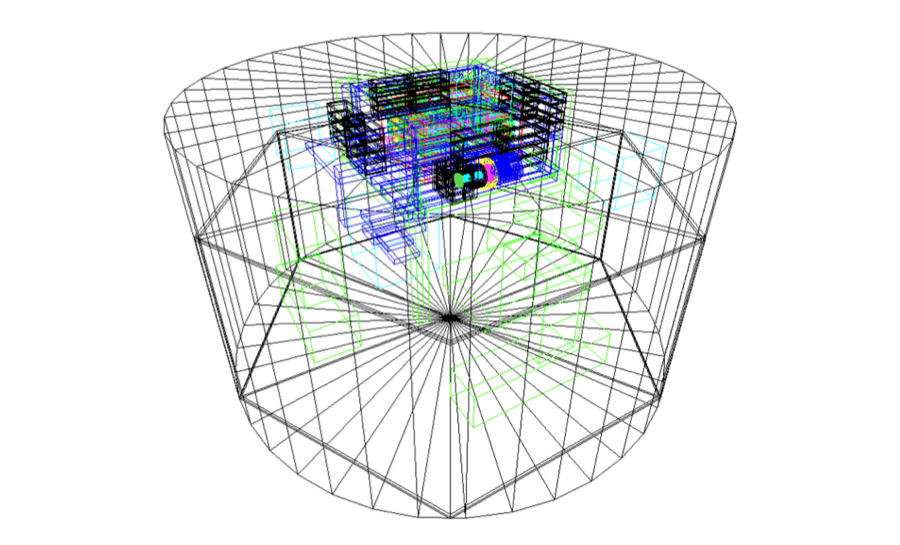}
\caption{COSI payload design (top) and a schematic of the mass model used in our simulation (bottom).}
\label{fig_COSI}
\end{figure}

To achieve these science goals, COSI employs a Compton telescope with excellent energy resolution.
The core of the COSI instrument consists of an array of 16 high-purity germanium detectors \citep[][Figure~\ref{fig_COSI}]{COSIofficial}.
Each is a double-sided strip detector with 64 strips per side with a strip pitch of 1.162 mm.
The detector dimensions are $8 \times 8 \times 1.5$ cm$^{3}$, and they are arranged in four stacks of four detectors.
They are cooled down to 80--90 K with a mechanical cryocooler.
The $x$ and $y$ positions of gamma-ray interactions are determined by the strip readouts, while the $z$ positions are measured by the time difference between anode and cathode signals.
This configuration acts as a Compton telescope, measuring deposited energies and positions of incident gamma rays via multiple Compton scattering.
The five sides of the Compton telescope are surrounded by bismuth germanium oxide (BGO) scintillators, which act as active anticoincidence shields and reduce background events in the satellite orbit.

With this detector design, COSI can instantaneously cover $>$ 25\% of the sky with a spectral resolution of 6.0 keV at 0.511 MeV and 9.0 keV at 1.157 MeV, while the angular resolution will be 4.1$^\circ$ at 0.511 MeV and 2.1$^\circ$ at 1.809 MeV.
COSI is expected to improve the gamma-ray line sensitivity by about one order of magnitude across its energy range during two years of operation.

\subsection{Data structure of COSI observations}

\begin{figure}
\centering
    \includegraphics[width = 0.95 \linewidth]{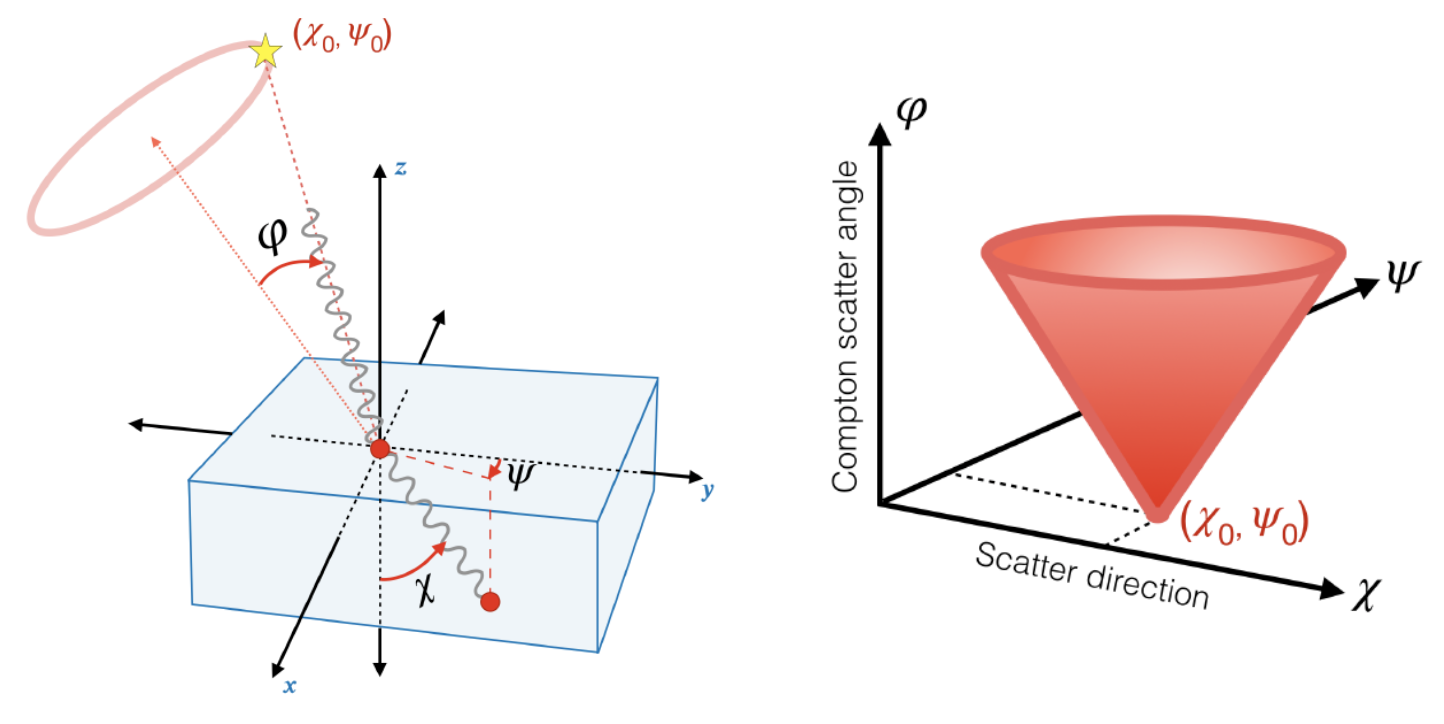}
\caption{Schematic of Compton Data Space \citep{kierans_detection_2018}. The left panel shows a photon scattering in a detector with the scattering angle $\phi$ and the scattered gamma-ray direction $(\psi, \chi)$. The source position $(\psi_0, \chi_0)$ and photon interaction points are shown. Photons from this source make a cone with a 90$^\circ$ opening angle in the CDS, as shown on the right panel. The apex of the cone corresponds to the source location.} 
\label{fig_CDS}
\end{figure}

Here, we outline the data structure of COSI for scientific analysis, including the image reconstruction \citep{Zoglauer2021}.
First, the raw detector output undergoes calibration to produce a list of detected energies and hit positions for each event \citep{Sleator2019,Beechert2022b}. Then, the path of the scattered gamma ray within the detector is determined using Compton scattering kinematics.
Finally, the reconstructed events are compiled into an event list, described by the Compton scattering angle ($\phi$), the direction of the scattered gamma-ray ($\psi$, $\chi$), and the measured gamma-ray energy ($E_{\mathrm{m}}$).
We refer to the data space described by these parameters as the Compton Data Space (CDS). 
Figure~\ref{fig_CDS} shows a schematic of the first three parameters in the CDS.
Note that $E_{\mathrm{m}}$ is generally different from the incident gamma-ray energy due to detector energy resolution, Compton event misreconstruction, and missing Compton interactions.
Most astrophysical analyses, including those described in this paper, typically begin with this processed data since it contains essential information about celestial source signals.

To apply the RL algorithm for image reconstruction, we first create a binned histogram ($D_{i}$) from the event list in the CDS\footnote{The unbinned analysis, or the photon list approach, is also possible but will be subject of future publications.}. Here, we denote the bin index in this space as $i$, representing a specific bin in the CDS, and $D_{i}$ is unitless.
Our goal is to estimate a model that best describes this data. In the context of all-sky image reconstruction, this model represents the gamma-ray flux in each pixel of the sky, denoted as $\lambda_j$. 
Here, $j$ is the bin index in the model space, which includes both spatial pixels and an energy axis ($E_i$), where $E_i$ represents the true initial gamma-ray energy.
In the cases of this paper, $\lambda_j$ has a unit of cm$^{-2}$ s$^{-1}$ sr$^{-1}$ with a single energy bin because we focus on the gamma-ray lines in our demonstration.
Note that generally, the energy axis can have multiple bins, and $\lambda_j$ can also represent a differential flux by using its corresponding unit, e.g., cm$^{-2}$ s$^{-1}$ sr$^{-1}$ keV$^{-1}$.
Furthermore, we introduce background models, $B_{ik}$, which we refer to as the background template matrix.
This matrix describes the event distribution pattern in the CDS for a certain background component, including instrumental backgrounds such as in-orbit activation. The index $k$ is a label for each background component.
The background template matrix does not account for the absolute normalization; thus, we introduce the background normalization $b_k$ for each component. Finally, we define a response matrix, $R_{ij}$, which connects the model ($\lambda_j$) to the observed data ($D_{i}$), essentially describing how a gamma-ray photon from a certain model space bin produces events in the CDS.
We will explain how we generated these histograms for our numerical tests in Section~\ref{sec_simulation}.

With this representation, each iteration of the conventional RL algorithm can be described mathematically as follows:
\begin{align}
& \text{(E-step)} & \epsilon_i & = \sum_j R_{ij} \lambda_j^{\mathrm{old}} +  \sum_k B_{ik} b_k \label{eq_expectation} & \\
& \text{(M-step)} & \lambda^{\mathrm{new}}_j & = \dfrac{\lambda^{\mathrm{old}}_j}{\displaystyle \sum_{i} R_{ij}} \displaystyle \sum_{i} \dfrac{D_{i}}{\epsilon_i} R_{ij}\label{eq_Mstep} & \\
&& b_k^{\mathrm{new}} & = \dfrac{b_k^{\mathrm{old}}}{\displaystyle \sum_i B_{ik}} \displaystyle \sum_i \dfrac{D_i}{\epsilon_i} B_{ik}\label{eq_Mstep_bkg_orig}~,&
\end{align}
where $\epsilon_i$ is an expected count at the bin index $i$ in the CDS.
Note that ``old'' and ``new'' indices refer to values before and after an iteration process of the RL algorithm.
Through sufficient iterations of this process, the RL algorithm maximizes the log-likelihood defined as:
\begin{align}
\log L(\vector{\lambda}, \vector{b}) = \sum_{i} D_{i} \log \epsilon_{i} - \sum_{i} \epsilon_{i}~.
\end{align}
Note that the bold letters represent the set of parameters, e.g., $\vector{\lambda} = \{\lambda_1, \lambda_2, \cdots\}$ and $\vector{b} = \{b_1, b_2, \cdots\}$.

\section{Modified RL algorithm}
\label{sec_MAP_RL}

This section presents a modified RL algorithm tailored for the COSI data analysis.\footnote{The presented algorithm is implemented in the cosipy library, a high-level analysis software for COSI. The codes are available at \url{https://github.com/cositools/cosipy}.}
Our approach introduces prior distributions to solve the image reconstruction as the maximum a posteriori estimation within a Bayesian framework.
This modification addresses several challenges for COSI and MeV gamma-ray observations: the ineffectiveness of L1 norm regularization, flexible incorporation of pixel-to-pixel correlations such as Total Square Variation \citep[TSV,][]{Rudin1992,Chambolle2004}, and simultaneous background optimization.

First, we introduce Bayesian prior distributions $P_{\mathrm{s}}(\vector{\lambda})$ and $P_{\mathrm{b}}(\vector{b})$ for the image we want to reconstruct and the normalizations of the background models, respectively.
In the framework of the maximum a posteriori estimation, the optimal solution can be defined as the set of parameters $(\vector{\lambda},  \vector{b})$ that maximizes the following log-posterior probability:
\begin{align}
\log Q(\vector{\lambda}, \vector{b}) = \log L(\vector{\lambda}, \vector{b}) + \log P_{\mathrm{s}}(\vector{\lambda}) + \log P_{\mathrm{b}}(\vector{b})~.
\end{align}
Then, using the EM algorithm, we can derive the solution iteratively.
While the E-step remains the same as Equation~\ref{eq_expectation},
the M-step is modified such that it finds $(\vector{\lambda}^{\mathrm{new}},  \vector{b}^{\mathrm{new}})$ that maximizes the following function:
\begin{align}
\begin{split}
&\sum_{ij} \dfrac{D_{i}}{\epsilon_i} R_{ij} \lambda_{j}^{\mathrm{old}} \log \left(R_{ij} \lambda_{j}^{\mathrm{new}}\right)
- \sum_{ij} R_{ij} \lambda_j^{\mathrm{new}} 
+ \log P_{\mathrm{s}}(\vector{\lambda}^{\mathrm{new}}) \\
+& \sum_{ik} \dfrac{D_{i}}{\epsilon_i} B_{ik} b_{k}^{\mathrm{old}} \log \left(B_{ik} b_{k}^{\mathrm{new}}\right)
- \sum_{ik} B_{ik} b_k^{\mathrm{new}}  
+ \log P_{\mathrm{b}}(\vector{b}^{\mathrm{new}})~.
\end{split}
\label{eq_Mstep_orig_function}
\end{align}
The derivation of the above equation is described in Section 3 of \cite{Wang1997}. In the following subsections, we explain how to derive $\vector{\lambda}^{\mathrm{new}}$ and $\vector{b}^{\mathrm{new}}$ separately while introducing the actual formulas for the prior distributions.

\subsection{Source}
\label{sec_rl_source}

Various methods are proposed to find the parameters that maximize Equation~\ref{eq_Mstep_orig_function}
\cite[e.g.,][]{Morii2019,Morii2024}.
In this paper, we follow the approach of \cite{Wang1997}. 
By taking the first derivative of Equation~\ref{eq_Mstep_orig_function} with respect to $\lambda^{\mathrm{new}}_j$ and setting it to zero, the M-step is modified to solving the following equation:
\begin{align}
\frac{\lambda^{\mathrm{old}}_j}{\lambda_j^{\mathrm{new}}} \sum_{i} \dfrac{D_{i}}{\epsilon_i} R_{ij} - \sum_{i} R_{ij} + \frac{\partial \log P_{\mathrm{s}}(\vector{\lambda}^{\mathrm{new}})}{\partial \lambda_{j}} = 0~.
\label{eq_modified_Mstep}
\end{align}
Note that it is identical to Equation~\ref{eq_Mstep} if the prior is constant as typically assumed for the conventional RL algorithm.
Generally, solving this equation explicitly is not straightforward for arbitrary prior distributions. 
In this paper, we separate the prior into a sparse term introduced by \citet{Ikeda2014} and other priors and then solve the above equation approximately. 

\cite{Ikeda2014} pointed out that the L1 norm, commonly used for sparsity in other contexts, is inefficient because the gamma-ray flux corresponds to the probability that a detected gamma-ray originated from a particular pixel and the sum of these probabilities (and also L1) will be conserved.
As an alternative, they introduced a prior distribution, which can be described in our case as:
\begin{align}
\log P_{\mathrm{s}}(\vector{\lambda}) = - \sum_{j} c^{\mathrm{SP}}_{j} \log \lambda_{j},
\end{align}
where $c^{\mathrm{SP}}_{j}$ is the coefficient for the sparse term.
They demonstrated that this prior function can work well as a sparsity term for Poisson-distributed data. This approach has been verified by \citet{Morii2019} using actual X-ray observational data. Given these previous works, we adopt it as our sparsity term.
In our case, this prior distribution corresponds to a special case of the gamma distribution with its scale parameter set to infinity:
\begin{align}
P_{\mathrm{s}}(\vector{\lambda}) = \prod_{j} \frac{\lambda_{j}^{\alpha_{\mathrm{s},j}-1} \exp(-\lambda_{j} / \beta_{\mathrm{s},j})}{\Gamma(\alpha_{\mathrm{s},j}) \beta_{\mathrm{s},j}^{\alpha_{\mathrm{s},j}}} \quad (\alpha_{\mathrm{s},j} = 1 - c^{\mathrm{SP}}_{j}, \beta_{\mathrm{s},j} \to \infty)~,
\label{eq_gamma_sparse}
\end{align}
where $\alpha_{\mathrm{s},j}$ and $\beta_{\mathrm{s},j}$ are shape and scale parameters of the gamma distribution for the flux at pixel $j$, and $\Gamma(x)$ is the gamma function.

The gamma distribution is the conjugate prior to the Poisson distribution \citep{Bishop2006}, and thus Equation~\ref{eq_modified_Mstep} can be solved explicitly if using only this sparsity term \citep{Wang1997}.
With this feature, we first solve Equation~\ref{eq_modified_Mstep} by ignoring other priors and then calculate a correction term.
By separating the prior distribution into the sparse term and other priors, we can express it as follows:
\begin{align}
\log P_{\mathrm{s}}(\vector{\lambda}) = - \sum_{j} c^{\mathrm{SP}}_{j} \log \lambda_{j} + f_{p}\left(\vector{\lambda}\right)~,
\end{align}
where $f_{p}\left(\vector{\lambda}\right)$ is the log probability of the other priors, such as the smoothness prior (TVS), which will be introduced in Section~\ref{sec_Al26}. Then, Equation~\ref{eq_modified_Mstep} is expanded as
\begin{align}
\frac{1}{\lambda_j^{\mathrm{new}}} \left( - c^{\mathrm{SP}}_{j} + \lambda^{\mathrm{old}}_j \sum_{i} \dfrac{D_{i}}{\epsilon_i} R_{ij} \right) 
- \sum_{i} R_{ij} + \frac{\partial f_{p}(\vector{\lambda}^{\mathrm{new}})}{\partial \lambda_{j}} = 0~.
\end{align}

To solve the above, we adopt a perturbation approach: we solve the equation by ignoring the last term and derive a correction term afterward. Let $\lambda^{\mathrm{EM}}_j$ be the solution when the last term is absent \citep[see][]{Ikeda2014}. Then, we introduce the correction term $\Delta \omega_j$ and describe the solution of Equation~\ref{eq_modified_Mstep} as
\begin{align}
\lambda_j^{\mathrm{new}} = \lambda^{\mathrm{EM}}_j \exp( \Delta \omega_j)~,
\label{eq_model_EM_w_correction_term}
\end{align}
where
\begin{align}
\lambda^{\mathrm{EM}}_j = \mathrm{max}\left(
\dfrac{1}{\displaystyle \sum_{i} R_{ij}}
\left(- c^{\mathrm{SP}}_{j}
+ \lambda^{\mathrm{old}}_j \displaystyle \sum_{i} \dfrac{D_{i}}{\epsilon_i} R_{ij} \right), 0\right)~.
\label{eq_model_EM}
\end{align}
In the first-order approximation, 
the correction term can be determined from the following equation:
\begin{align}
\Delta \omega_j &= \dfrac{1}{\displaystyle \sum_{i} R_{ij} } \times 
\biggl(\dfrac{\partial f_{p}(\vector{\lambda}^{\mathrm{EM}})}{\partial \lambda_{j}} 
+ \sum_{j'} \dfrac{\partial^2 f_{p}(\vector{\lambda}^{\mathrm{EM}})}{\partial \lambda_{j} \partial \lambda_{j'}}  \lambda^{\mathrm{EM}}_{j'} \Delta \omega_{j'}\biggr)~.
\label{eq_Mstep_correction_term}
\end{align}

Since Equation~\ref{eq_Mstep_correction_term} includes a Hessian matrix,
it requires resolving a multidimensional simultaneous system of equations, which is computationally expensive. Thus, we neglect the Hessian term to simplify the solution process and reduce computational cost. This simplification finally leads us to the following equation.
\begin{align}
\Delta \omega_j \simeq \dfrac{1}{\displaystyle \sum_{i} R_{ij}} \times \dfrac{\partial f_{p}(\vector{\lambda}^{\mathrm{EM}})}{\partial \lambda_{j}}~.
\label{eq_correction_term}
\end{align}
Given that the second term in Equation~\ref{eq_Mstep_correction_term} should be negligible compared to the first term, our solution can be valid when the following condition is satisfied.
\begin{align}
\left| \dfrac{ \displaystyle \sum_{j'} \dfrac{1}{\sum_{i} R_{ij'}} \dfrac{\partial^2 f_{p}(\vector{\lambda})}{\partial \lambda_{j} \partial \lambda_{j'}} \dfrac{\partial f_{p}(\vector{\lambda})}{\partial \lambda_{j'}} \lambda_{j'} }{ \dfrac{\partial f_{p}(\vector{\lambda})}{\partial \lambda_{j}}} \right| \ll 1~.
\label{eq_approx_condition}
\end{align}
This condition does not pose significant problems in most cases relevant to this paper. We discuss more details of this point in Appendix~\ref{sec_condition_approx_Mstep}.

\subsection{Background}

Similar to the approach we took for the source distribution, the normalization for each background component can be obtained by solving the following equation:
\begin{eqnarray}
\label{eq_Mstep_bkg}
\frac{b^{\mathrm{old}}_k}{b_k^{\mathrm{new}}} \sum_{i} \dfrac{D_{i}}{\epsilon_i} B_{ik} - \sum_{i} B_{ik} + \frac{\partial \log P_{\mathrm{b}}(\vector{b}^{\mathrm{new}})}{\partial b_{k}} = 0~.
\end{eqnarray}

We assume that the prior knowledge for the background normalizations is informed by initial values and associated uncertainties derived by preliminary background modeling.
To incorporate this information while solving the above equation explicitly, we again introduce the gamma distribution as the prior distribution:
\begin{align}
P_{\mathrm{b}}(\vector{b}) = \prod_{k} \frac{b_{k}^{\alpha_{\mathrm{b},k}-1} \exp(-b_{k} / \beta_{\mathrm{b},k})}{\Gamma(\alpha_{\mathrm{b},k}) \beta_{\mathrm{b},k}^{\alpha_{\mathrm{b},k}}}~,
\end{align}
where $\alpha_{\mathrm{b},k}$ and $\beta_{\mathrm{b},k}$ are the shape and scale parameters for each background component $k$. These parameters can be determined from the preliminary background estimation results, considering that
its mean and variance are $\alpha_{\mathrm{b},k}~\beta_{\mathrm{b},k}$ and $\alpha_{\mathrm{b},k}~\beta_{\mathrm{b},k}^2$, respectively \citep{Bishop2006}.
For example, if the background normalization is derived as $1$ with a 1\% accuracy as 1 $\sigma$ level, we can set $\alpha_{\mathrm{b},k}$ and $\beta_{\mathrm{b},k}$ as $10^4$ and $10^{-4}$, respectively.
On the other hand, we can use $(\alpha_{\mathrm{b},k}, \beta_{\mathrm{b},k}) = (1, \infty)$ to use a flat distribution when we do not have such prior information.
Since the gamma distribution is the conjugate prior to the Poisson distribution, Equation~\ref{eq_Mstep_bkg} can be solved explicitly as:
\begin{align}
b_k^{\mathrm{new}} = \dfrac{\alpha_{\mathrm{b},k}-1 + b^{\mathrm{old}}_{k} \displaystyle \sum_{i} \dfrac{D_{i}}{\epsilon_i} B_{ik}}{\displaystyle \sum_{i} B_{ik} + \frac{1}{\beta_{\mathrm{b},k}}}~.
\label{eq_RL_bkg}
\end{align}

\subsection{Summary of the proposed algorithm}

Summarizing the above, we update the source distribution and background normalization simultaneously for each iteration as follows.
First, we calculate the expected counts in the CDS as shown in Equation~\ref{eq_expectation}.
Then, for the source distribution, we calculate the intermediate solution as Equation~\ref{eq_model_EM} and then apply the correction term following Equations~\ref{eq_model_EM_w_correction_term} and \ref{eq_correction_term}.
The background normalization can be updated simply as Equation~\ref{eq_RL_bkg}.
Then, we can maximize the log-posterior probability $\log Q(\vector{\lambda}, \vector{b})$ iteratively.
It continues until a certain convergence criterion is satisfied, such as when the increase of the log-posterior probability falls below a threshold, or the maximum number of iterations is reached.

\section{Simulation setup for algorithm tests}
\label{sec_simulation}

To evaluate the effectiveness of our proposed algorithm for the upcoming COSI observations, we apply it to simulation data spanning three months. We have prepared three datasets with distinct spatial features: $^{44}$Ti (point source), $^{26}$Al (smooth distribution over the Galactic disk), and e$^{+}$/e$^{-}$ (combination of sparse and smooth components). These datasets allow us to test different aspects of our algorithm: the sparseness prior ($^{44}$Ti), the smoothness prior ($^{26}$Al), and the combination of multiple priors (e$^{+}$/e$^{-}$). Additionally, we conduct background simulations assuming the conditions of the COSI satellite orbit and generate a response matrix using large-scale simulations. The following sections provide details on each source model, the background simulations, and the response matrix generation process.

\subsection{Source model}

For our source simulations, we employ the Medium-Energy Gamma-ray Astronomy library (MEGAlib), a comprehensive software package designed for gamma-ray detectors specializing in Compton telescopes \citep{MEGAlib}. Our simulated observations span three months, assuming an equatorial orbit at an altitude of 550 km with zenith pointing, considering Earth occultation.
After performing the Monte Carlo simulations, we generated a binned histogram in the CDS for each source distribution within the Galactic coordinate system. 
In the following, we briefly describe the source models for $^{44}$Ti, $^{26}$Al, and 0.511 MeV emission.\footnote{These datasets are publicly available as a part of the second data challenge of the COSI project \citep{MC2023}. The details are described at \url{https://github.com/cositools/cosi-data-challenge-2}.}

\subsubsection{$^{44}$Ti}

$^{44}$Ti is primarily produced during the alpha-rich freeze-out phase in the explosion mechanisms of core-collapse supernovae and Type Ia supernovae \citep{Thielemann1990,Nomoto2006,Maeda2010}. It has a lifetime of about 85 years, decaying to $^{44}$Sc, which subsequently decays to $^{44}$Ca with a lifetime of approximately 3.9 hours, producing a gamma-ray line at 1.157 MeV \citep{Ahmad2006}. Due to its relatively short lifetime, $^{44}$Ti can be a tracer for young supernova remnants in our Galaxy, typically those with ages up to a few hundred years.

To investigate the behavior of the sparse prior under high background events,
we model the $^{44}$Ti distribution as a point source at Cas A \citep{Grefenstette2014}.
Gamma-ray and X-ray lines from $^{44}$Ti were firmly detected by previous observations \citep{Iyudin1997,Grefenstette2014,Siegert2015}, and thus Cas A is the primary target for COSI in the early phase of its operation.
While the flux of $^{44}$Ti is slightly different among the previous results, we adopted the flux from the INTEGRAL/SPI analysis \citep{Siegert2015}, which is $3.5 \times 10^{-5}~\mathrm{ph~cm^{-2}~s^{-1}}$.
We note that while this flux is detectable in our simulated three-month observation, the line sensitivity of COSI is expected to improve significantly with longer observation times. For instance, over its planned 2-year mission, COSI will achieve approximately a few $\times 10^{-6}~\mathrm{ph~cm^{-2}~s^{-1}}$, allowing for studying fainter $^{44}$Ti sources.

\subsubsection{$^{26}$Al}

$^{26}$Al is a long-lived radioactive isotope with a lifetime of about 1 Myr, decaying to $^{26}$Mg via $\beta^+$ decay and producing a 1.809 MeV gamma-ray line \citep{Norris1983}. It is generated and injected into our Galaxy through stellar winds and supernova explosions \citep{Diehl2006}. COMPTEL provided the first flux map of $^{26}$Al emission across the Galactic plane \citep{Diehl1995}. While localized hotspots may exist, the overall $^{26}$Al distribution is relatively smooth over the Galactic plane due to its long lifetime, tracing the cumulative effects of stellar nucleosynthesis and Galactic chemical evolution over timescales $\gtrsim$ 1 Myr.

For the algorithm test, we modeled $^{26}$Al as a smooth spatial distribution. It is represented as a doubly exponential disk model in our Galaxy as a three-dimensional model, given by:
$$
\rho(R,z,\phi) = \frac{L}{4\pi R_s^2 z_s} \exp(-R/R_s) \exp(-|z|/z_s)~,
$$
where $L$ is the total luminosity of $^{26}$Al in our Galaxy, i.e., $L = \frac{M}{m} \times \frac{p}{\tau}$, with $m = 26$u the atomic mass of $^{26}$Al nuclei, $p = 0.9976$ the branching ratio to emit a photon, and $\tau = 1.05$ Myr the lifetime of $^{26}$Al. This gives a quasi-persistent luminosity for a living $^{26}$Al mass $M$. The values $R_s$ and $z_s$ are the scale radius and scale height, respectively. The radial coordinate is given by $R = \sqrt{(x-x_0)^2 + (y-y_0)^2}$, and $z$ is the vertical coordinate, $z = z' - z_0$, where $x_0 = 8.178$, $y_0 = 0$, and $z_0 = -0.019$ are the coordinates of the Galactic center seen from Earth. All distance and size units are in kpc. The line of sight integration is performed so that the flux per pixel with 3 deg resolution is in units of $\mathrm{ph/cm^{2}/s/sr}$. We adopt $M = 6$ $M_\odot$, $R_s = 5.0$, and $z_s = 1.0$, which results in a total flux of $1.8 \times 10^{-3}~\mathrm{ph/cm^{2}/s}$, compatible with previous observations \citep[e.g.,][]{Pleintinger2023}.

\subsubsection{e$^{+}$/e$^{-}$}

The 0.511 MeV gamma-ray line, produced by the annihilation of positrons with electrons, has been detected by various missions over five decades \citep[see][for review]{Siegert2023}. Despite extensive observations, including recent balloon observations \citep{Kierans2020,Siegert2020}, the origin of these positrons remains unclear. However, it is known that positron annihilation occurs prominently close to the Galactic center region, with a flux comparable to that of the Galactic disk 0.511 MeV emission. While the exact spatial distribution of the 0.511 MeV emission is still under investigation, several models have been proposed. For our simulation, we implement two different models derived from INTEGRAL/SPI analysis: \cite{Skinner2015} and \cite{Siegert2016}.

\cite{Skinner2015} proposed a model consisting of a point source at the Galactic center, two Gaussian components for the Galactic bulge emission slightly offset from the center, and a Galactic disk emission with a scale height of 3 degrees, explains the SPI data the best.
In contrast, \cite{Siegert2016} claims that the Galactic disk has a larger scale height of 10.5 degrees vertically and a longitudinal extent of 60 degrees. 
We refer to the former and latter as the thin and thick disk models, respectively.
The two different spatial distributions are derived from different data treatments. This originated because it is challenging to distinguish a spatially broad feature from the instrumental backgrounds with the coded aperture mask instruments adopted by SPI.
Since these models have several spatial components featuring both sparse and smooth characteristics, we can use them as test cases to evaluate the performance of the proposed algorithm.
We summarized the parameters for these 0.511 MeV spatial models in Table~\ref{tab_511keV_models}.
Additionally, based on \cite{Siegert2016}, the line shape of 0.511 MeV emission is assumed to be a Gaussian with FWHM of 2 keV and 3 keV for the bulge and disk components, respectively.
Note that the intrinsic line width is ignored for the $^{44}$Ti and $^{26}$Al cases.

\begin{table}
\centering
\caption{Spatial distribution model parameters for the 0.511 MeV emission.}
\label{tab_511keV_models}
\begin{tabular}{lccccc}
\hline
Component & \multicolumn{2}{c}{$\sigma$} & \multicolumn{2}{c}{Centre} & Flux $\times 10^{-4}$ \\
 & $l$ ($^\circ$) & $b$ ($^\circ$) & $l$ ($^\circ$) & $b$ ($^\circ$) & (ph cm$^{-2}$ s$^{-1}$) \\
\hline
Disk (Thin) & 90 & 3 & 0 & 0 & 17.0 \\
Disk (Thick) & 60 & 10.5 & 0 & 0 & 17.0 \\
\hline
Broad bulge & 8.7 & 8.7 & 0 & 0 & 7.3 \\
Narrow bulge & 2.5 & 2.5 & $-$1.25 & $-$0.25 & 2.8 \\
Central Point & 0 & 0 & 0 & 0 & 1.2 \\
\hline
\end{tabular}
\end{table}

\subsection{Background simulations}

For the background simulations, we employed MEGAlib to simulate three months of instrumental and astrophysical backgrounds, assuming the same orbit as the source simulations.
The background components include albedo emission, extragalactic gamma-ray background, and instrumental backgrounds caused by cosmic rays bombarding the instrument. The instrumental backgrounds, which have both prompt and delayed components, are mainly due to primary protons, primary alpha particles, atmospheric neutrons, primary electrons, primary positrons, and secondary protons. The input spectra are based on the model from \cite{Cumani2019}, with considering the changing geomagnetic cutoff along the orbit. We note that charged particles trapped in the South Atlantic Anomaly (SAA) are not included in our current simulations.\footnote{SAA particles could cause non-negligible background events, which will be addressed in future work. The simulation includes the veto shield by the BGO detectors. More details about the background simulations can be seen at \url{https://github.com/cositools/cosi-data-challenge-2/tree/main/backgrounds}.}

After simulations, we produced a binned histogram by adding all components, applied Gaussian smoothing on the $\psi\chi$ plane in the CDS to mitigate the Poisson fluctuation in the background model, and created the total background model in the CDS histogram. 
Background events were then re-sampled in the CDS according to the Poisson distribution to simulate data.
For simplicity in this study, we assumed a single background component, and the proposed algorithm optimized its normalization parameter.
In this paper, we focus on evaluating the performance of the RL algorithm under this idealized condition, while background modeling typically involves multiple components and potential model uncertainties in real observations.
Addressing these complexities is beyond the scope of the current study, but the optimization of multiple background components and the handling of model uncertainties could potentially lead to further refinements in our data analysis. We discuss these aspects in more detail in Section~\ref{sec_discussion}.
The number of extracted events for source and background of each target is shown in Table~\ref{tab_num_events}.

In our implementation, the background is modeled in the CDS. While some astrophysical backgrounds, such as extragalactic gamma-ray background, could theoretically be modeled in the model space through the response matrix $R_{ij}$ like source components, we choose to incorporate all backgrounds as the background template matrix $B_{ik}$ for algorithmic consistency. This approach simplifies the implementation by treating all non-source events uniformly and also conceptually separates the signal of interest from all other contributions. In principle, an astrophysical background component that is initially representable through $R_{ij}$ can be converted into the $B_{ik}$ representation by pre-convolving it with the response. Therefore, from the algorithm's perspective, the distinction between different background types is not necessary, and all background events are treated equivalently as matrices in the CDS.

\begin{table}
\centering
\caption{Number of events filled in a histogram for the three-month simulation of each source distribution.}
\label{tab_num_events}
\begin{tabular}{cccc}
\hline
Target & Source (S) & Background (B) & S/B (\%)\\
\hline
$^{44}$Ti & $1.77 \times 10^{3}$ & $4.92 \times 10^{5}$ & 0.36 \\
$^{26}$Al & $1.00 \times 10^{5}$ & $2.26 \times 10^{5}$ & 44 \\
e$^{+}$/e$^{-}$ & $2.29 \times 10^{5}$ & $2.09 \times 10^{6}$ & 11 \\
\hline
\end{tabular}
\tablefoot{The bin size of each axis in the event histogram is the same as shown in Table~\ref{tab_response_matrix}. Here, the total event counts in the CDS are compared. However, for a known point source, it is also important to consider the number of background events that overlap with the source event distribution in CDS. This is because point source events are distributed in a cone shape in CDS, and primarily, only the nearby background events affect the analysis.
When counting only the background events close to the source distribution for Cas A, the S/B ratio is about 3--5\% in our $^{44}$Ti simulation.} 
\end{table}

\subsection{Response matrix generation}

\begin{figure}
\centering
    \includegraphics[width = 0.9 \linewidth]{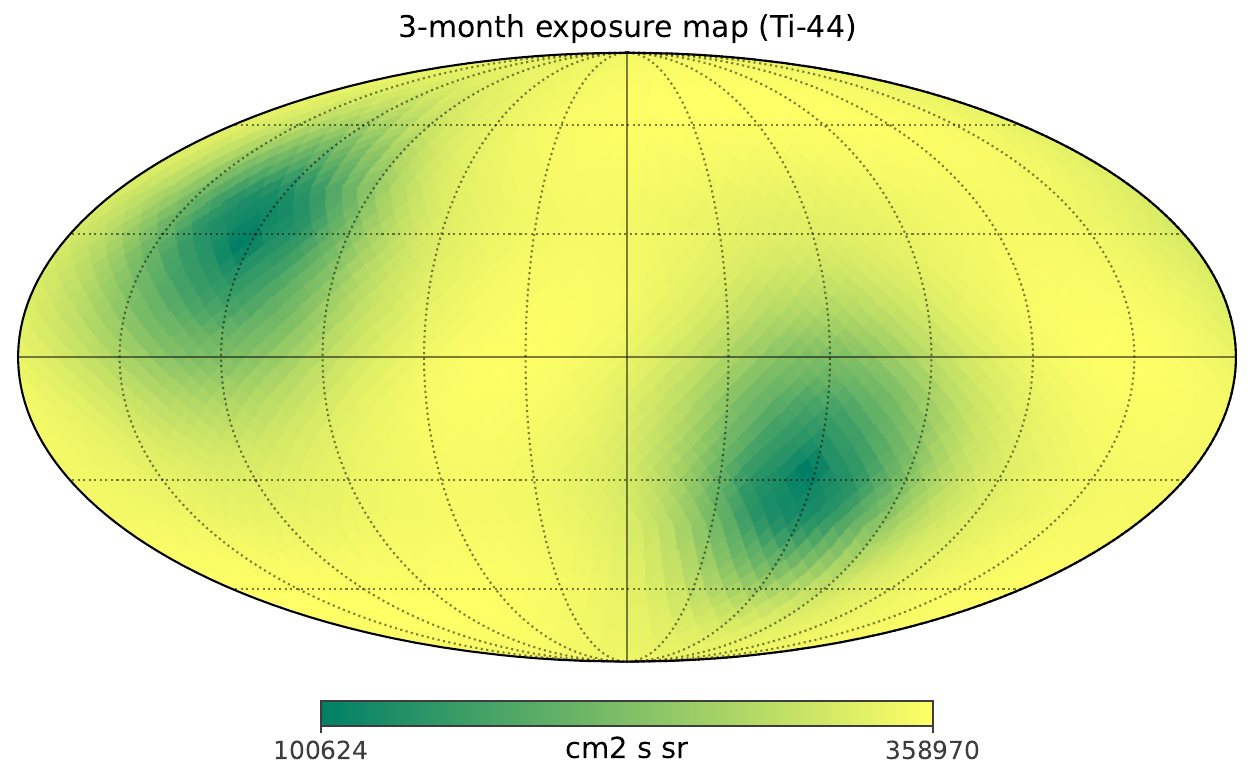}
    \includegraphics[width = 0.9 \linewidth]{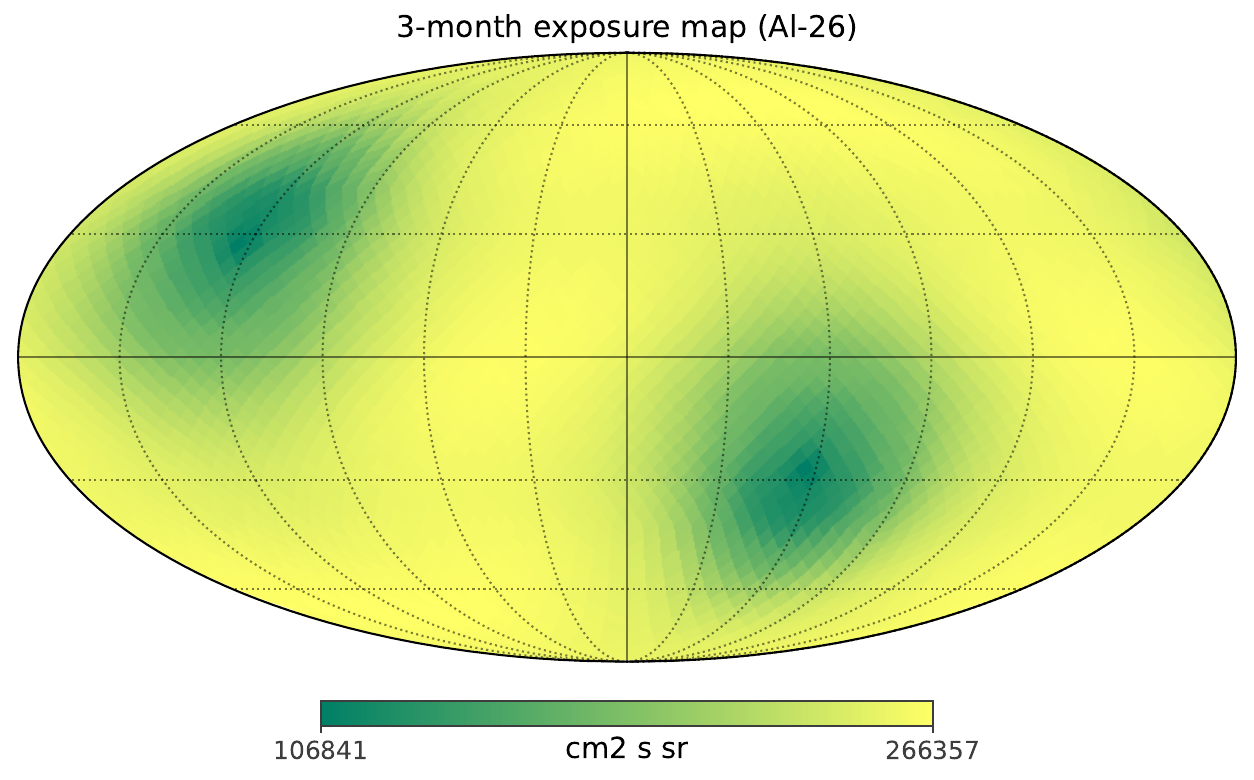}
    \includegraphics[width = 0.9 \linewidth]{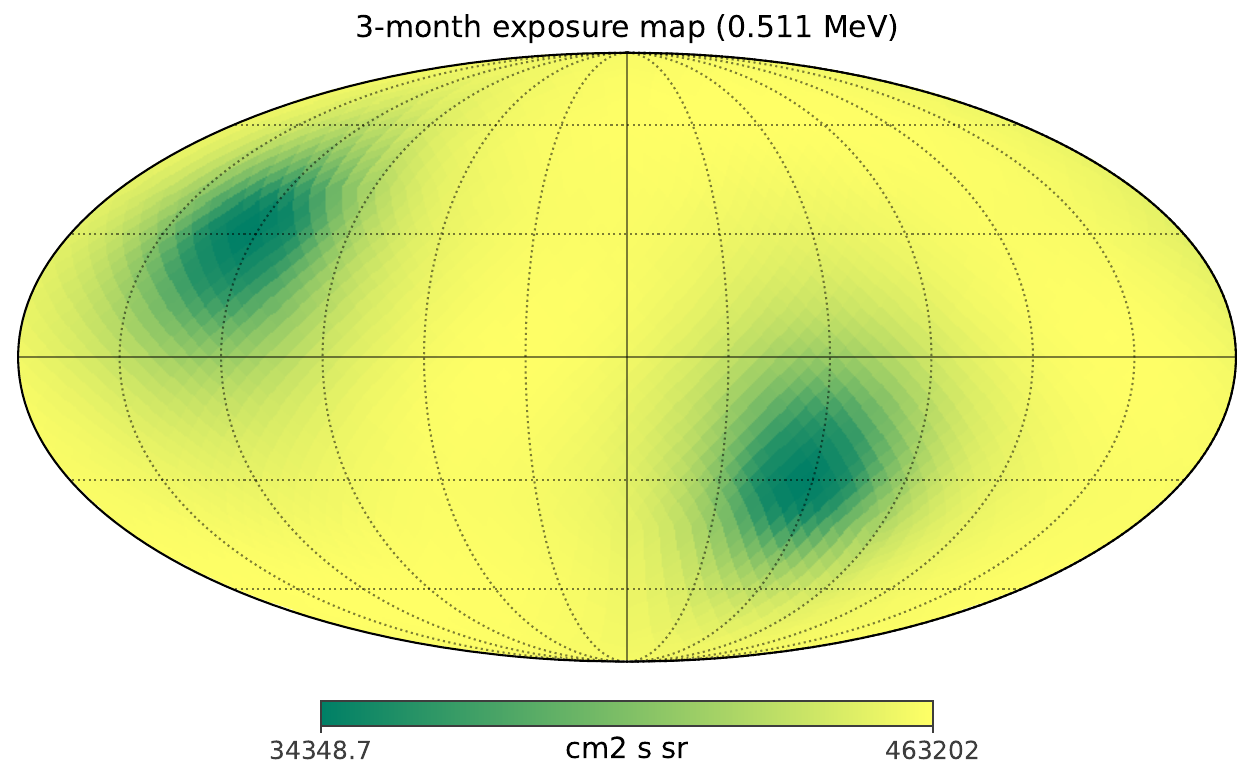}
\caption{Exposure maps for the three targets with three-month observations in Galactic coordinates. The exposures for $^{44}$Ti, $^{26}$Al, and 0.511 MeV are shown from top to bottom. The values represent the effective area integrated over both the observation time and the solid angle of each pixel. The solid angle per pixel is $4.1 \times 10^{-3}$ sr, and the number of pixels is 3072. The range of the color bar is different in each case. Note that the color bars of the images shown in this paper are in the log scale.} 
\label{fig_exposure_map}
\end{figure}

The response matrix for our analysis was generated using large-scale simulations with MEGAlib. We simulated more than $10^{12}$ events uniformly distributed over 4$\pi$ steradians with the same initial gamma-ray energy as each source distribution and created a response matrix in the detector coordinates.
Then, we integrated the response matrix over time while weighting it with the time-dependent satellite attitude and Earth occultation effects over the three months, and produced the final response matrix in Galactic coordinates.
The number of bins for each element in the response matrix is shown in Table~\ref{tab_response_matrix}.
The actual numbers of simulated events are also presented in this table.
For the Galactic sky and $\psi\chi$ plane, we used HEALPix pixelization \citep{Healpix}.
Figure~\ref{fig_exposure_map} shows the effective area convolved with the exposure time for each case.
Two regions with relatively short exposure times can be observed, which is because we assumed zenith pointing for simplicity in this simulation. COSI is planned to observe the sky by pointing 22$^\circ$ north, then south, of zenith with approximately a 12-hour period, which will result in a more uniform exposure map.

\begin{table*}
\centering
\caption{Binning of the response matrix for each target.}
\label{tab_response_matrix}
\begin{tabular}{ccccccc}
\hline
Target & $E_{\mathrm{i}}$ & $E_{\mathrm{m}}$ & Galactic Sky & $\phi$ & $\psi\chi$ & num. of simulated events\\
 & (MeV) & (MeV) & (nside, bins) & (deg, bins) & (nside, bins) \\
\hline
$^{44}$Ti & 1.157 & 1.150--1.164 & 16, 3072 & 0--180, 60 & 16, 3072 & $6.82 \times 10^{12}$\\
$^{26}$Al & 1.809 & 1.805--1.812 & - & - & - & $8.74 \times 10^{12}$
\\
e$^{+}$/e$^{-}$ & 0.511 & 0.509--0.513 & - & - & - & $5.39 \times 10^{12}$
\\
\hline
\end{tabular}
\end{table*}

\section{Results}
\label{sec_results}

Here, we present the results of applying our algorithm to the three science objectives. We will first evaluate the effect of the sparseness term using the $^{44}$Ti simulation and then test the smoothness term with $^{26}$Al. Finally, we will analyze the combined effect of both terms using the 0.511 MeV dataset.

The stopping criterion for the RL algorithm was set to when the increase in the log-posterior probability becomes less than $10^{-2}$.
We performed our data analysis on a workstation equipped with an Intel Core i9-10980XE processor and 256 GB memory.
To accelerate computational time related to the response matrix, we utilized an NVIDIA A6000 GPU and the CuPy library \citep{cupy_learningsys2017}. 
Typically, we were able to execute about 20 iterations per second. While the total computation time and number of iterations varied depending on the prior distribution parameters, image reconstruction for a single parameter set was computed within 1-3 minutes.

\subsection{$^{44}$Ti: a point source $+$ a sparsity prior}

To evaluate the sparse imaging analysis, we set the prior distribution of the source as the following equation:
\begin{align}
   \log P_{\mathrm{s}}(\vector{\lambda}) = - c^{\mathrm{SP}} \sum_{j} \log \lambda_{j}~.
\end{align}
Here we used the same coefficient $c^{\mathrm{SP}}$ for all of the pixels and varied it from $10^{-4}$ to $3 \times 10^{-1}$. For the background model, we initially assumed a flat distribution ($\alpha_{\mathrm{b}} = 1$ and $\beta_{\mathrm{b}} = \infty$), which results in Equation~\ref{eq_RL_bkg} becoming the conventional RL algorithm for the background part. The effect of the background optimization will be discussed later.

\subsubsection{Reconstructed image and flux}

\begin{figure*}
    \centering

    \begin{subfigure}{0.33\textwidth}
      \includegraphics[width=\linewidth]{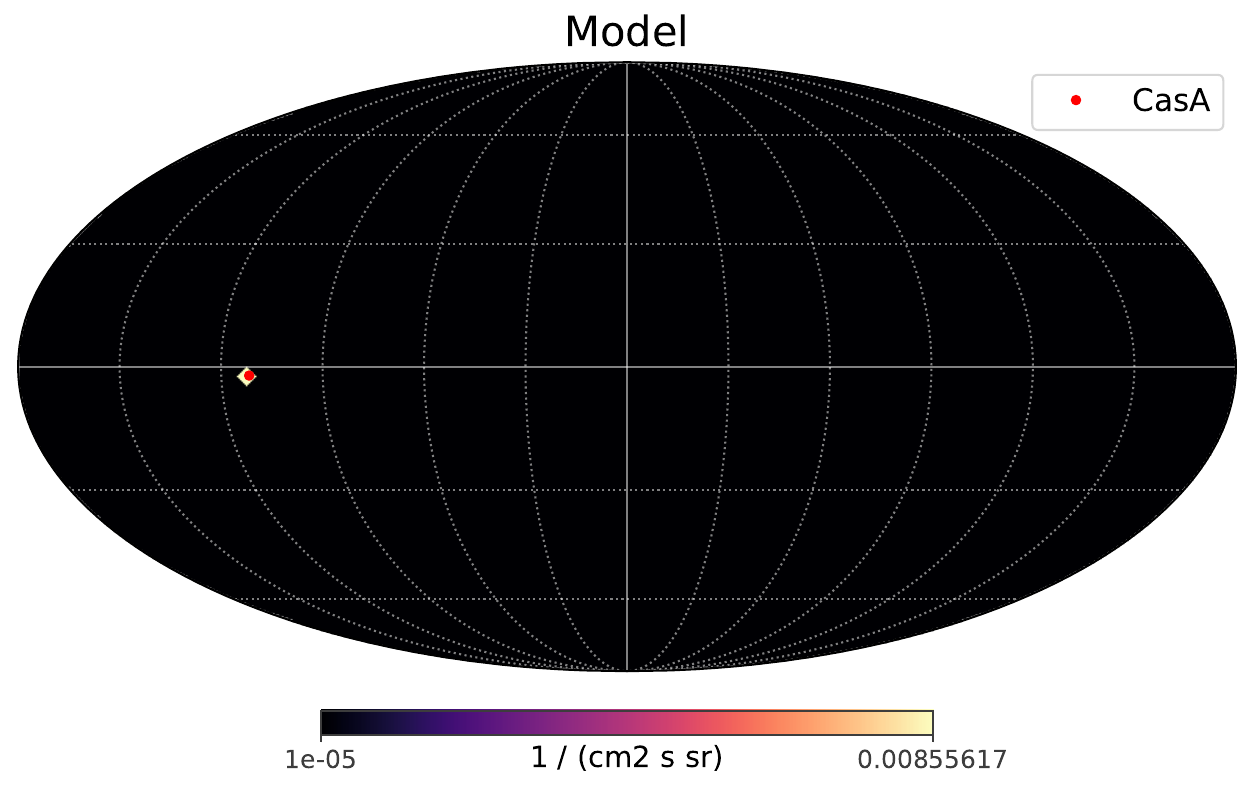}
    \end{subfigure}
    \begin{subfigure}{0.33\textwidth}
      \includegraphics[width=\linewidth]{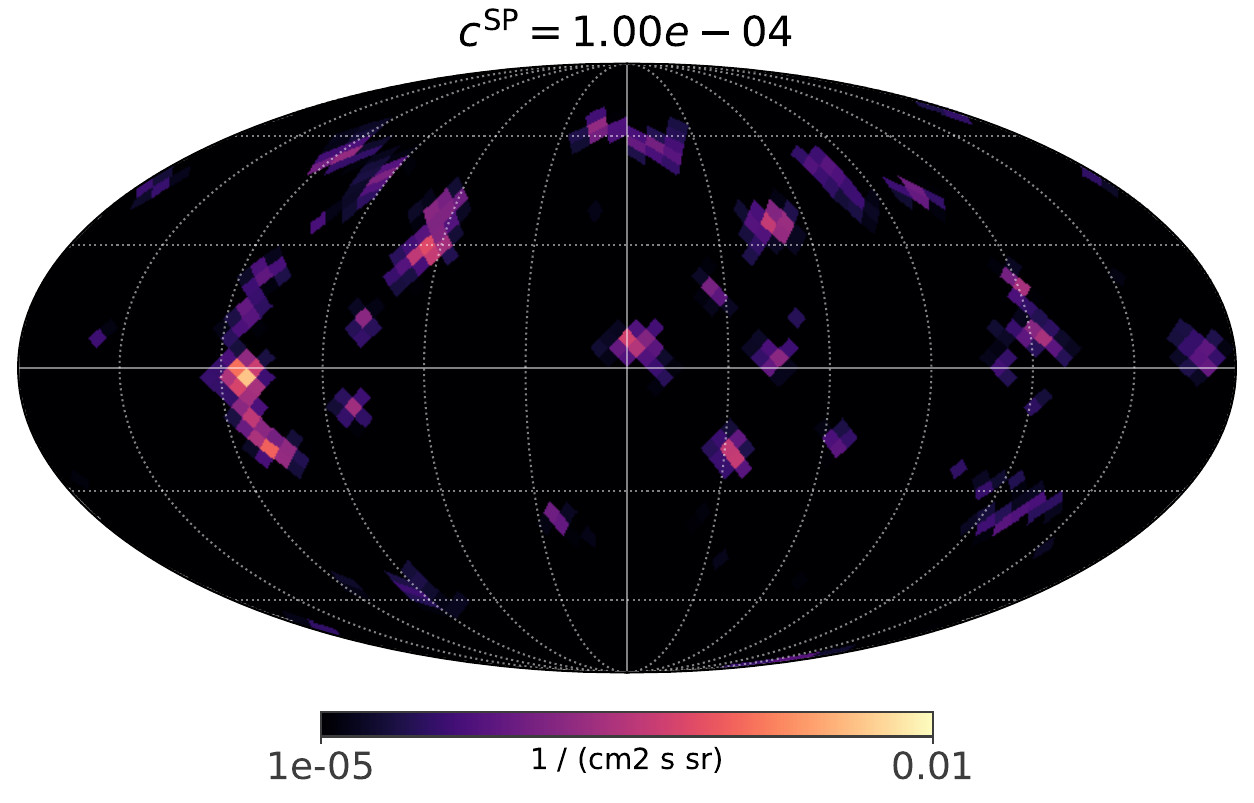}
    \end{subfigure}
    \begin{subfigure}{0.33\textwidth}
      \includegraphics[width=\linewidth]{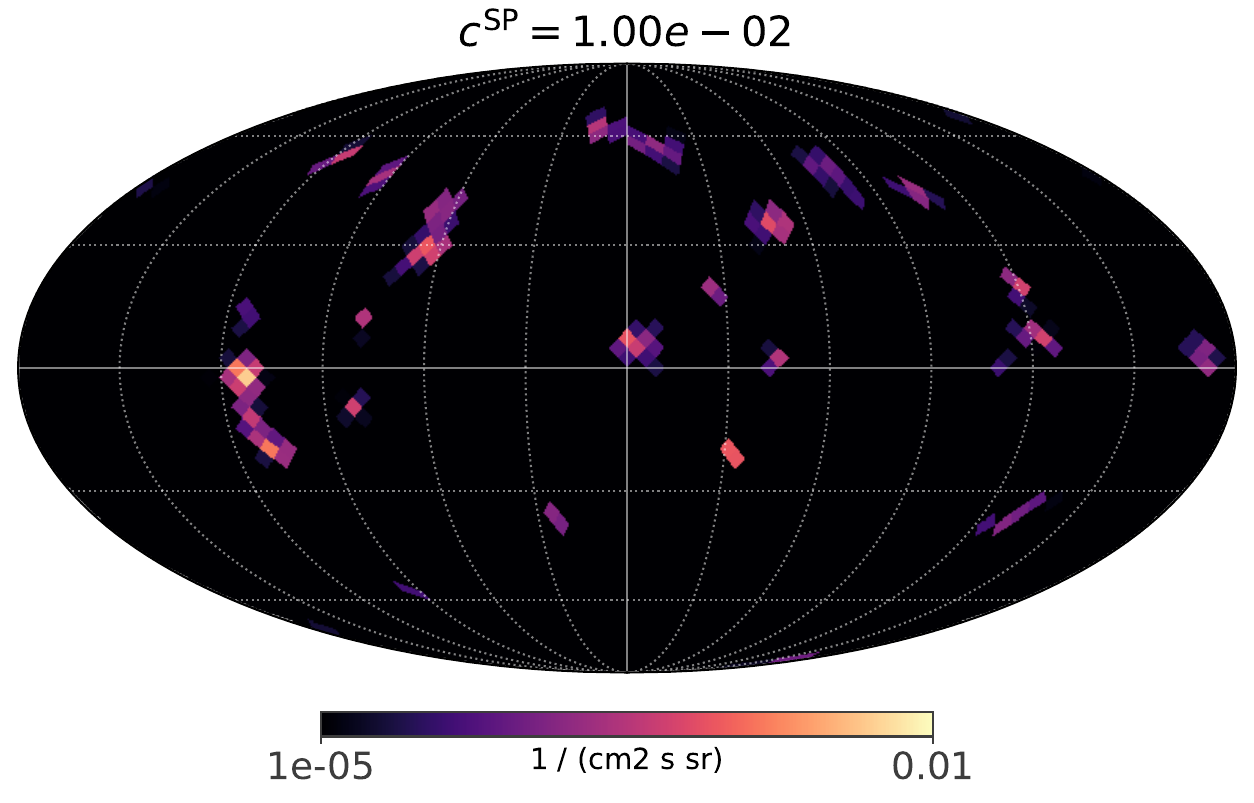}
    \end{subfigure}
    
    \begin{subfigure}{0.33\textwidth}
      \includegraphics[width=\linewidth]{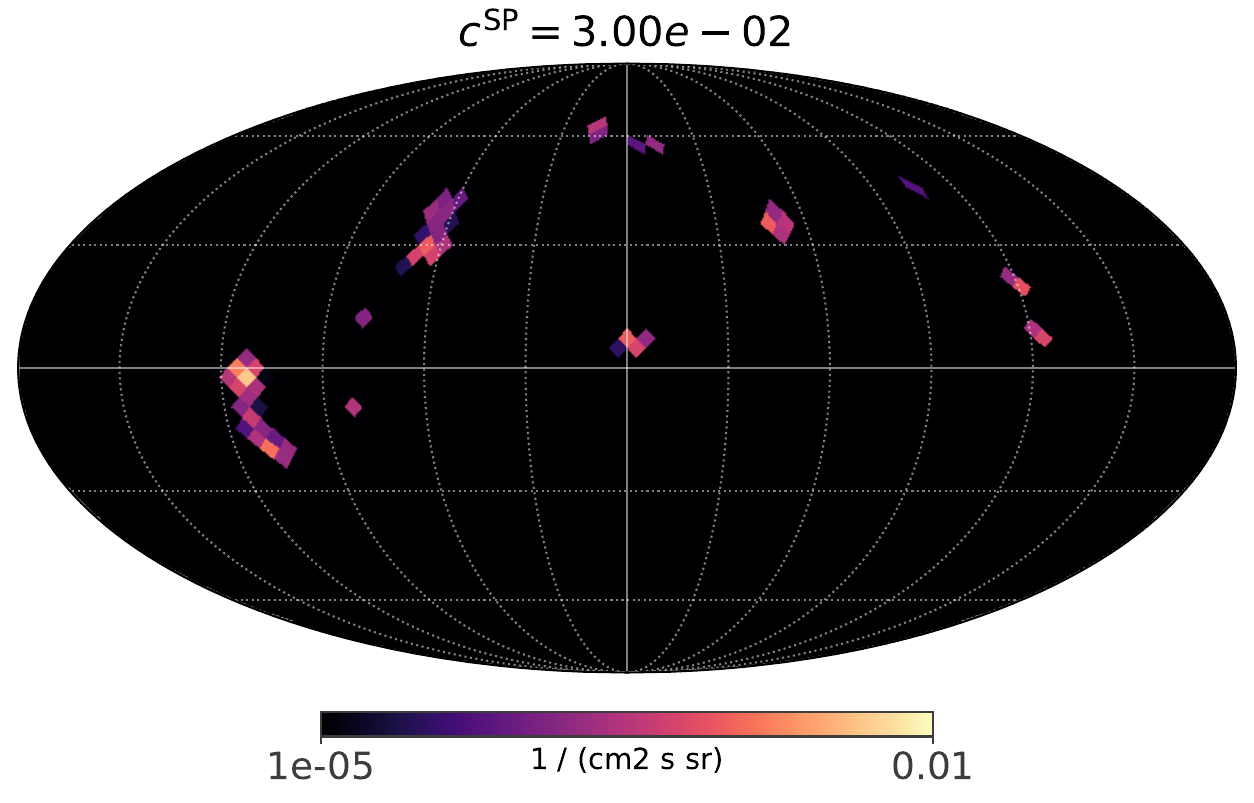}
    \end{subfigure}
    \begin{subfigure}{0.33\textwidth}
      \includegraphics[width=\linewidth]{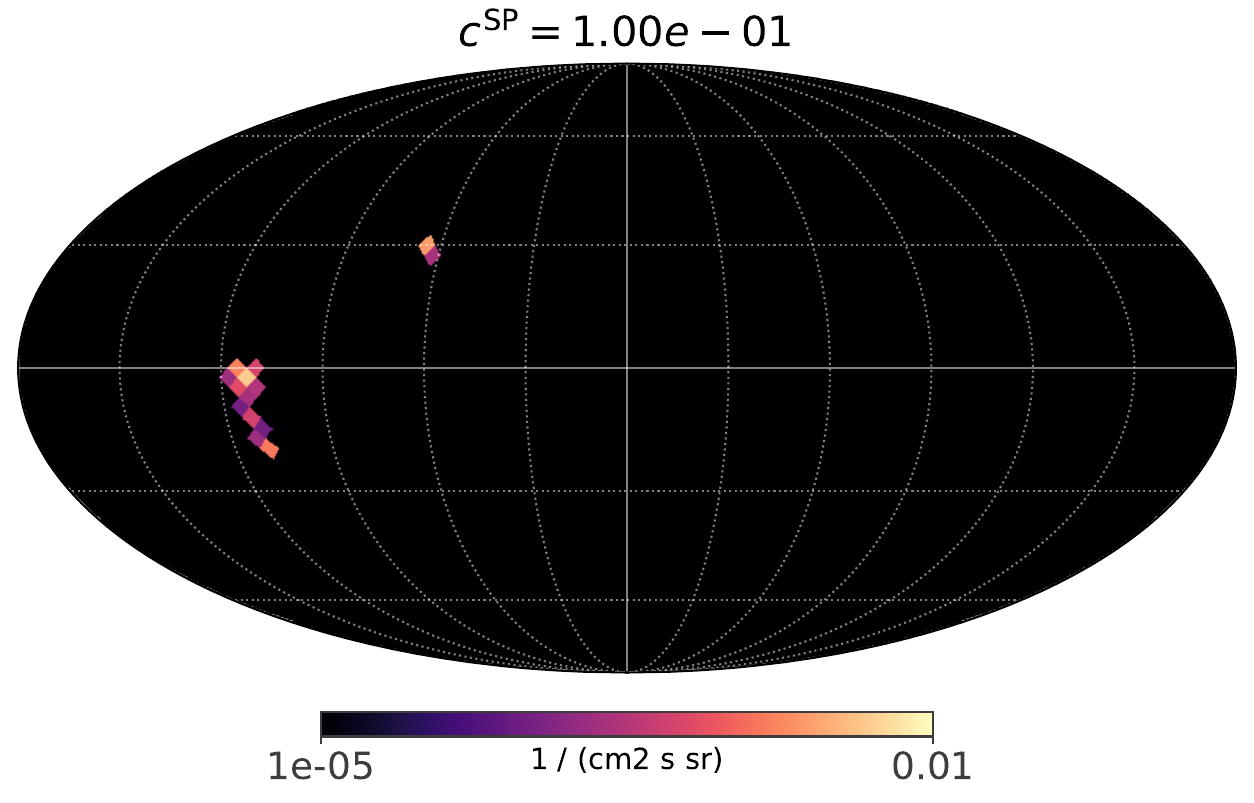}
    \end{subfigure}
    \begin{subfigure}{0.33\textwidth}
      \includegraphics[width=\linewidth]{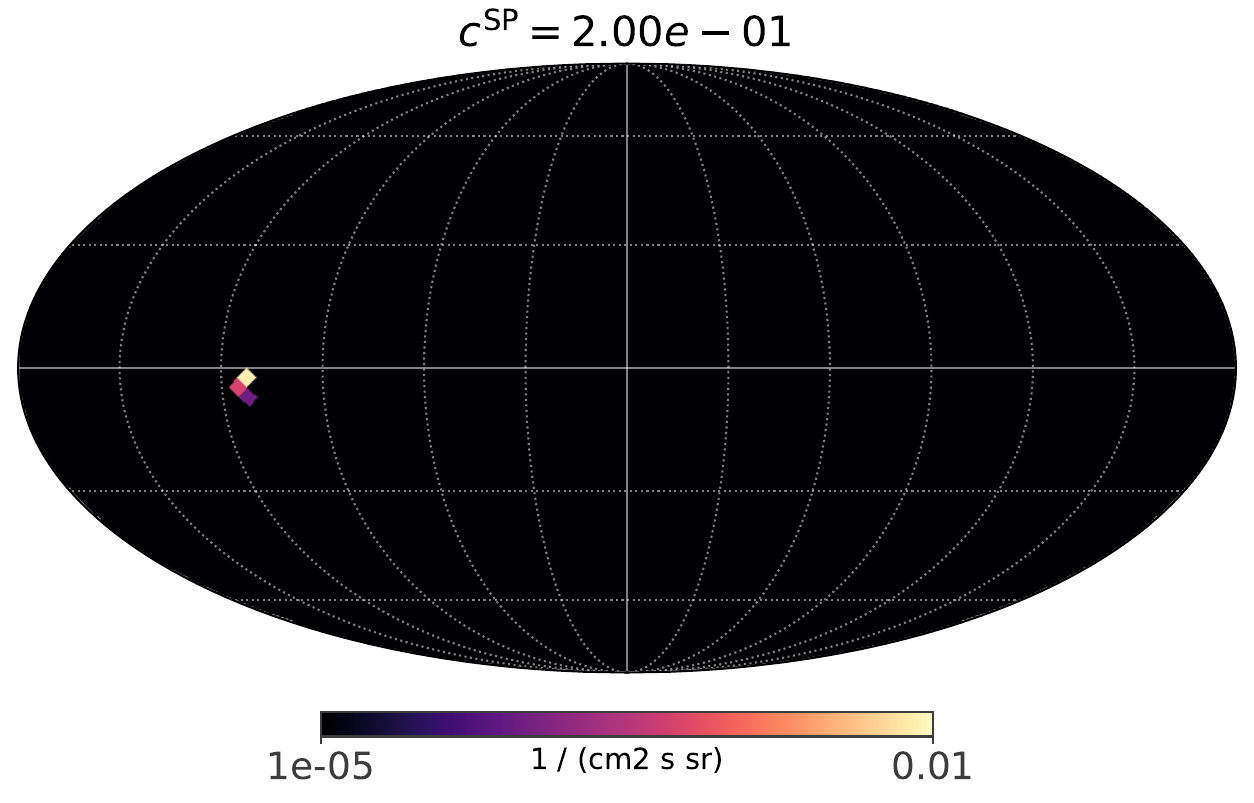}
    \end{subfigure}

    \caption{Reconstructed images of $^{44}$Ti with different values for $c^{\mathrm{SP}}$. The left top panel shows the injected model for $^{44}$Ti.
    From the next to bottom right, $c^{\mathrm{SP}}$ is set to $10^{-4}$, $10^{-2}$, $3 \times 10^{-2}$, $1 \times 10^{-1}$, and $2 \times 10^{-1}$.
    }
    \label{fig_Ti44_images}
\end{figure*}

\begin{figure}
\centering
    \includegraphics[width = 0.95 \linewidth]{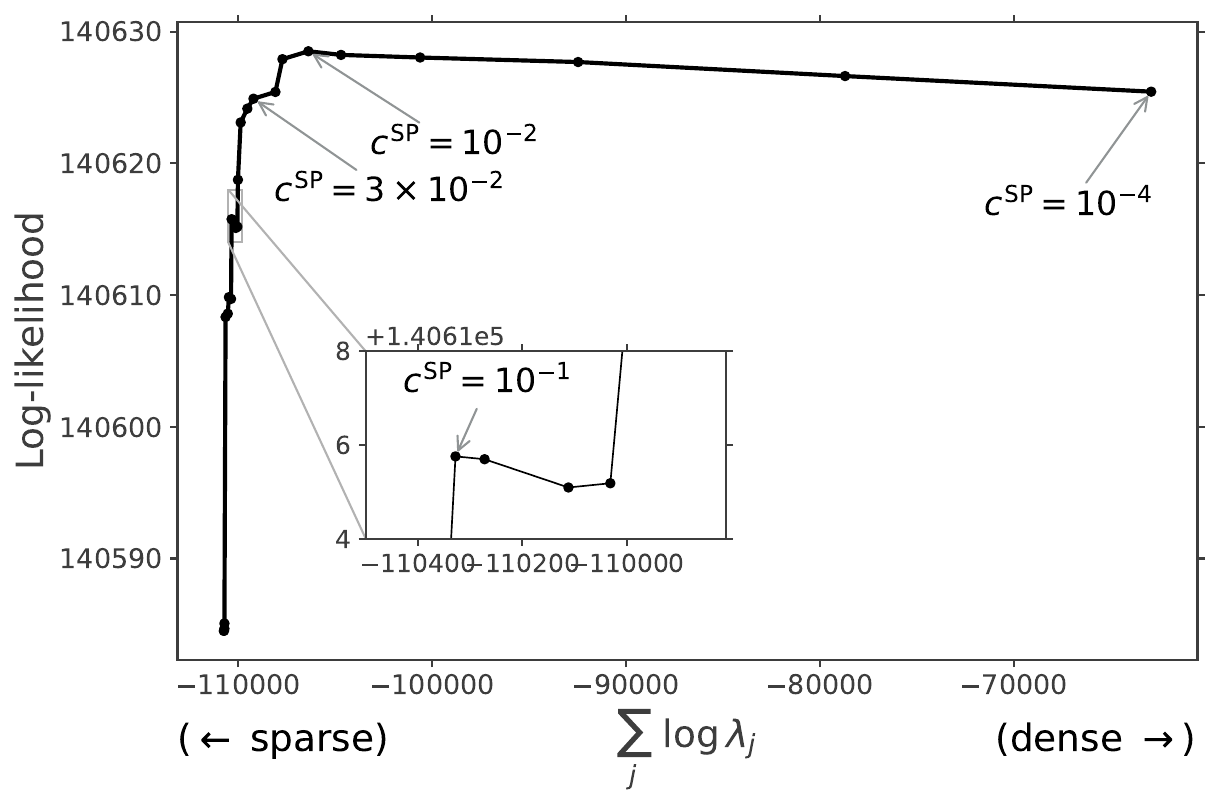}
\caption{L-curve for the sparse prior. The x-axis shows the sparseness defined as $\sum_j \log \lambda_{j}$, while the y-axis shows the resulting log-likelihood.}
\label{fig_graph_Ti44_lcurve}
\end{figure}

\begin{figure}
\centering
    \includegraphics[width = 0.9 \linewidth]{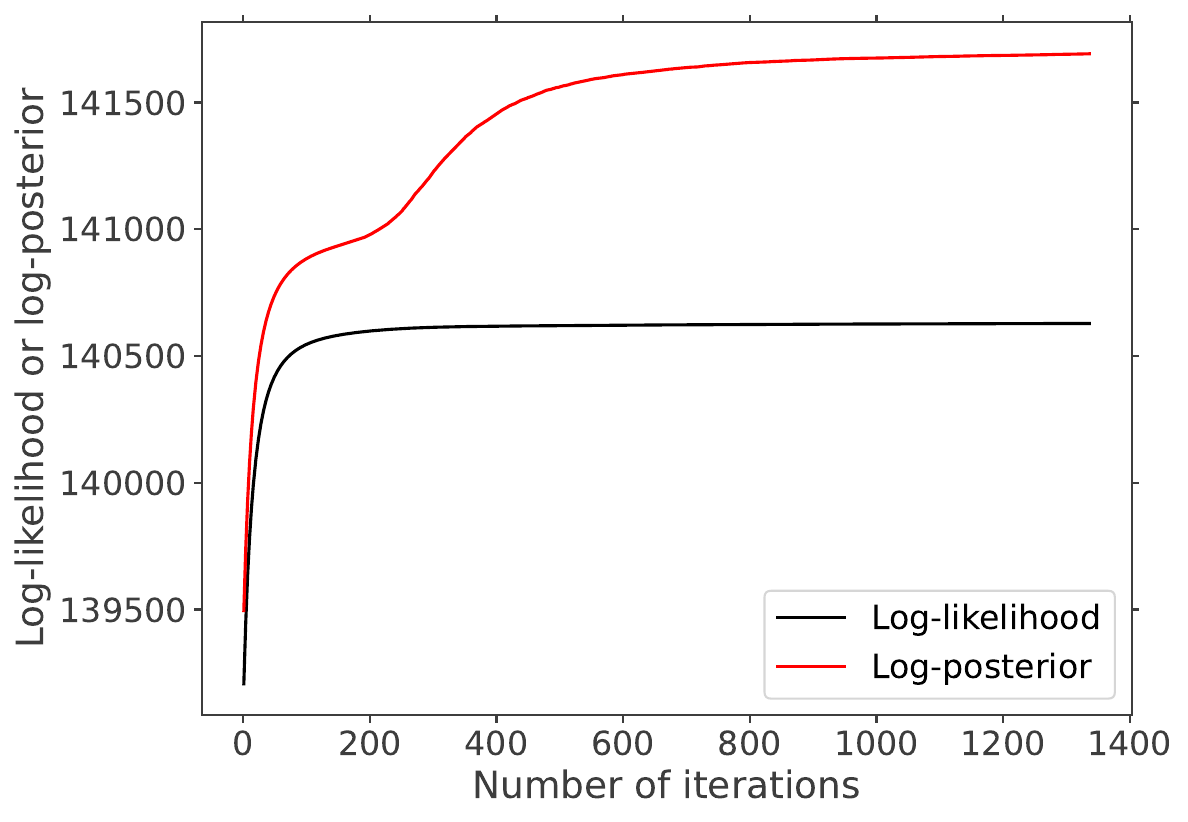}
\caption{Log-posterior and log-likelihood for each iteration. Here, $c^{\mathrm{SP}}$ is set to $10^{-2}$.}
\label{fig_graph_Ti44_loglikelihood_curve}
\end{figure}

The reconstructed images are shown in Figure~\ref{fig_Ti44_images}. As $c^{\mathrm{SP}}$ increases, the diffuse structure around point sources is gradually suppressed. We can see that, at $c^{\mathrm{SP}} \approx 2 \times 10^{-1}$, the diffuse structure around point sources disappears completely. At this point, only the point source remains at the location of the injected source.
When $c^{\mathrm{SP}}$ is increased more, the image flux finally becomes zero.
We can investigate this feature more by plotting the sparse term against the log-likelihood, as shown in Figure~\ref{fig_graph_Ti44_lcurve}. 
The log-likelihood initially increases from $c^{\mathrm{SP}} = 10^{-4}$ to $c^{\mathrm{SP}} = 10^{-2}$ as the diffuse structure around the point source is suppressed. 
Then, above $c^{\mathrm{SP}} \approx 10^{-2}$, the log-likelihood begins to drop. It reaches some local minima, for instance, around $c^{\mathrm{SP}} = 10^{-1}$.
As $c^{\mathrm{SP}}$ is increased, point-source artifacts disappear, as seen from $c^{\mathrm{SP}} = 3\times10^{-2}$ to $c^{\mathrm{SP}} = 1\times10^{-1}$ in Figure~\ref{fig_Ti44_images}.
The discrete change in the image corresponds to the local minimum observed in the log-likelihood curve.
We also show the log-likelihood and log-posterior distribution values for each iteration at $c^{\mathrm{SP}} = 10^{-2}$ in Figure~\ref{fig_graph_Ti44_loglikelihood_curve}.
This plot indicates that the algorithm optimizes the log-likelihood and then maximizes the sparse term.

\begin{figure}
\centering
    \includegraphics[width = 0.95 \linewidth]{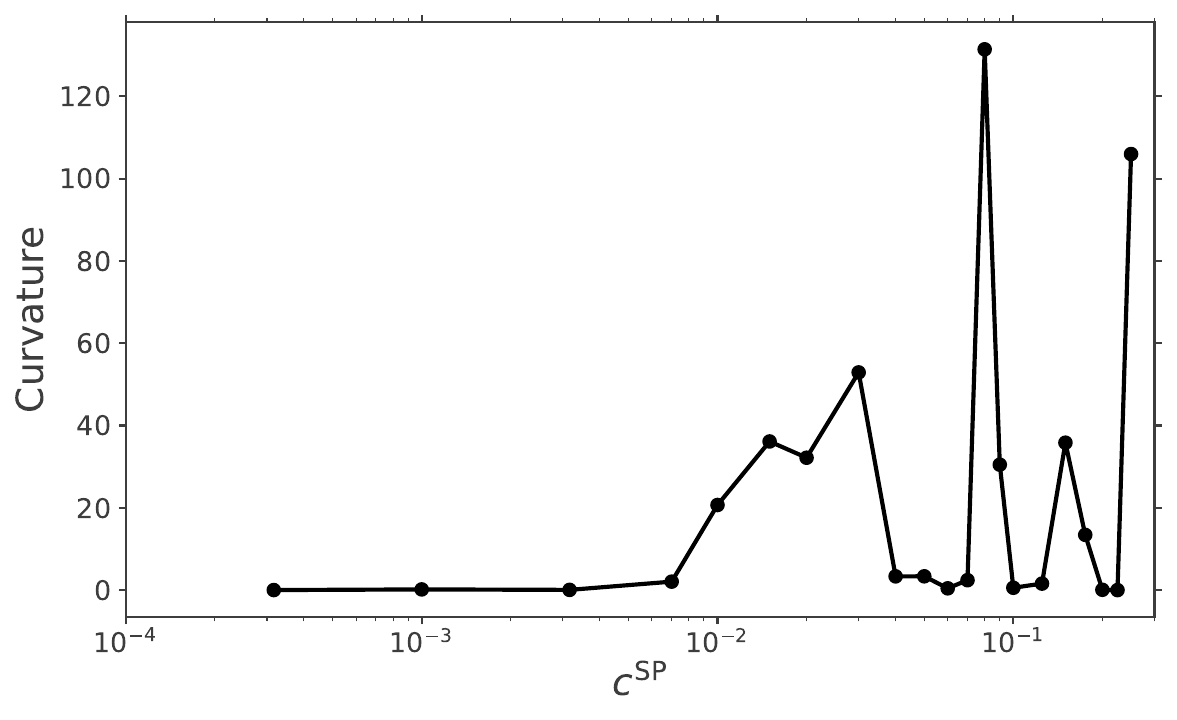}
\caption{Curvature of the L-curve shown in Figure~\ref{fig_graph_Ti44_lcurve} plotted against $c^{\mathrm{SP}}$.}
\label{fig_curvature_Ti44}
\end{figure}

\begin{figure}
\centering
    \includegraphics[width = 0.9 \linewidth]{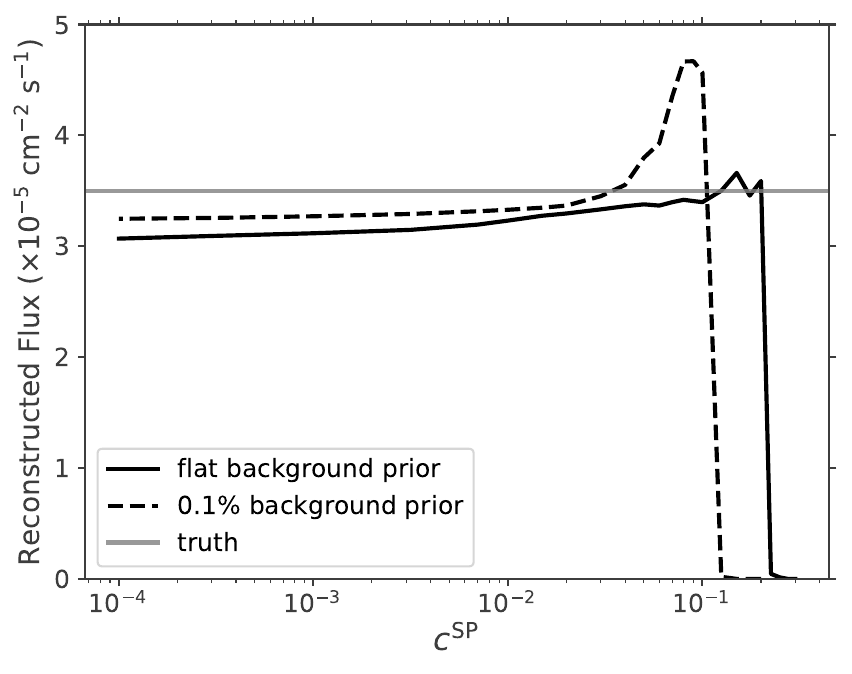}
\caption{Reconstructed flux compared with the simulated one.
The solid line corresponds to using the flat prior distribution for the background normalization, while the dashed line is for the prior distribution with 0.1\% accuracy.}
\label{fig_Ti44_reconstructed_flux}
\end{figure}

\begin{figure}
\centering
    \includegraphics[width = 0.9 \linewidth]{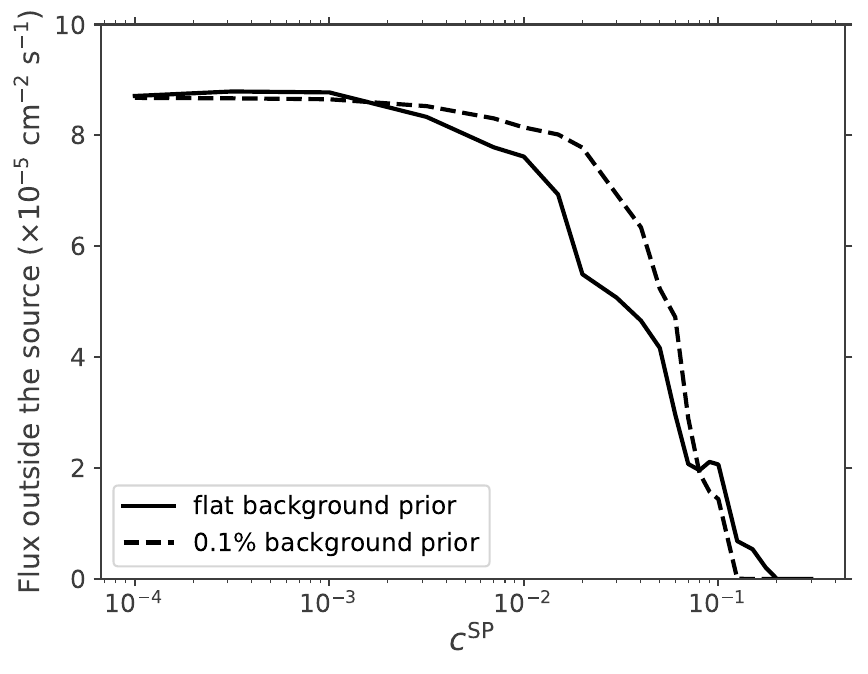}
\caption{Flux outside the source.}
\label{fig_Ti44_darkregion_flux}
\end{figure}

Figure~\ref{fig_Ti44_reconstructed_flux} shows the reconstructed flux depending on different $c^{\mathrm{SP}}$.
Here, the reconstructed flux of the source is defined as the sum of flux values in a pixel at the source location and its adjacent 8 pixels. 
In this case, the source flux was reconstructed within 3--10\%, depending on $c^{\mathrm{SP}}$.
Furthermore, we show the reconstructed flux outside the source, which should be ideally zero but is bound to be positive due to the Poisson fluctuations (Figure~\ref{fig_Ti44_darkregion_flux}).
With increasing $c^{\mathrm{SP}}$, the flux outside the source becomes smaller, and finally almost zero after $c^{\mathrm{SP}} \sim 10^{-1}$.
Notably, the flux outside the source is not negligible when $c^{\mathrm{SP}}$ is small.
It can be a few $\times 10^{-5}~\mathrm{ph~cm^{-2}~s^{-1}}$, which is of the same order as the injected source flux.
In this case, the Poisson fluctuation of the background events ($\sim 10^3$) is comparable to the number of photon events collected from the source (see Table~\ref{tab_num_events}).
Then, the conventional RL algorithm assigns some of the fluctuations to gamma-ray emission, which leads to a significant flux outside the source location.
With a time-integrated effective area of about $10^{8}$ cm$^{2}$ s, the background causes an uncertainty of $O(10^{-5})$ ph cm$^{-2}$ s$^{-1}$.
Introducing the sparseness prior can mitigate such a problem by suppressing the artifacts, resulting in better flux estimation.

\subsubsection{L-curve method}

It should be noted that determining the optimal $c^{\mathrm{SP}}$ is not always straightforward and may depend on various factors, including specific scientific goals, the nature of the dataset, and the particular features of interest in the analysis.
While several methods are proposed to determine the optimal prior distribution, here we adopt the L-curve method by following \cite{Allain2006}.
In this method, an optimal solution can be derived as the maximum curvature point in a graph between statistics and penalty values, corresponding to Figure~\ref{fig_graph_Ti44_lcurve}.
Note that the graph in the figure takes an inverted L shape since the log-likelihood is plotted on the y-axis.
When applying the L-curve method, we normalized both the x-axis and y-axis to have the same range.
This method suggests $c^{\mathrm{SP}} \approx 10^{-1}$ as the optimal value (see Figure~\ref{fig_curvature_Ti44}), but it is likely affected by the sudden change in the L-curve due to several local maximum curvature as seen in Figure~\ref{fig_graph_Ti44_lcurve}.
Considering the global trend of the L-curve and the maximum value of the likelihood, 
$c^{\mathrm{SP}} \approx 10^{-2}$ would also be a candidate for the solution.
In this particular case, it would be necessary to determine the optimal solution by combining other methods, such as visual inspection and cross-validation, according to a specific purpose, e.g., the flux from a specific region, the flux from the entire region, or the identification of the location of a known point source, etc.
We will further discuss this point in Section~\ref{sec_discussion}.

\subsubsection{Background normalization}

We also investigated the effect of different prior distributions on the background model normalization. In addition to the flat distribution, we used the gamma distribution, assuming a precise determination of the normalization parameter with 0.1\% accuracy. Specifically, we set $(\alpha_{\mathrm{b}}, \beta_{\mathrm{b}}) = (10^6, 10^{-6})$. 
This value was chosen to be comparable to the Poisson fluctuation of the number of background events.

\begin{figure}
\centering
    \includegraphics[width = 0.9 \linewidth]{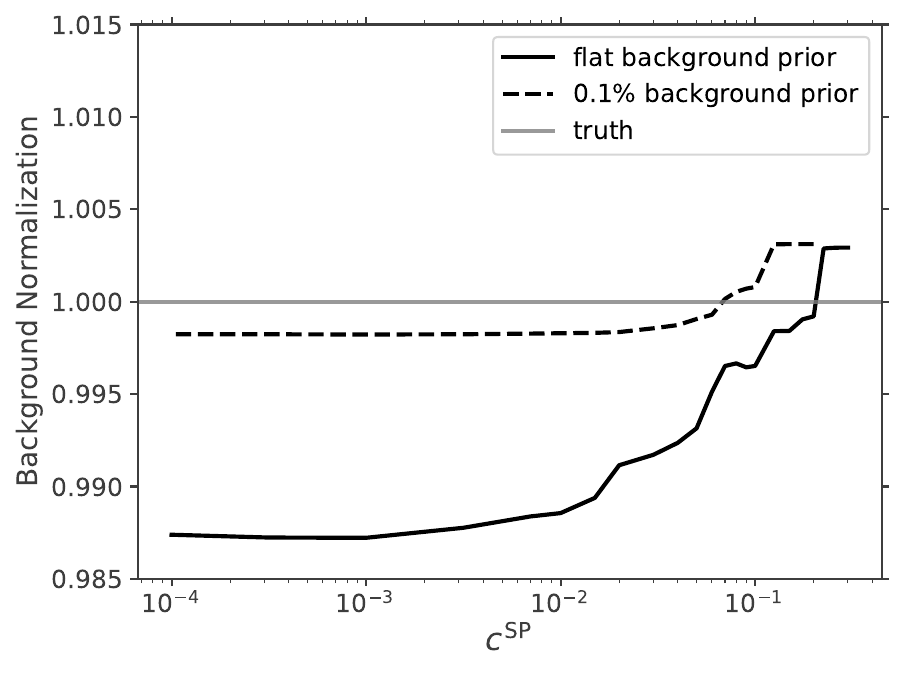}
\caption{Optimization of the background normalization.}
\label{fig_Ti44_bkg_norm}
\end{figure}

We added the resulting fluxes from the source and outside its region as the dashed lines in Figures~\ref{fig_Ti44_reconstructed_flux} and \ref{fig_Ti44_darkregion_flux}.
It can be seen that the flux from the source region is better reconstructed, especially around $c^{\mathrm{SP}} \approx 10^{-2}$, where the likelihood reaches the maximum in both cases.
Note that the flux becomes larger after that, which likely corresponds to overfitting, and then sharply drops to zero since the sparsity prior became too strong.
We also show the optimized background normalization in Figure~\ref{fig_Ti44_bkg_norm}.
While the background normalization converges well in both cases, it converges closer to 1 with the 0.1\% prior distribution, indicating the strong influence of the prior information on the estimation.
Although it might be unlikely that the background normalization is determined within 0.1\% accuracy in advance,
this approach can, in principle, incorporate such background information, potentially improving the estimation of both the source and background.
In real observations, 
such a constraint on the background would be helpful when the number of detected events is small, like transient event analysis. Also, it may be possible to optimize the background splitting into different components using detector information.
We will discuss it again in Section~\ref{sec_discussion}.

\subsection{$^{26}$Al: a diffuse emission $+$ a smoothness prior}
\label{sec_Al26}

Next, we evaluate the performance of the algorithm on smooth, extended sources.
Here, we applied our method to the simulated $^{26}$Al data. 
For this case, we introduced TSV as a smoothness prior.
It is described as follows:
\begin{align}
\log P_{\mathrm{s}}(\vector{\lambda}) = - c^{\mathrm{TSV}} \sum_j \sum_{j' \in \sigma_{j}} (\lambda_{j} - \lambda_{j'})^2~,
\end{align}
where $c^{\mathrm{TSV}}$ is the coefficient controlling the strength of the smoothness constraint, and $\sigma_j$ represents the set of indices for pixels adjacent to pixel $j$.
We varied $c^{\mathrm{TSV}}$ from $10^4$ to $10^7$ to examine its effect on the reconstructed images. The background model was treated the same as in the $^{44}$Ti case, assuming a flat distribution.

\subsubsection{Reconstructed image}
\label{subsubsec_Al26_image}

\begin{figure*}
    \centering

    \begin{subfigure}{0.33\textwidth}
      \includegraphics[width=\linewidth]{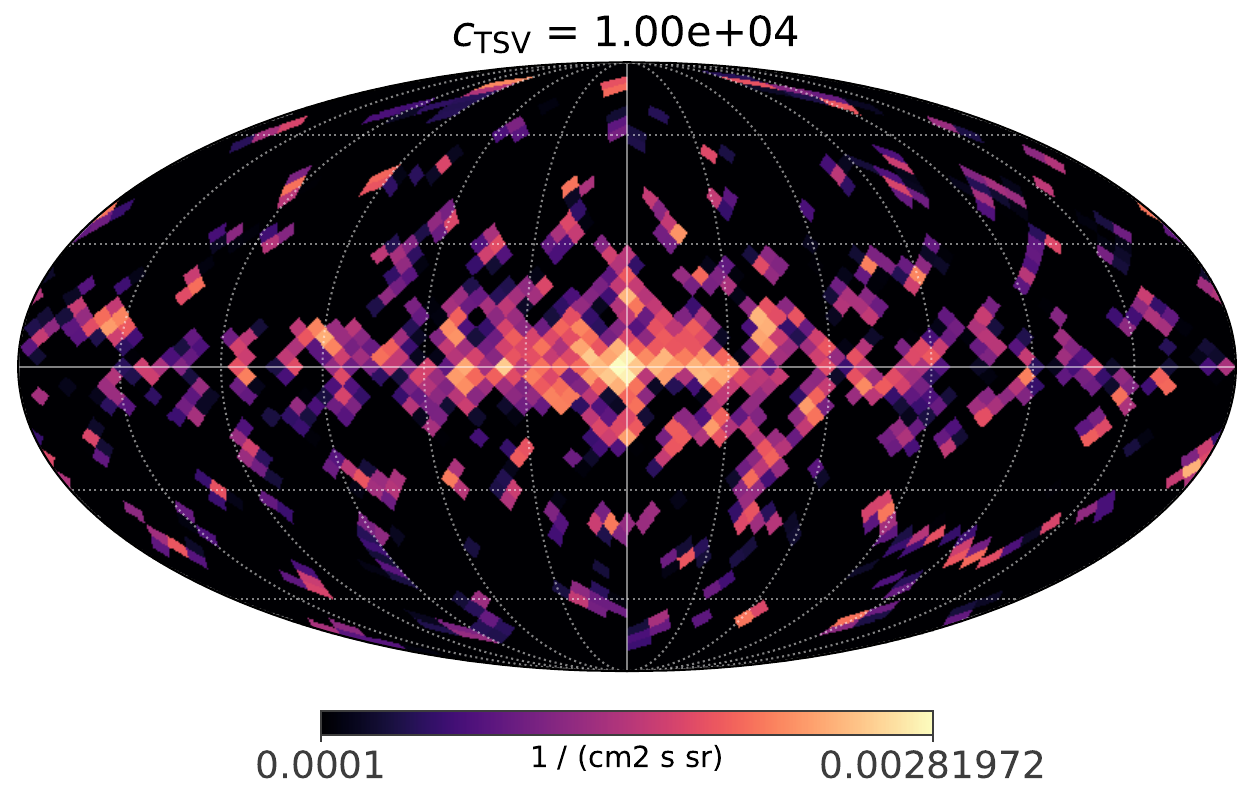}
    \end{subfigure}
    \begin{subfigure}{0.33\textwidth}
      \includegraphics[width=\linewidth]{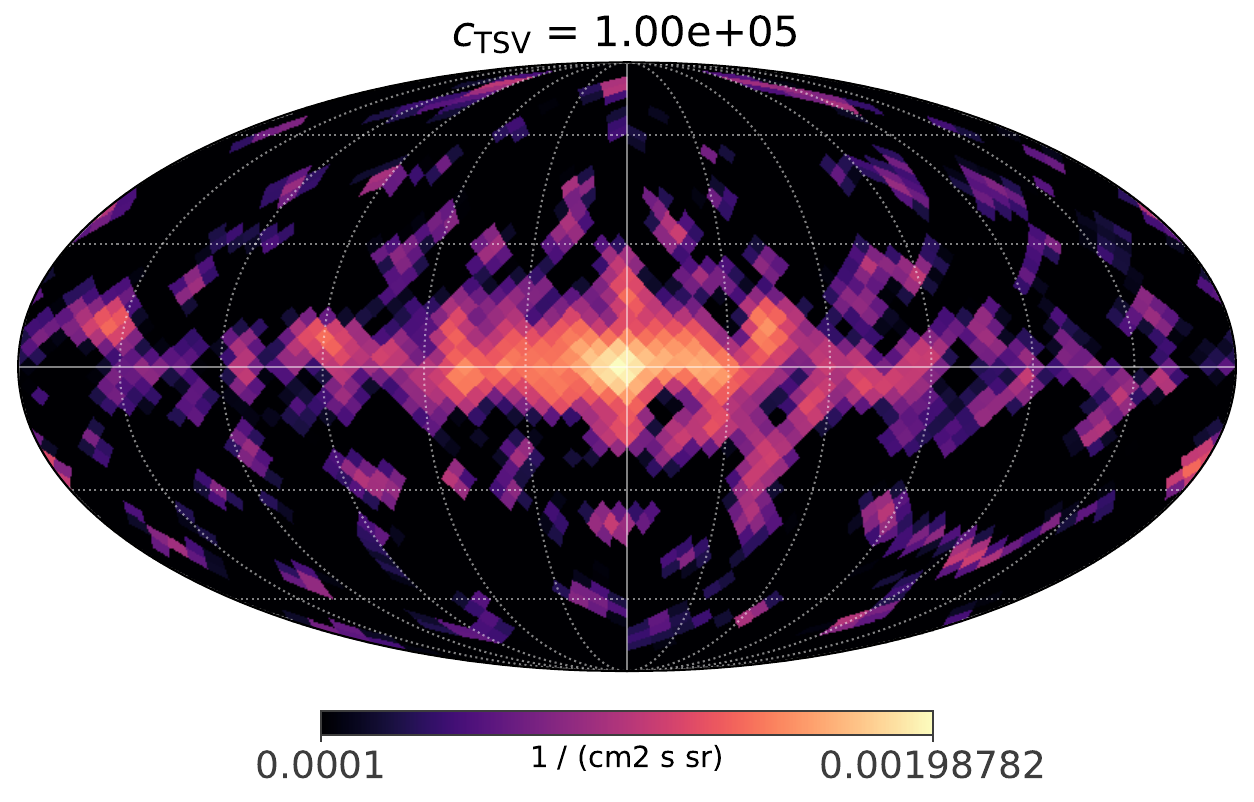}
    \end{subfigure}
    \begin{subfigure}{0.33\textwidth}
      \includegraphics[width=\linewidth]{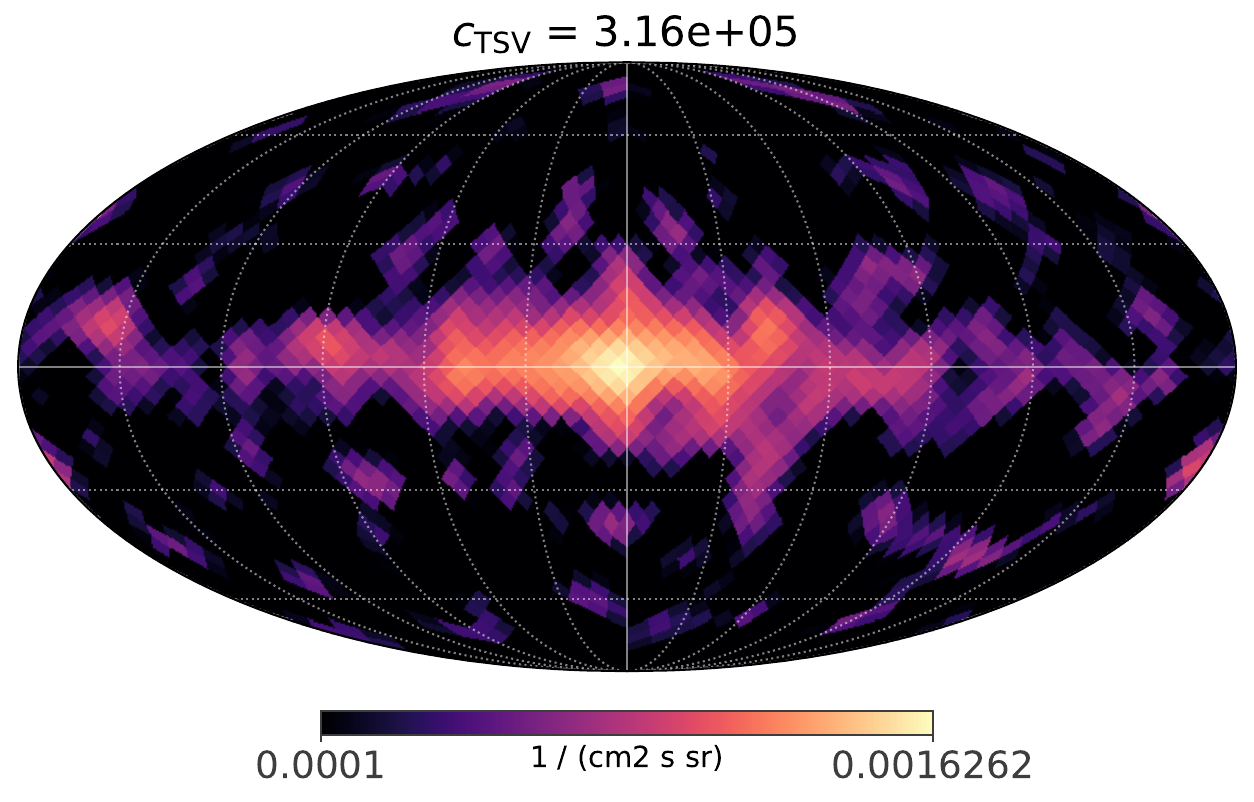}
    \end{subfigure}

    \begin{subfigure}{0.33\textwidth}
      \includegraphics[width=\linewidth]{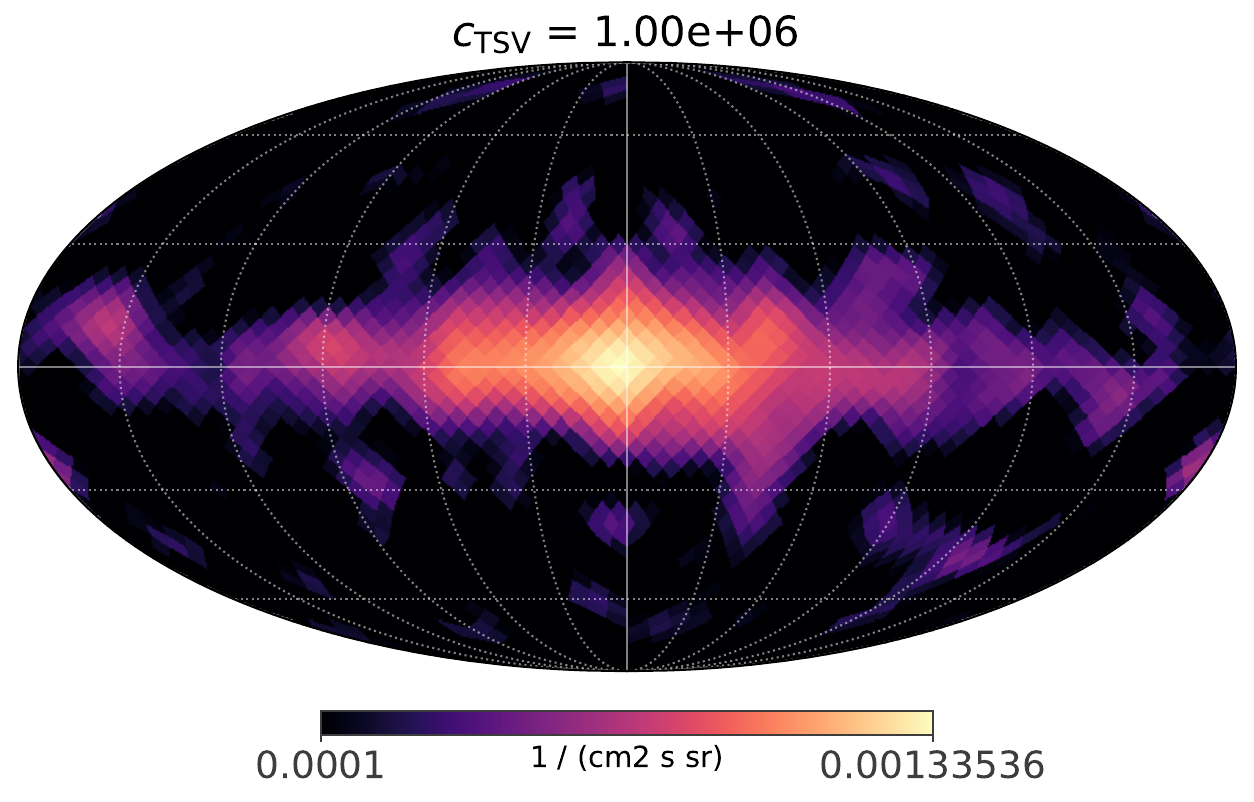}
    \end{subfigure}
    \begin{subfigure}{0.33\textwidth}
      \includegraphics[width=\linewidth]{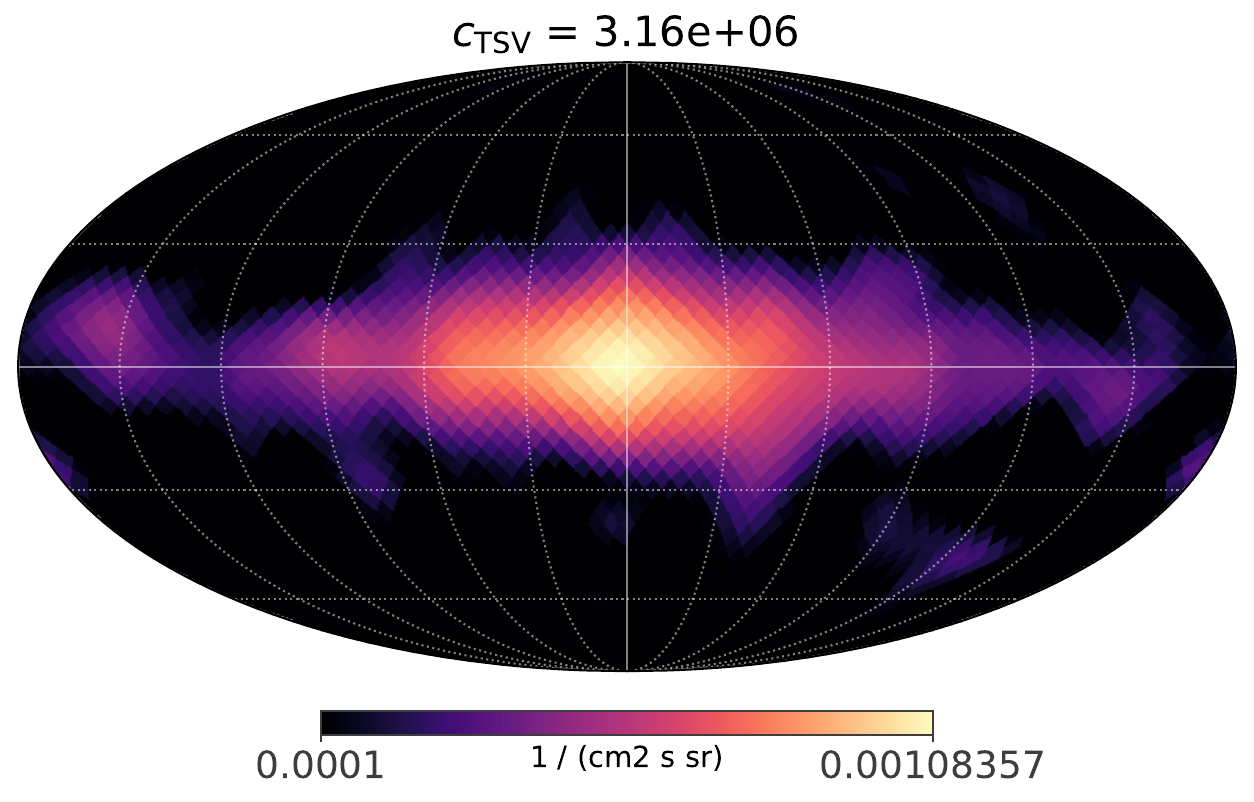}
    \end{subfigure}
    \begin{subfigure}{0.33\textwidth}
      \includegraphics[width=\linewidth]{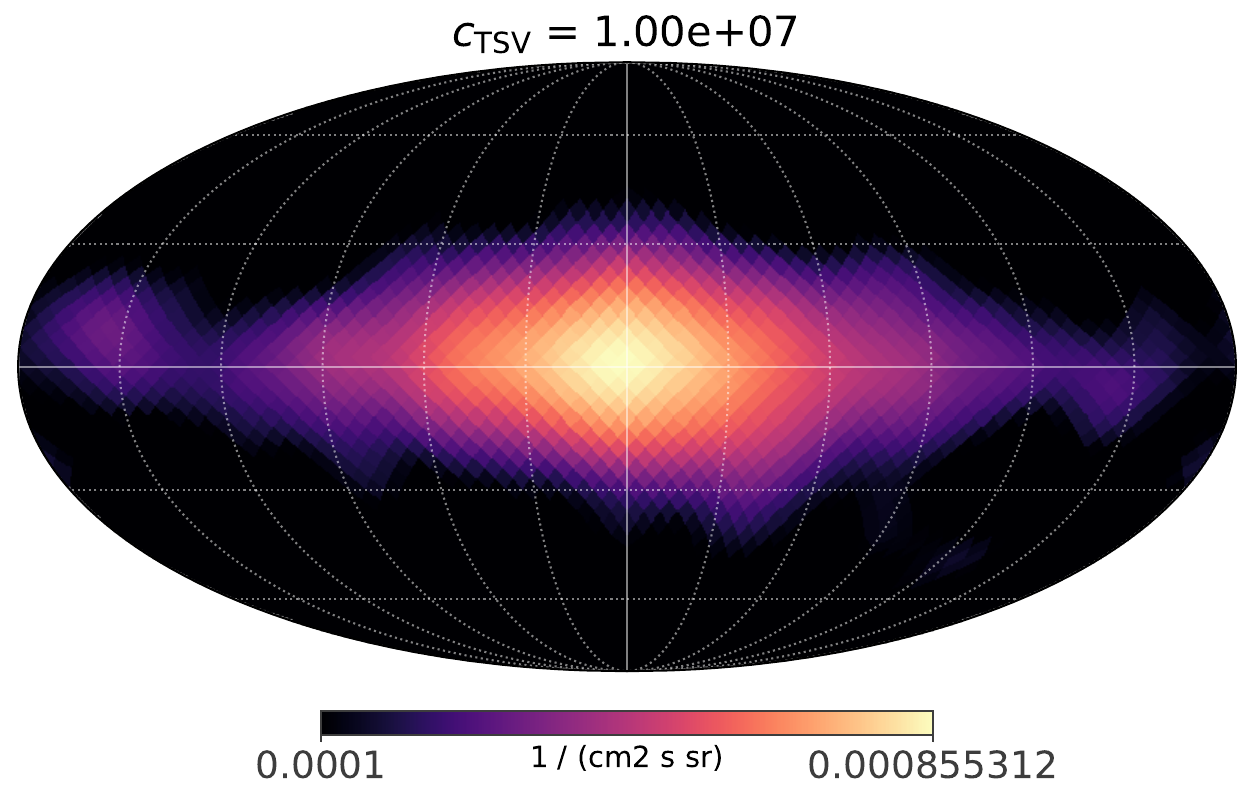}
    \end{subfigure}

    \caption{Reconstructed images of $^{26}$Al with different values for $c^{\mathrm{TSV}}$. From top left to bottom right, $c^{\mathrm{TSV}}$ is set to $10^{4}$, $10^{5}$, $3.16 \times 10^{5}$, $10^{6}$, $3.16 \times 10^{6}$, and $10^{7}$. The injected model is shown in Figure~\ref{fig_Al26_image_comp}.}
    \label{fig_Al26_images}
\end{figure*}

The reconstructed images for various $c^{\mathrm{TSV}}$ values are presented in Figure~\ref{fig_Al26_images}. At smaller $c^{\mathrm{TSV}}$ values, the images show more noise and artificial structures. As $c^{\mathrm{TSV}}$ increases, the distribution becomes smoother, more closely resembling the expected extended emission from $^{26}$Al.
Here, we set the minimum value in the images to $10^{-4}$ cm$^{-2}$ s$^{-1}$ sr$^{-1}$ from a rough estimation of the noise level. Assuming the background events affect an image uniformly over the sky, each pixel suffers from $\sim 70$ background events (= $2.26 \times 10^5 / 3072$), which means a Poisson fluctuation of $\sim 8$ events per pixel. Given the typical exposure of $2 \times 10^6$ cm$^2$ s sr per pixel, this fluctuation corresponds to $\sim 10^{-4}$ ph cm$^{-2}$ s$^{-1}$ sr$^{-1}$.

\begin{figure}
\centering
    \includegraphics[width = 0.9 \linewidth]{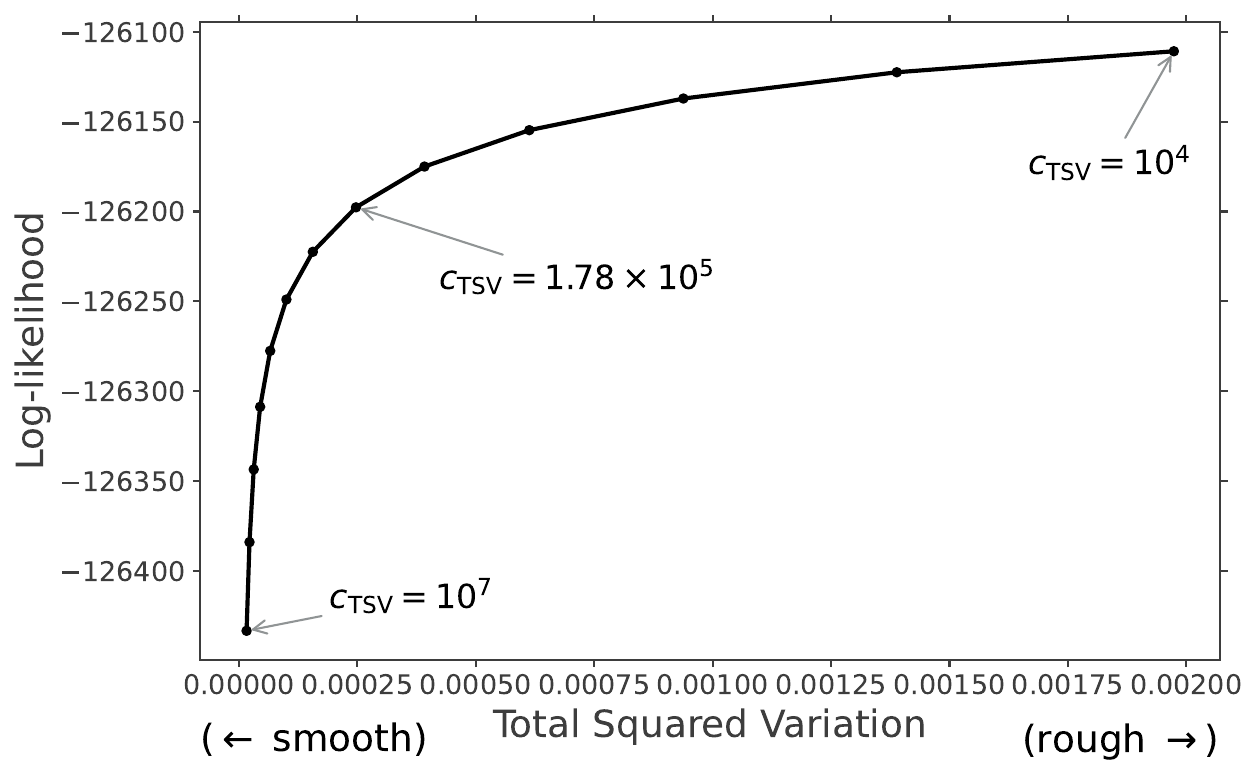}
\caption{L-curve for the smooth prior. The x-axis shows the total squared variation, while the y-axis shows the resulting log-likelihood.}
\label{fig_Al26_lcurve}
\end{figure}

\begin{figure}
\centering
    \includegraphics[width = 0.95 \linewidth]{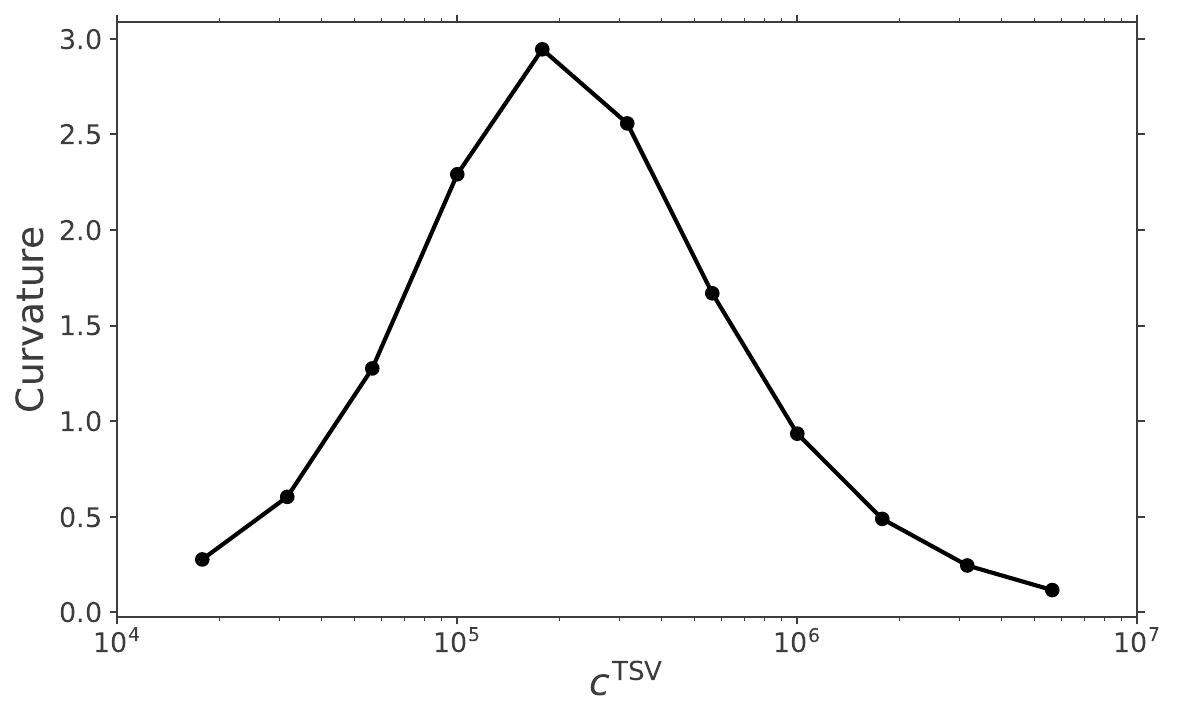}
\caption{Curvature of the L-curve shown in Figure~\ref{fig_Al26_lcurve} plotted against $c^{\mathrm{TSV}}$.}
\label{fig_curvature_Al26}
\end{figure}

Figure~\ref{fig_Al26_lcurve} shows the relationship between the TSV and the log-likelihood for different values of $c^{\mathrm{TSV}}$. As $c^{\mathrm{TSV}}$ increases, we observe a clear trade-off between image smoothness and likelihood optimization.
Visual inspection suggests that $c^{\mathrm{TSV}} \approx 10^{5-6}$ provides a good balance between noise suppression and preserving the overall structure of the $^{26}$Al distribution. At this value, the large-scale features of the Galactic plane emission are well-recovered, while smaller-scale noise is effectively suppressed.
To quantitatively assess the optimal $c^{\mathrm{TSV}}$, we again employed the L-curve method \citep{Allain2006}. 
This analysis suggests an optimal value of $c^{\mathrm{TSV}} = 1.78 \times 10^5$ (see Figure~\ref{fig_curvature_Al26}).
In this case, the optimal solution obtained by the L-curve method is consistent with the visual evaluation. 

\subsubsection{Comparison with the conventional algorithm}

\begin{figure}
\centering
    \includegraphics[width = 0.9 \linewidth]{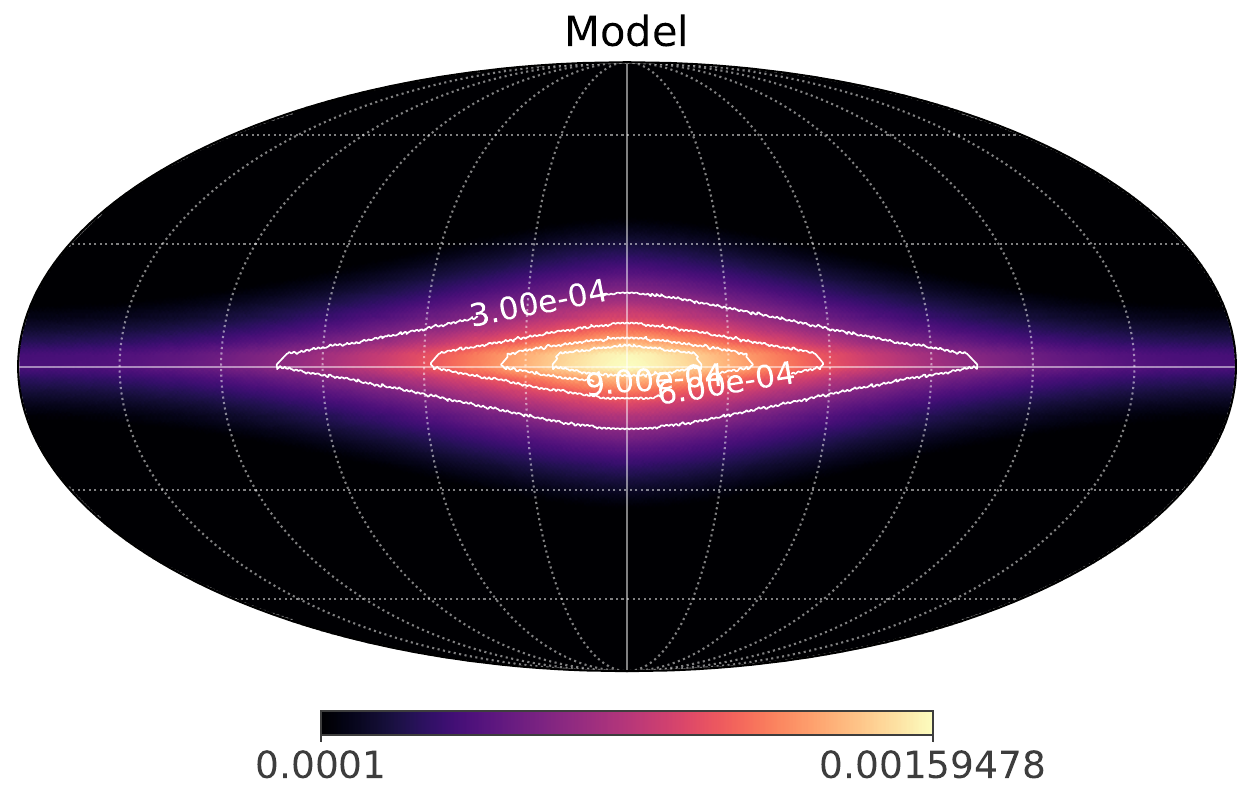}
    \includegraphics[width = 0.9 \linewidth]{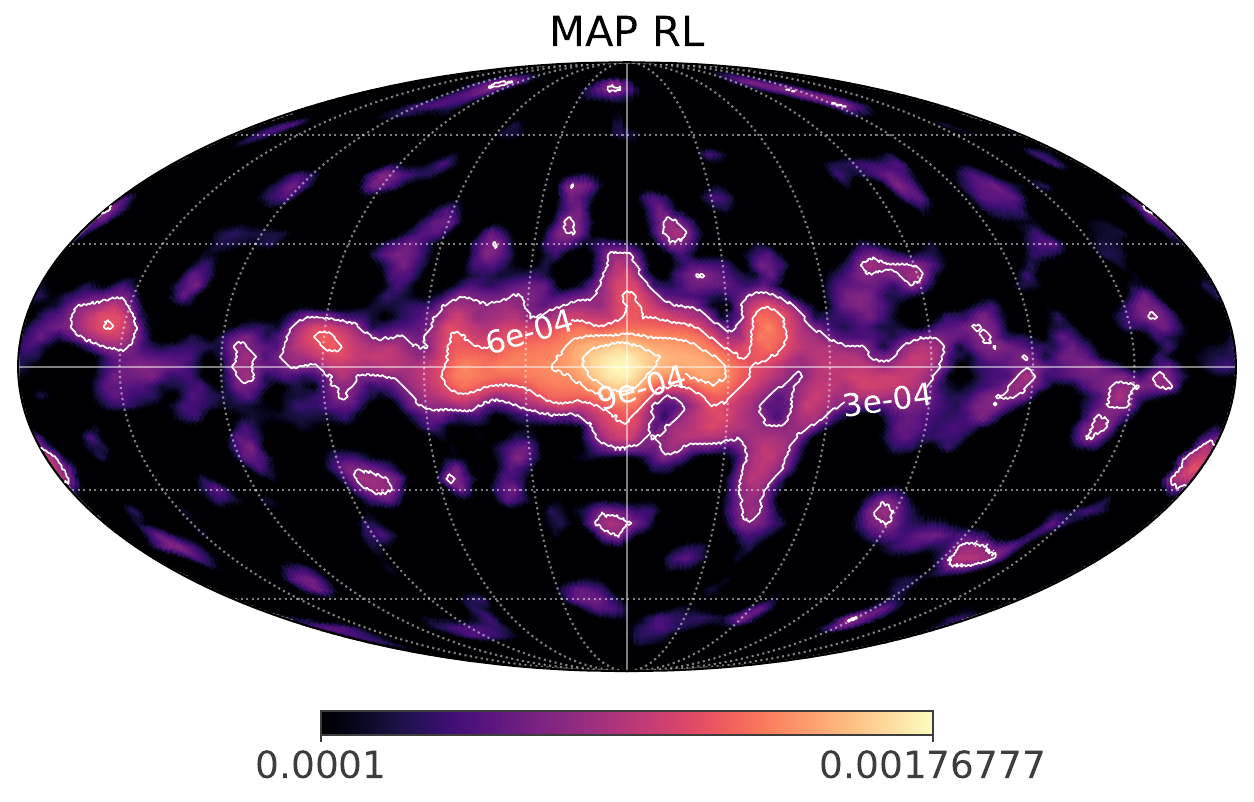}
    \includegraphics[width = 0.9 \linewidth]{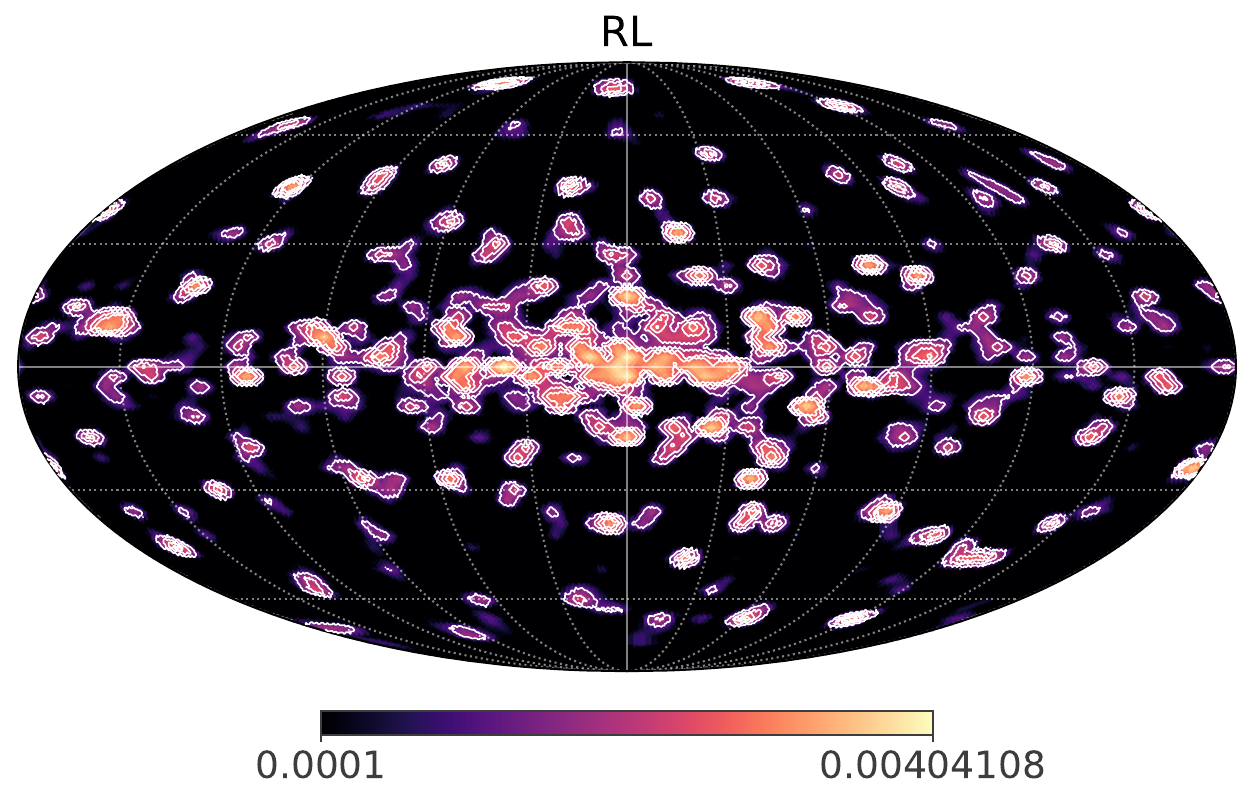}
\caption{Comparison between the reconstructed images obtained with different methods. The images are in the units of $\mathrm{ph/cm^{2}/s/sr}$. The top shows the injected model for the $^{26}$Al map, while the middle and bottom show the reconstructed image with the proposed algorithm and the conventional RL algorithm, respectively.}
\label{fig_Al26_image_comp}
\end{figure}

\begin{figure}
\centering
    \includegraphics[width = 0.9 \linewidth]{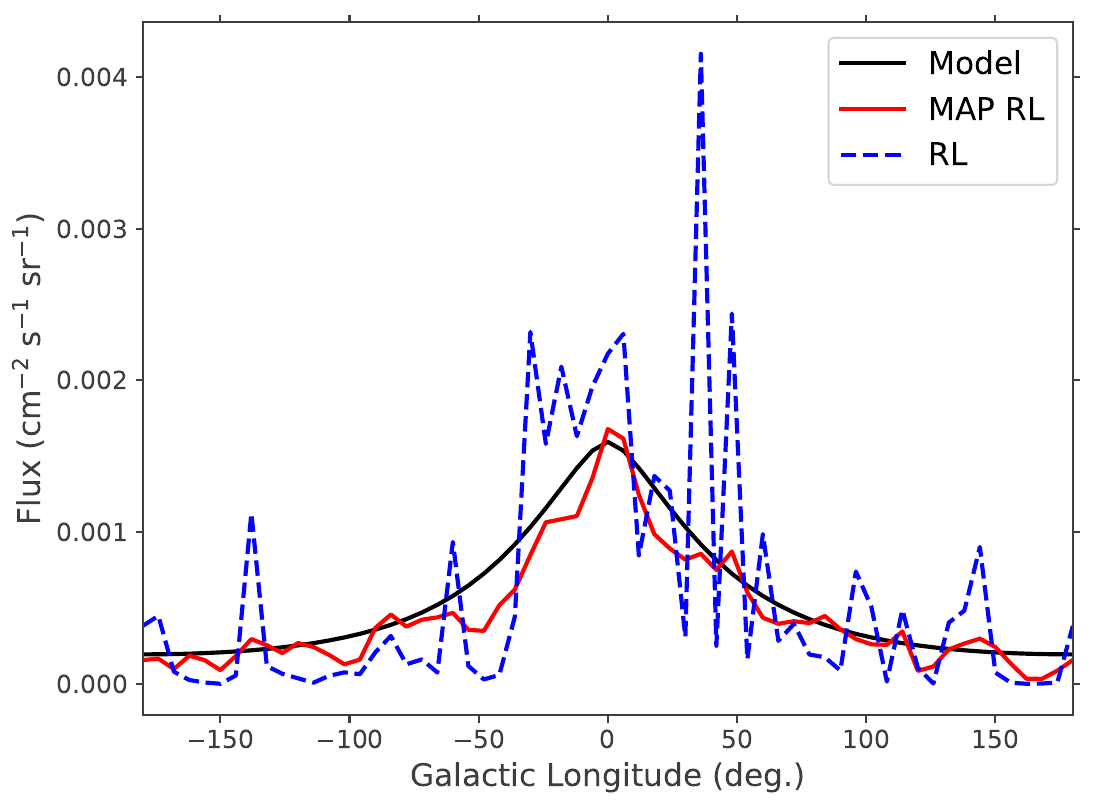}
    \includegraphics[width = 0.9 \linewidth]{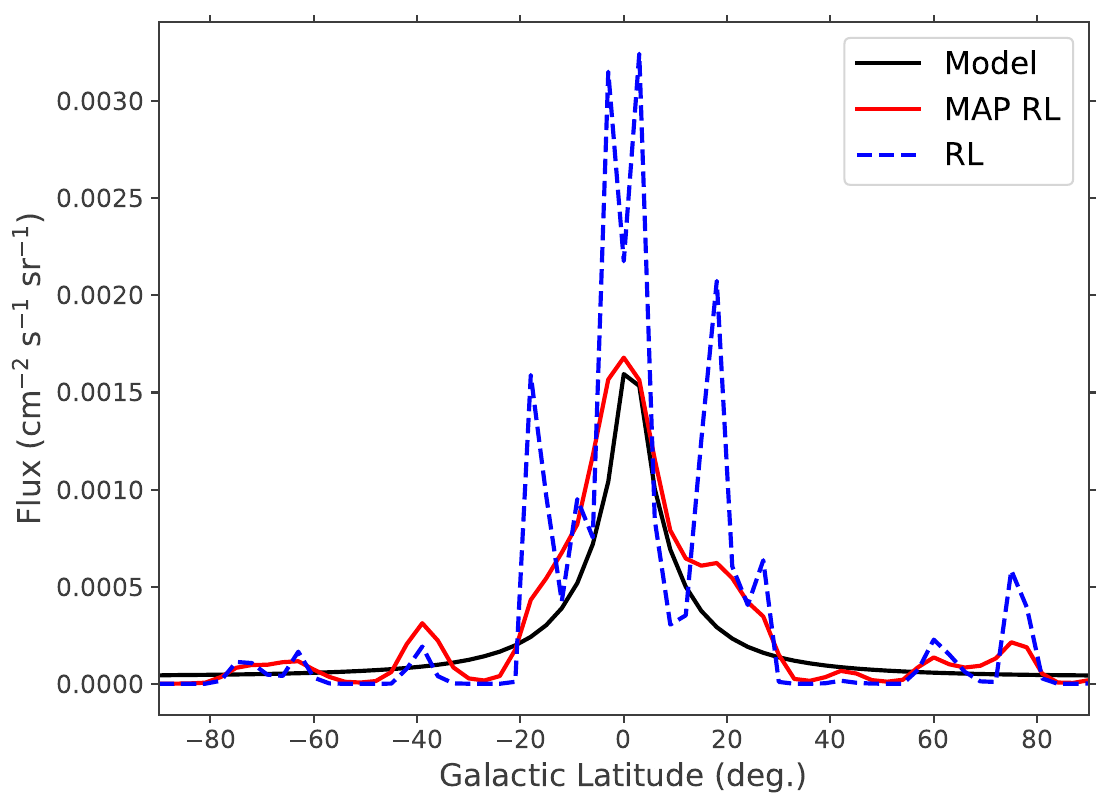}
\caption{Reconstructed $^{26}$Al fluxes along the Galactic longitude (top) and latitude (bottom).}
\label{fig_Al26_sliced_image}
\end{figure}

Figure~\ref{fig_Al26_image_comp} compares the input model, the results with the proposed algorithm ($c^{\mathrm{TSV}} = 1.78 \times 10^5$), and the conventional RL algorithm. 
The conventional approach shows that its solution converges to a distribution composed of numerous point-like structures, enhancing statistical fluctuations in the data.
Such a result makes it challenging to discuss the spatial scales and morphology of the diffuse emission that are crucial for $^{26}$Al science.
In contrast, the proposed algorithm maintains a continuous spatial distribution in its solution. 
The resulting image allows us to discuss the spatial characteristics of the $^{26}$Al emission. 
For instance, we can see a contour line with a flux of $3 \times 10^{-4} \mathrm{ph/cm^2/s/sr}$, and claim that it extends $\pm \sim 90$ degrees along the Galactic plane.
This distribution closely matches the input model whose extension is $\pm \sim 100$ degrees.
We also show the fluxes obtained along the Galactic longitude and latitude for each algorithm in Figure~\ref{fig_Al26_sliced_image}.
The fluctuations seen in the conventional method are clearly suppressed in the proposed one,
and the flux and structure around the Galactic center are well recovered.
The total flux was obtained as $1.82 \times 10^{-3}~\mathrm{cm^{-2}~s^{-1}}$, which is consistent with the injected model within $\sim$ 1\%.
This demonstrates the effectiveness of our method with the introduced prior in reconstructing extended smooth sources like $^{26}$Al emission.

\subsection{e$^{+}$/e$^{-}$: multiple spatial components} 

Finally, we evaluate the performance of our algorithm for the 0.511 MeV emission, which presents a complex scenario combining several spatial components. For this case, we incorporate both the sparseness and smoothness priors:
\begin{align}
\begin{split}
\log P_{\mathrm{s}}(\vector{\lambda}) = - c^{\mathrm{TSV}} \sum_j \sum_{k \in \sigma_{j}} (\lambda_{j} - \lambda_{k})^2 - c^{\mathrm{SP}} \sum_{j} \log \lambda_{j}.
\end{split}
\end{align}
We varied $c^{\mathrm{TSV}}$ from $10^2$ to $10^6$ and $c^{\mathrm{SP}}$ from $10^{-2}$ to $5 \times 10^{-1}$ to examine their effects on the reconstructed images.

First, we focus on the thin disk model. This model includes a point source at the Galactic center, two Gaussian components for the Galactic bulge emission slightly offset from the center, and a Galactic disk emission with a scale height of 3 degrees. 
In Appendix~\ref{sec_511keV_image_all}, we show the reconstructed images for various combinations of $c^{\mathrm{TSV}}$ and $c^{\mathrm{SP}}$ for the thin disk model. As expected, we observe a smoother distribution over the Galaxy as we increase $c^{\mathrm{TSV}}$. Increasing $c^{\mathrm{SP}}$, on the other hand, suppresses the point-like structures.

\begin{figure}
\centering
    \includegraphics[width = 0.95 \linewidth]{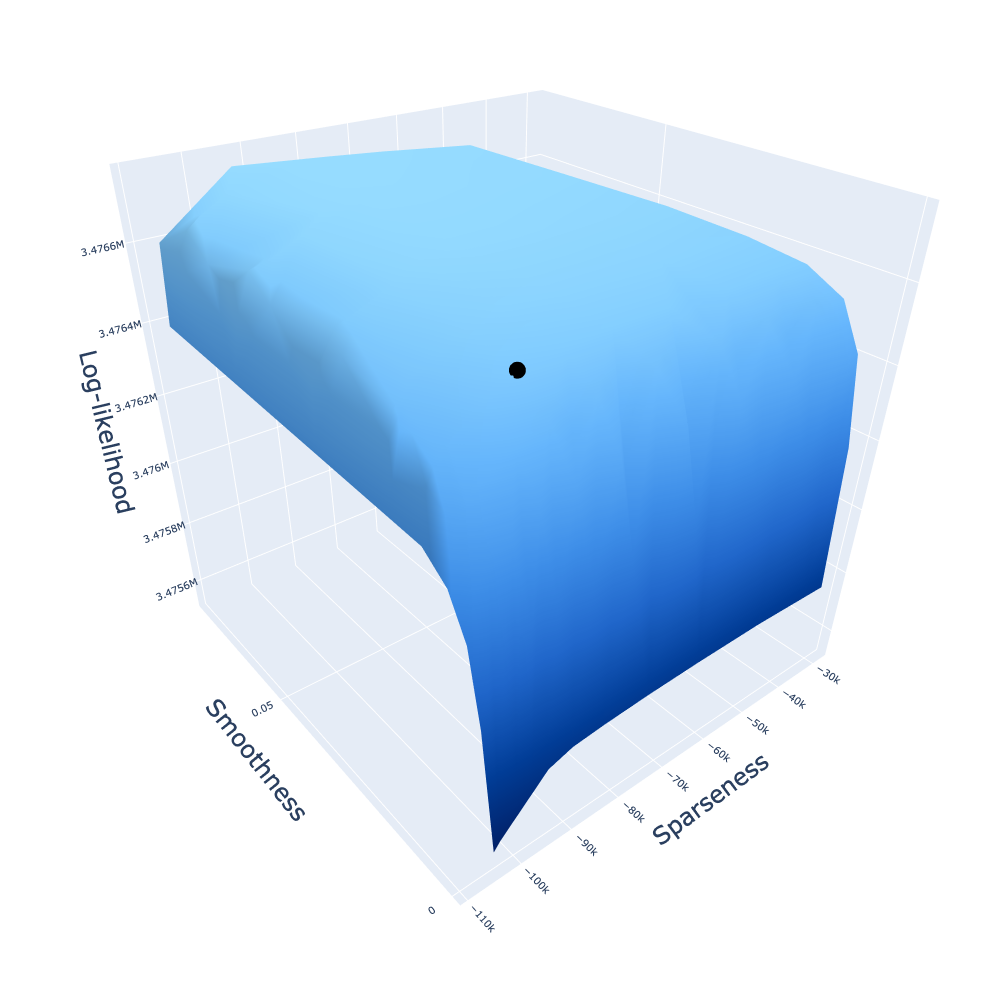}
\caption{Two-dimensional L-curve for the sparse and smooth priors. The bottom plane corresponds to the smoothness and sparseness, while the z-axis shows the resulting log-likelihood. The black point corresponds to the optimal solution we presented in the main text.}
\label{fig_511keV_lcurve}
\end{figure}

\begin{figure}
\centering
    \includegraphics[width = 0.95 \linewidth]{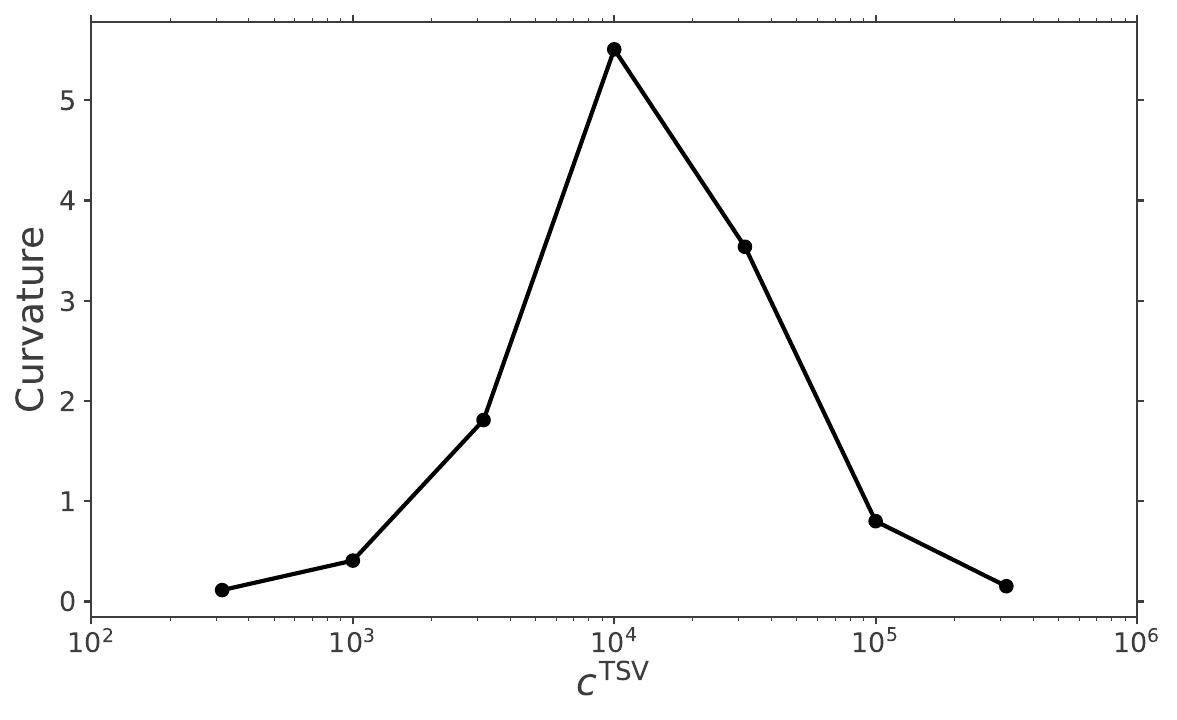}
    \includegraphics[width = 0.95 \linewidth]{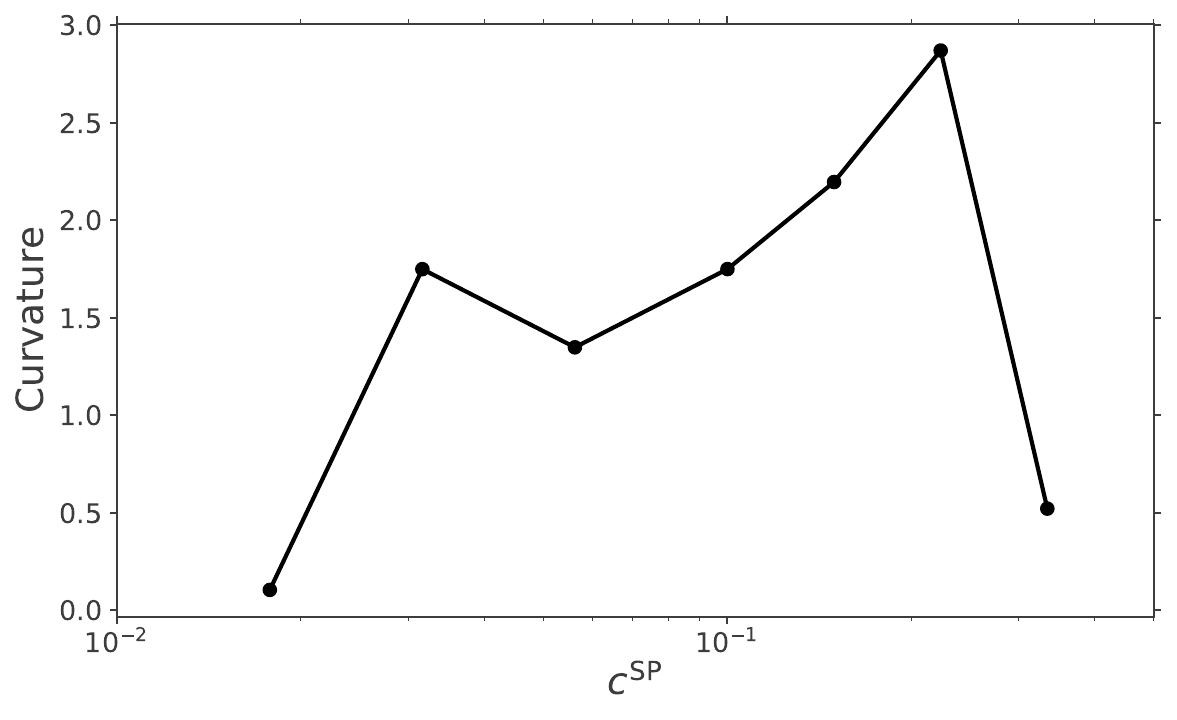}
\caption{Curvature plots for the two-dimensional L-curve shown in Figure~\ref{fig_511keV_lcurve}. Top: Curvature along $c^{\mathrm{TSV}}$ with $c^{\mathrm{SP}}$ fixed at its optimal value ($3.2 \times 10^{-2}$). Bottom: Curvature along $c^{\mathrm{SP}}$ with $c^{\mathrm{TSV}}$ fixed at its optimal value ($10^4$).}
\label{fig_curvature_511keV}
\end{figure}

Figure~\ref{fig_511keV_lcurve} shows the relationship between the log-likelihood, TSV, and sparseness terms. 
We applied the L-curve method two-dimensionally. 
First, we scanned the curvature along the $c^{\mathrm{TSV}}$ axis for each value of $c^{\mathrm{SP}}$.
For all values of $c^{\mathrm{SP}}$, we found the maximum curvature at $c^{\mathrm{TSV}} = 10^{4}$.
Next, we scanned the curvature along the $c^{\mathrm{SP}}$ axis while fixing $c^{\mathrm{TSV}}$ at $10^4$.
Then, we found two local maxima in the curvature of the graph (see Figure~\ref{fig_curvature_511keV}). Thus, two primary candidates for optimal solutions were identified: ($c^{\mathrm{TSV}}$, $c^{\mathrm{SP}}$) = ($10^4$, $3.2\times10^{-2}$) and ($c^{\mathrm{TSV}}$, $c^{\mathrm{SP}}$) = ($10^4$, $2.2\times10^{-1}$). 
The red frames in Figure~\ref{fig_511keV_images} indicate the two candidates.
Upon visually inspecting the reconstructed images, we found that the latter solution resulted in a truncated disk with unnatural structures. 
Therefore, we selected the former solution of ($c^{\mathrm{TSV}}$, $c^{\mathrm{SP}}$) = ($10^4$, $3.2\times10^{-2}$) as the most plausible reconstruction for this analysis.
We also show the expected counts from the selected model compared with the data in the CDS in Appendix~\ref{sec_comp_model_CDS}.

\begin{figure*}
\centering
    \includegraphics[width = 0.48 \linewidth]{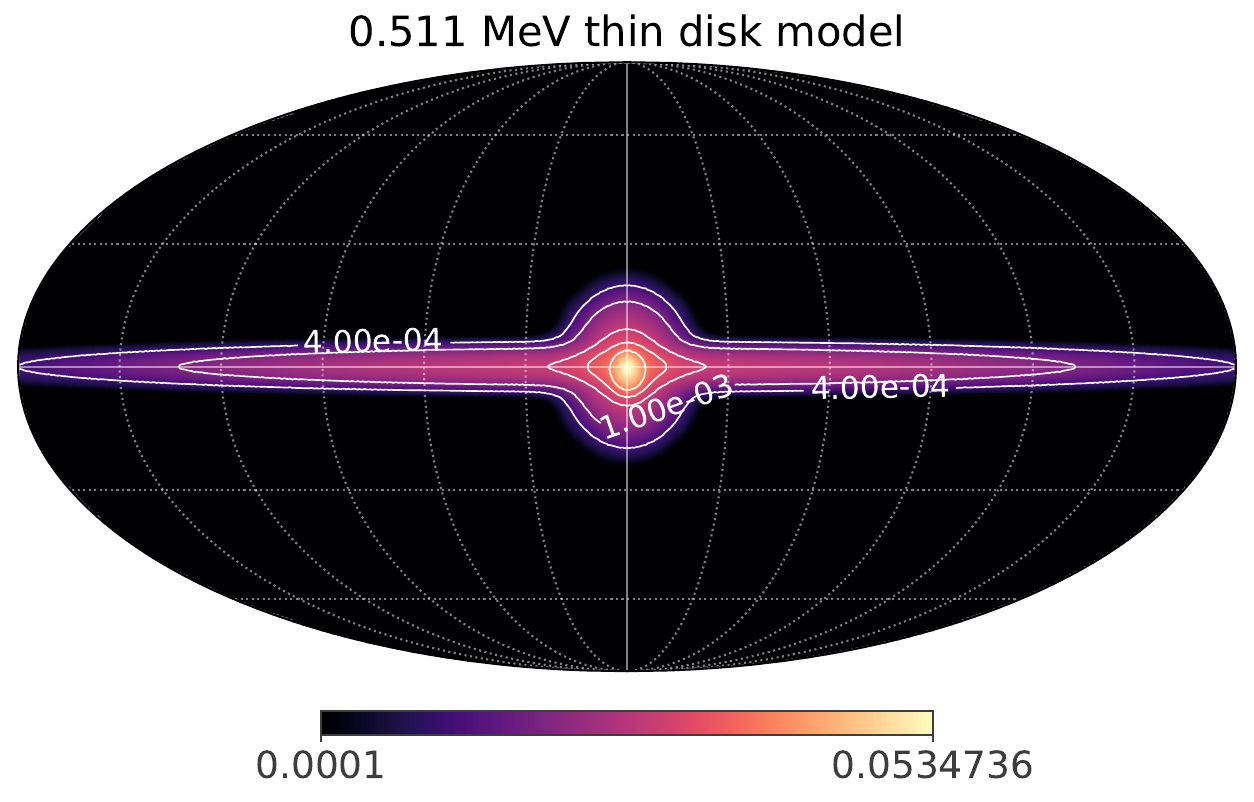}
    \includegraphics[width = 0.48 \linewidth]{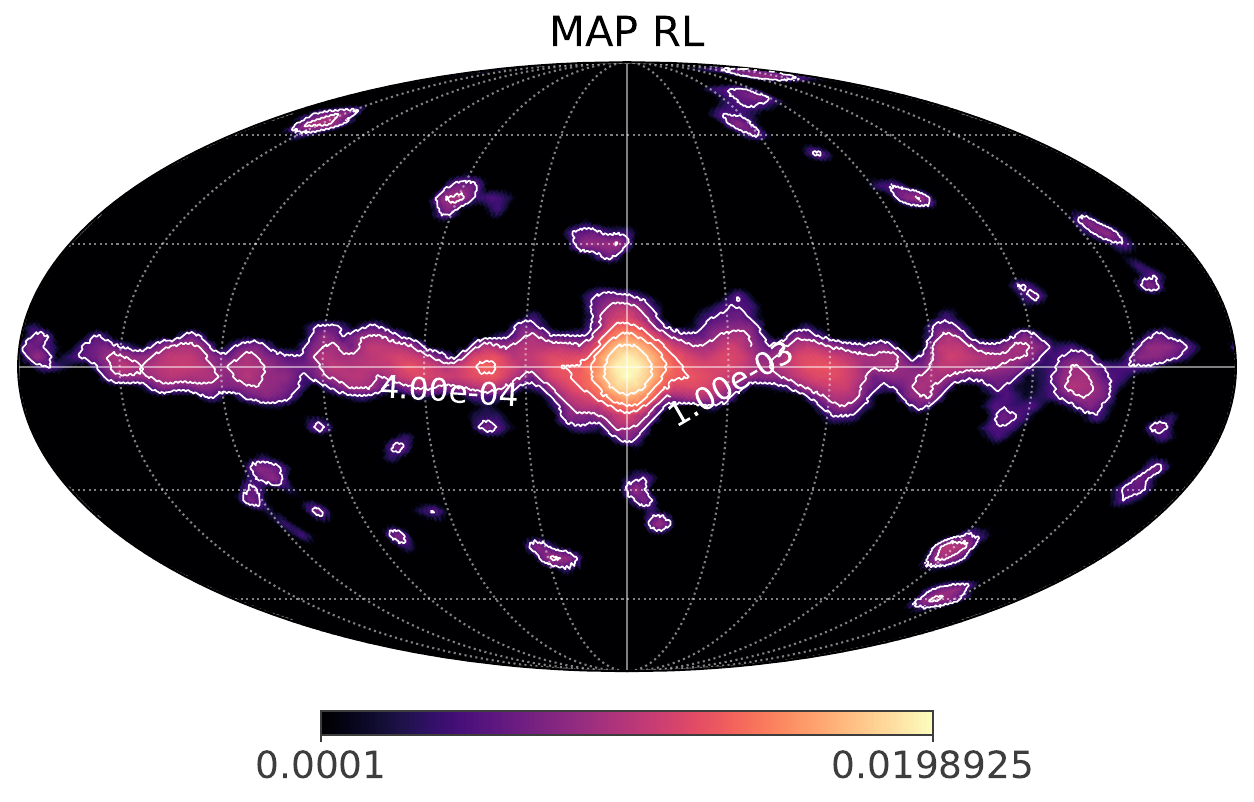}
    \includegraphics[width = 0.48 \linewidth]{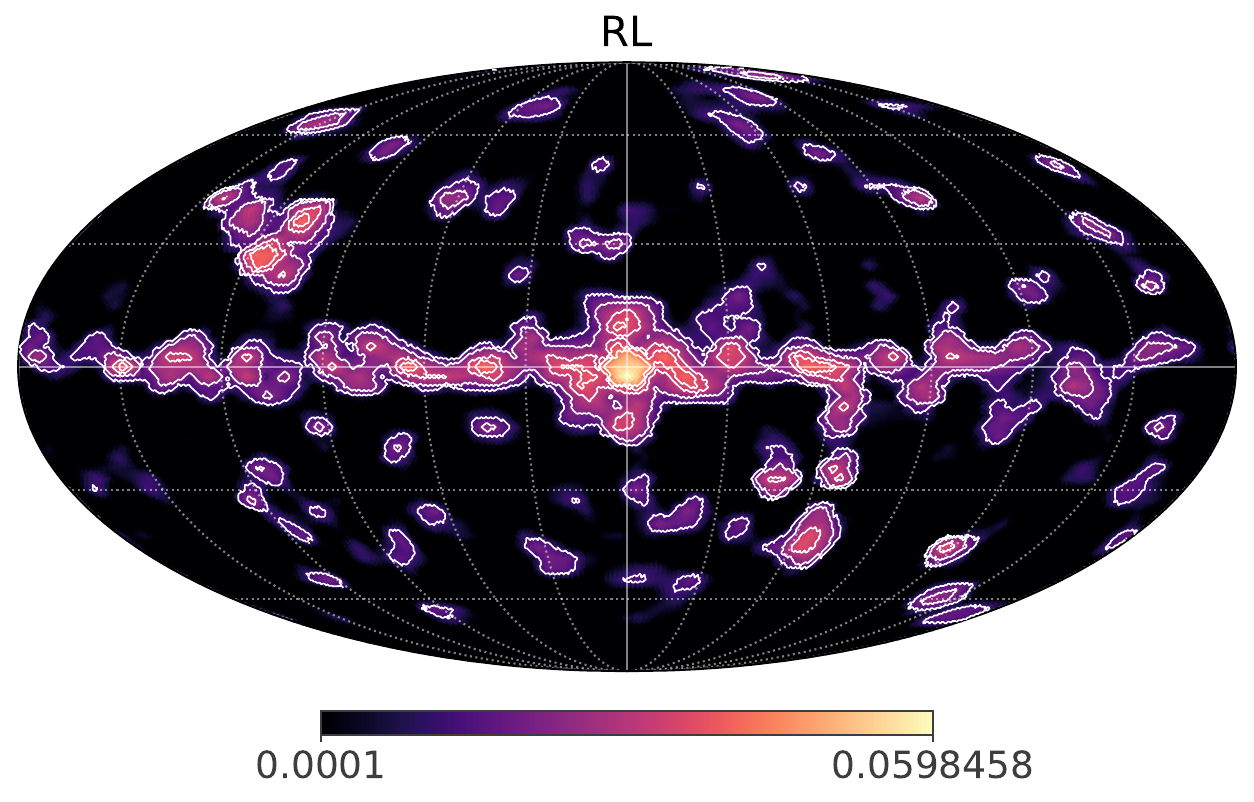}
    \includegraphics[width = 0.48 \linewidth]{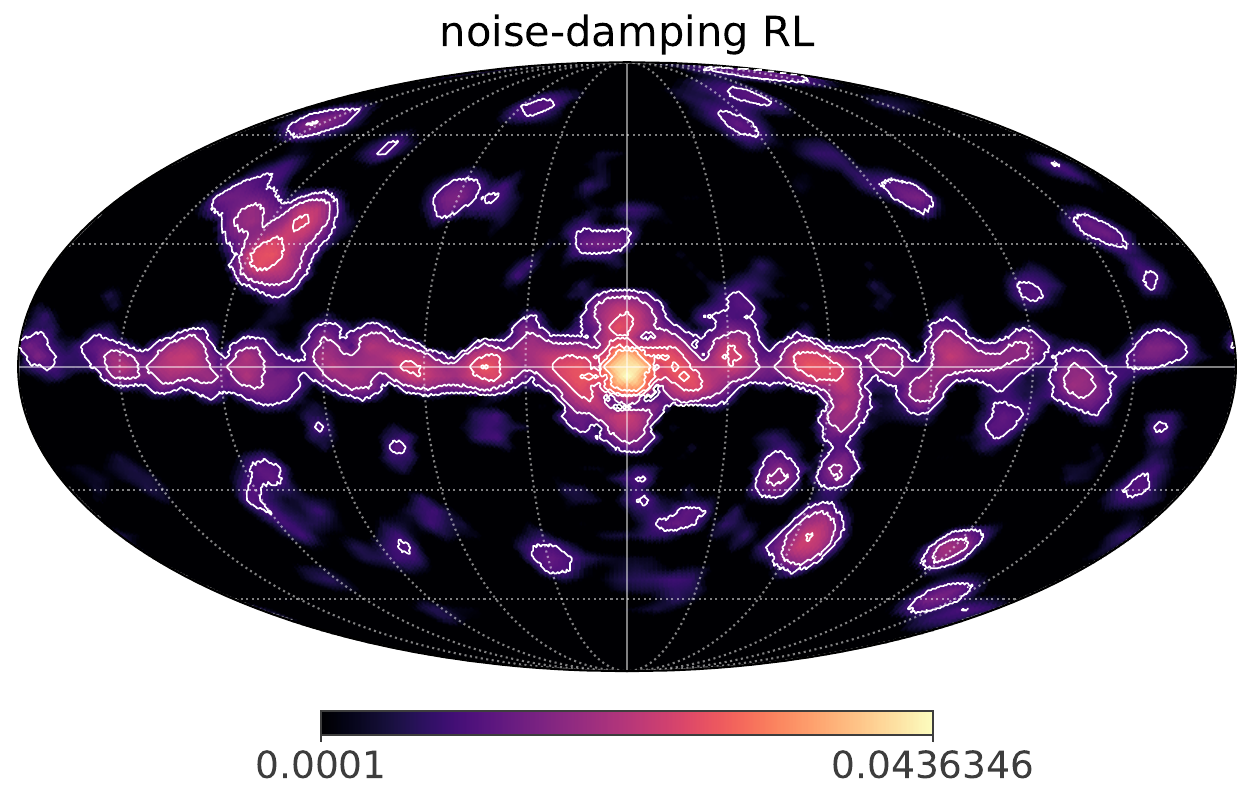}
\caption{Comparison between the reconstructed 0.511 MeV images with different methods. The images are in the unit of $\mathrm{ph/cm^{2}/s/sr}$. The top left shows the injected thin model, while the reconstructed images with the proposed method, the conventional RL, and the RL using a Gaussian filter are shown on the top right, bottom left, and bottom right panels, respectively.
Following the same discussion in Section~\ref{subsubsec_Al26_image}, each pixel suffers from the Poisson fluctuation of $\sim 1000$ background events (= $2.09 \times 10^6 / 3072$) with a typical exposure of $3.5 \times 10^5$ cm$^2$ s sr on a pixel, which corresponds to a noise level of $\sim 10^{-4}$ ph cm$^{-2}$ s$^{-1}$ sr$^{-1}$ in the reconstructed image.}
\label{fig_511keV_image_comp}
\end{figure*}

Figure~\ref{fig_511keV_image_comp} compares the input model, the results from our Bayesian method (using the selected optimal $c^{\mathrm{TSV}}$ and $c^{\mathrm{SP}}$ values), the conventional RL algorithm, and the RL with a noise damping approach used in \cite{Knoedlseder2005, Siegert2020}.
In the last case, we adopted a 5-degree Gaussian filter.
One of the key advantages of our proposed method is the suppression of artifacts commonly observed in other approaches. 
The bright artifacts shown in the upper left and lower right regions of the conventional RL and noise-damping RL results are significantly reduced in our method. 
These artifacts correspond to regions with shorter exposure times (see Figure~\ref{fig_exposure_map}), where low photon statistics lead to enhanced fluctuations in the conventional methods.
Our approach successfully mitigates this issue, resulting in a substantial improvement in flux estimation, as also seen in the $^{44}$Ti case.
Table~\ref{tab_algorithm_flux_comparison} shows the obtained total fluxes with different methods.
While the other methods overestimate the total flux by approximately 60\% due to these artifacts, our method achieves an accuracy within 10\% of the true value.

Another improvement is that the continuous disk structure is preserved. 
Traditional methods tend to emphasize point-like structures, as observed in the $^{26}$Al case, breaking up the inherently diffuse, connected structure into discrete segments. 
Our method maintains the continuity of the disk, enabling more discussion of its spatial structure.
This would be crucial for analyzing the large-scale distribution of the 0.511 MeV emission in the Galaxy.

\begin{table}
\centering
\caption{Total fluxes of the reconstructed 0.511 MeV images with different algorithms.}
\label{tab_algorithm_flux_comparison}
\begin{tabular}{ccc}
\hline
Algorithm & \multicolumn{2}{c}{Total Flux (cm$^{-2}$ s$^{-1}$)} \\
& Thin Disk Model & Thick Disk Model\\
\hline
MAP RL & $2.91 \times 10^{-3}$ & $2.72 \times 10^{-3}$ \\
RL & $4.26 \times 10^{-3}$ & $4.33 \times 10^{-3}$ \\
Noise-damping RL & $4.11 \times 10^{-3}$ & $4.17 \times 10^{-3}$ \\
\hline
Model & \multicolumn{2}{c}{$2.61 \times 10^{-3}$} \\
\hline
\end{tabular}
\tablefoot{The model flux shown here is the integrated flux from 0.509 MeV to 0.513 MeV of the assumed 0.511 MeV models. Since our assumed model has intrinsic line widths (FWHM) of 2 keV and 3 keV for the bulge and disk components, respectively,
the reconstructed flux was compared with the model flux within the bin width of the response matrix.}
\end{table}

\begin{figure*}
\centering
    \includegraphics[width = 0.48 \linewidth]{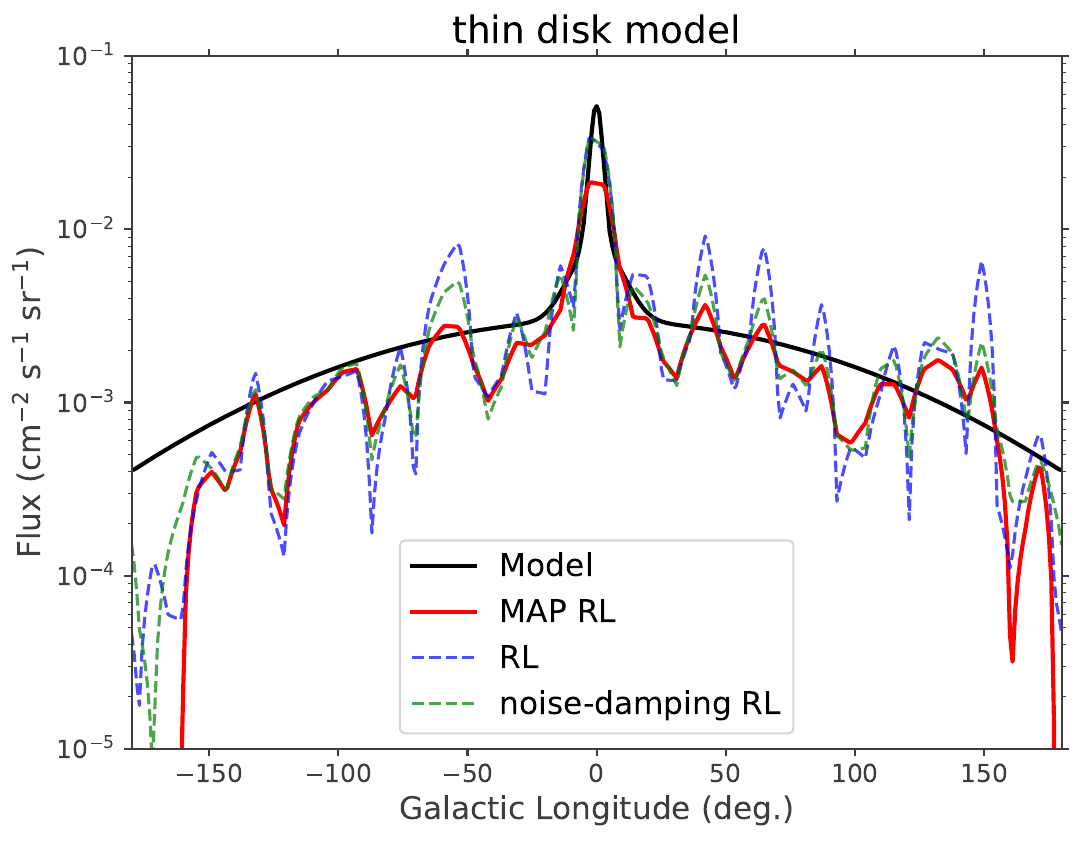}
    \includegraphics[width = 0.48 \linewidth]{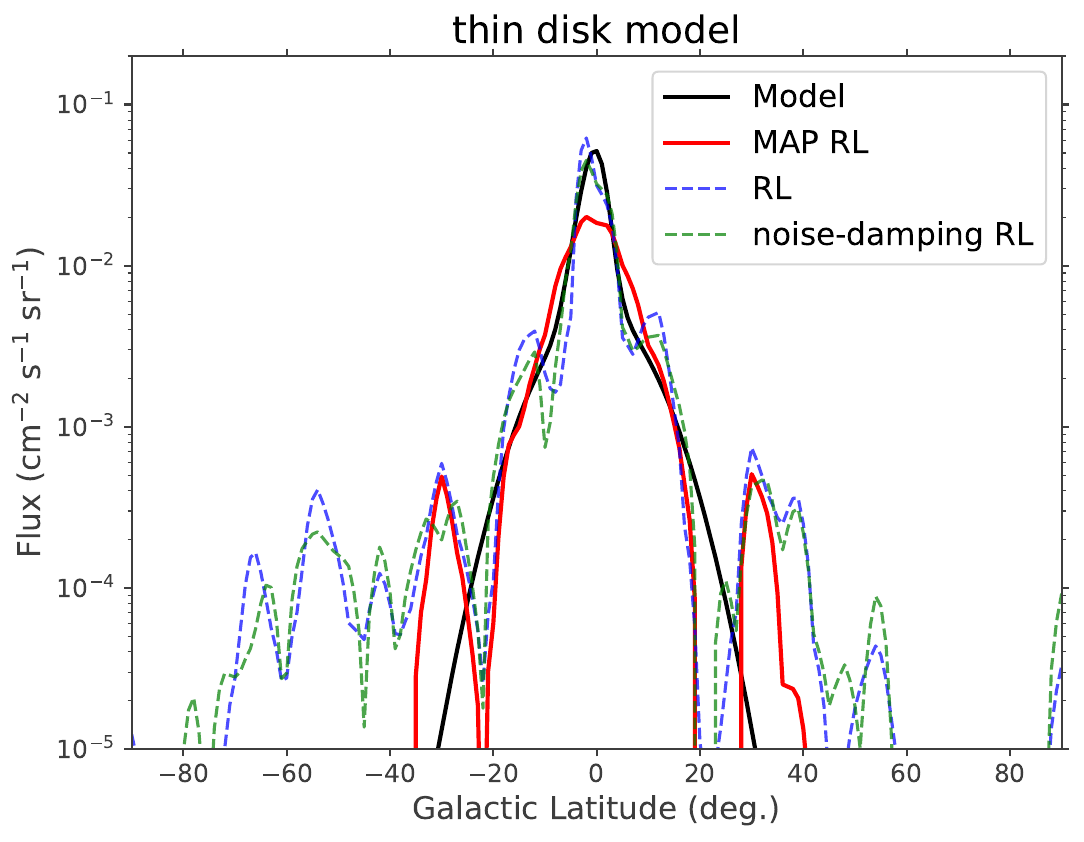}
\caption{Reconstructed 0.511 MeV fluxes along the Galactic longitude (left) and latitude (right). When plotting the model line, the flux from the point source in Table~\ref{tab_511keV_models} is blurred with the pixel size of the image ($\sim$ 3.7 degrees).}
\label{fig_511keV_sliced_image}
\end{figure*}

It is important to note a potential limitation of our method: the over-smoothing of the Galactic center structure.
When examining the flux along Galactic longitude and latitude, as shown in Figure~\ref{fig_511keV_sliced_image}, we observe that our reconstruction of the central region is broader than the input model.
This over-smoothing likely results from applying a uniform smooth prior across the entire sky.
While this approach works well for the disk structure, it may be excessively strong for the brighter, more sharply defined central region.
Future improvements could involve varying the strength of the prior based on the region, potentially using adaptive priors that account for the diverse structural characteristics across the Galactic plane.
Despite this limitation, the overall performance of the proposed method in reconstructing the 0.511 MeV emission demonstrates a significant advancement over conventional techniques, particularly in suppressing point-like artifacts and preserving extended structures.

\begin{figure*}
\centering
    \includegraphics[width = 0.48 \linewidth]{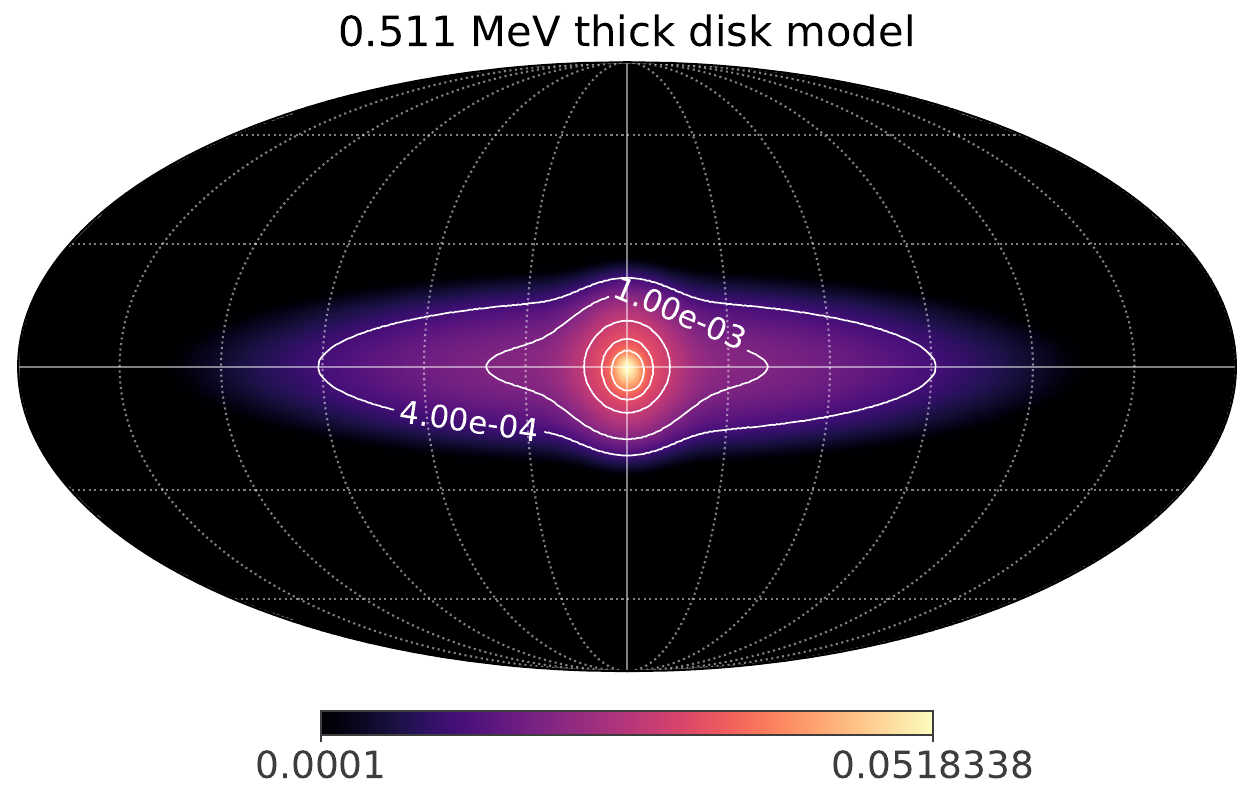}
    \includegraphics[width = 0.48 \linewidth]{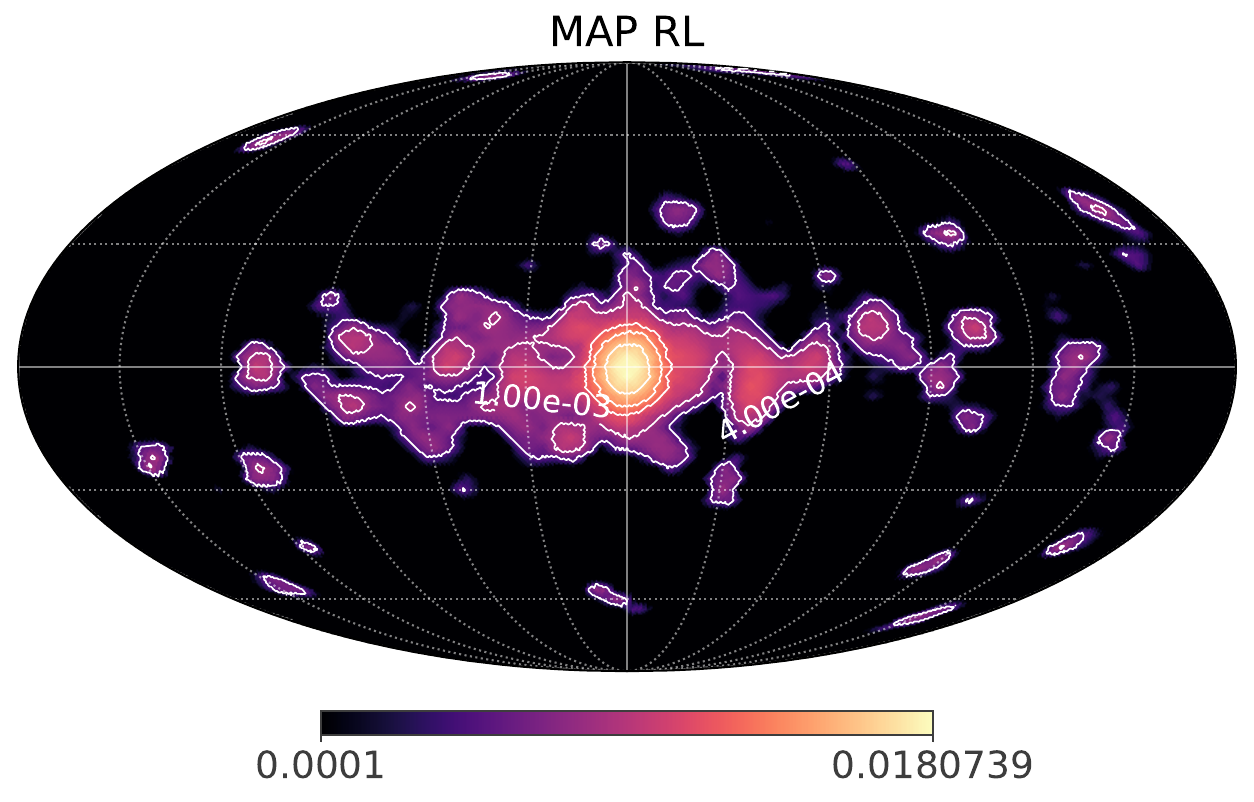}
    \includegraphics[width = 0.48 \linewidth]{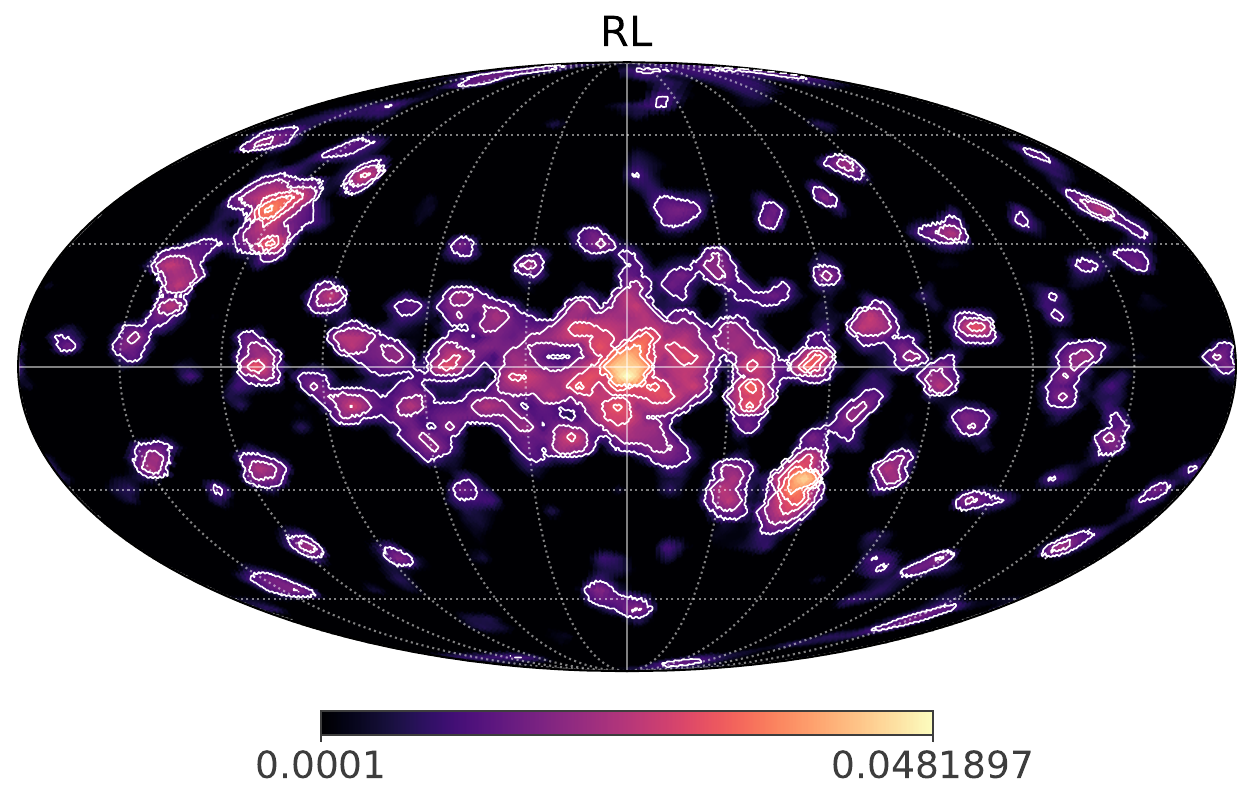}
    \includegraphics[width = 0.48 \linewidth]{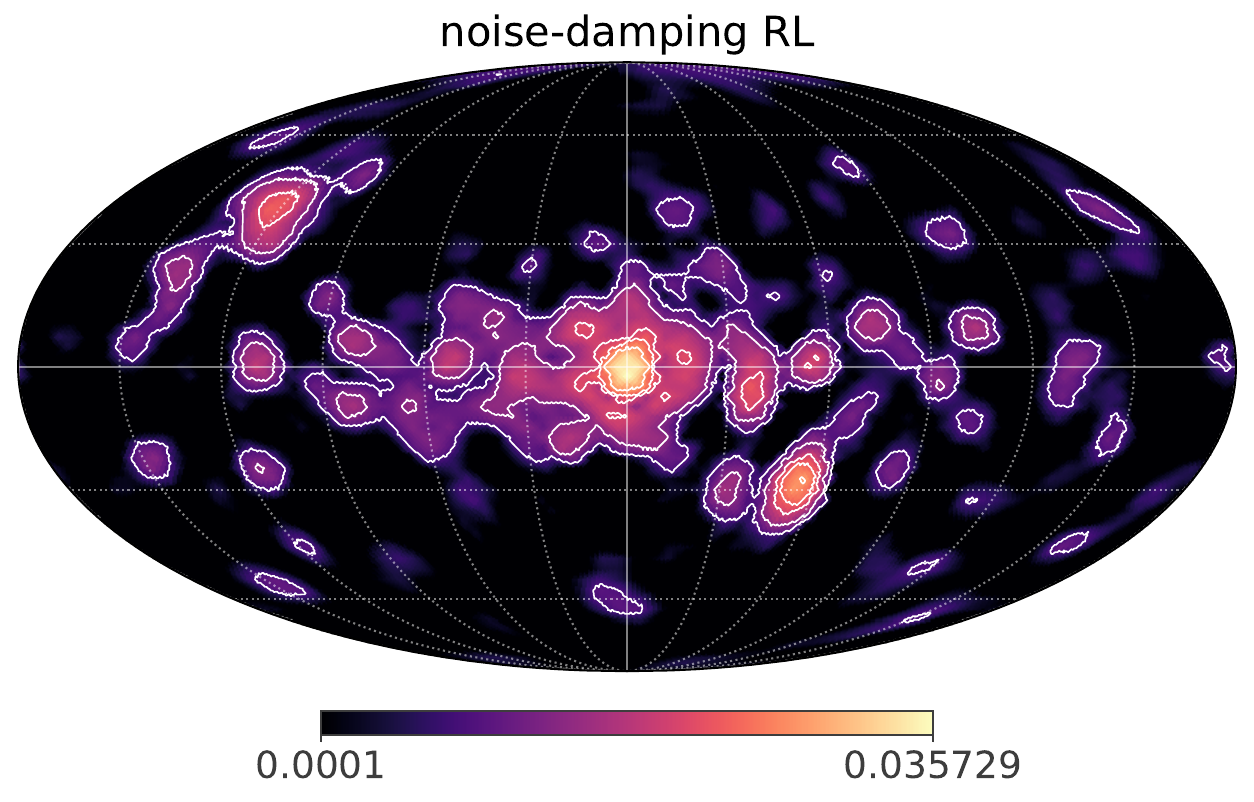}
\caption{Same as Figure~\ref{fig_511keV_image_comp} but for the thick disk model.}
\label{fig_511keV_thick_image_comp}
\end{figure*}

\begin{figure*}
\centering
    \includegraphics[width = 0.48 \linewidth]{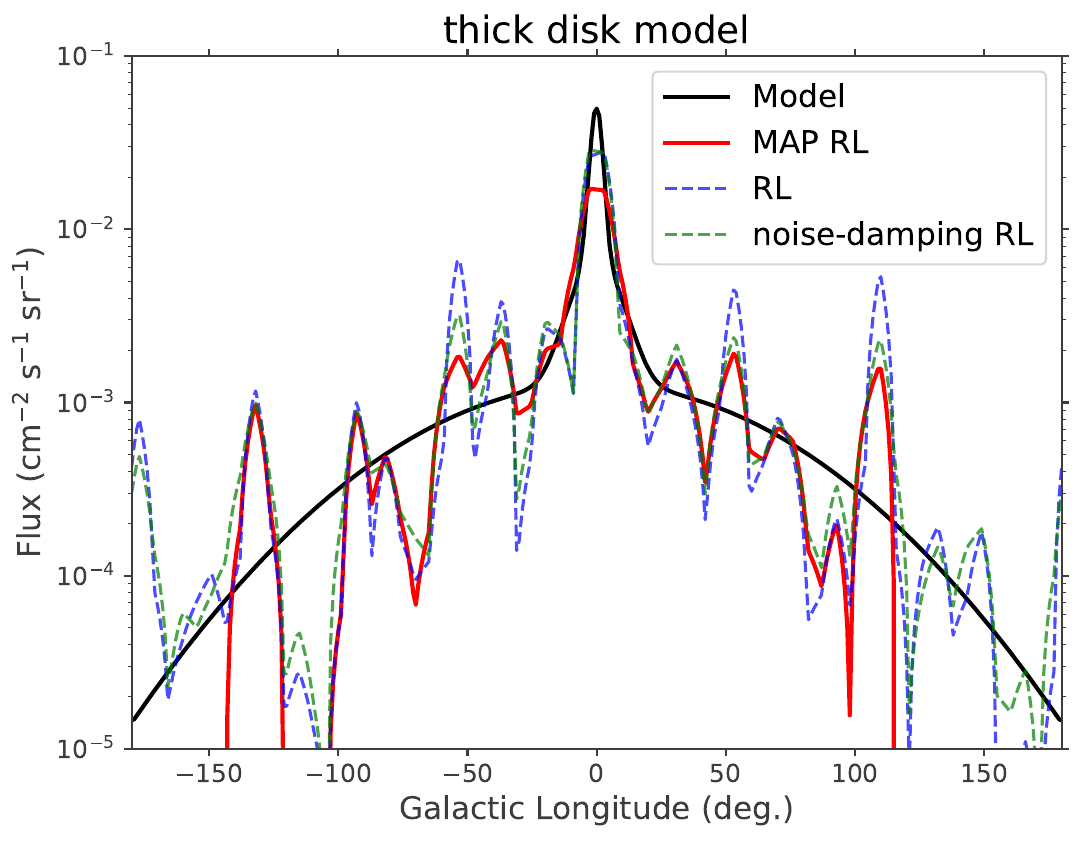}
    \includegraphics[width = 0.48 \linewidth]{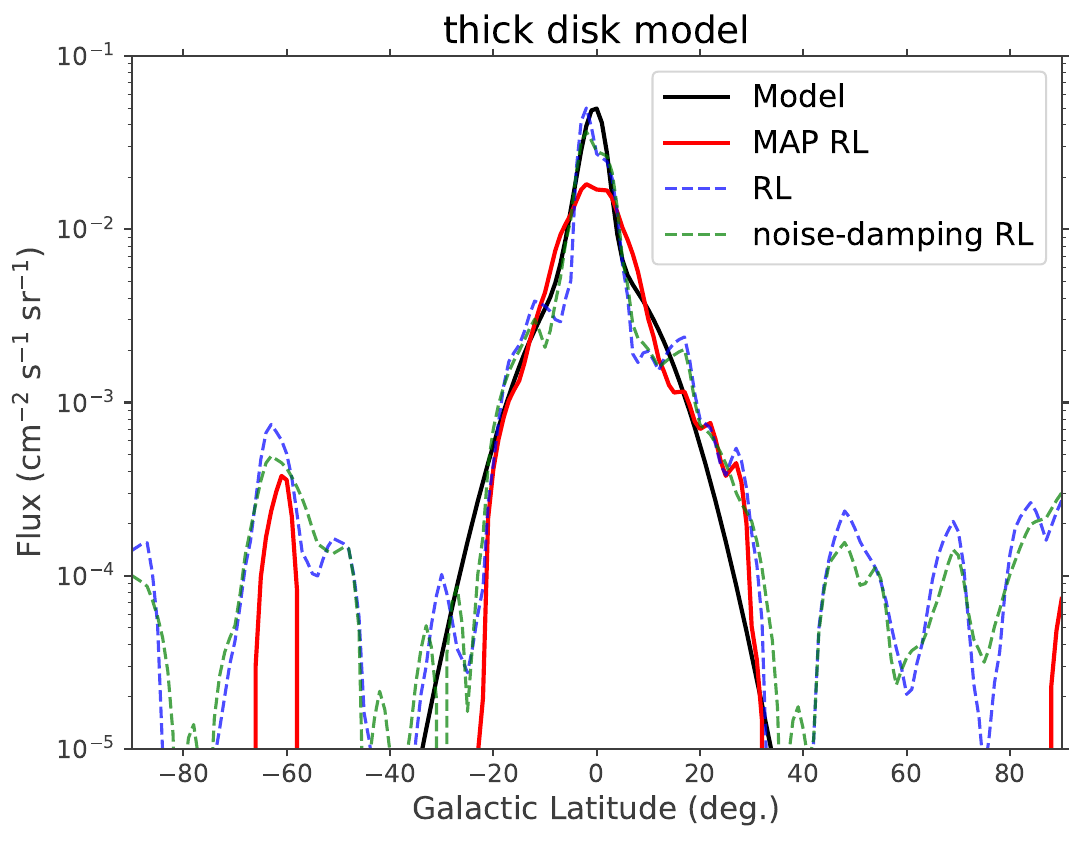}
\caption{Same as Figure~\ref{fig_511keV_sliced_image} but for the thick disk model.}
\label{fig_511keV_thick_sliced_image}
\end{figure*}

We also applied our method to the thick disk model. The results, shown in Figures~\ref{fig_511keV_thick_image_comp} and \ref{fig_511keV_thick_sliced_image}, demonstrate that our method performs well for this alternative model as well. Our algorithm successfully reconstructs the smooth distribution of the bulge and the thicker disk emission without introducing significant spurious point sources, while the conventional RL again shows a fragmented distribution.
The successful application of our method to both the thin disk and thick disk models demonstrates its robustness to different underlying source structures. This case illustrates the power of combining sparseness and smoothness priors in our modified RL algorithm, enabling accurate reconstruction of complex emission scenarios that include both point-like and diffuse components with varying scale heights.
These results suggest that three months of COSI observations might be sufficient to discriminate between competing 0.511 MeV emission models, which would advance our understanding of positron annihilation distribution in the Galaxy.

\section{Discussion}
\label{sec_discussion}

The results presented in this paper demonstrate the effectiveness of our modified RL algorithm for image reconstruction in COSI observations. Our method successfully combines the sparseness and smoothness priors, enabling accurate reconstruction of complex gamma-ray spatial distributions. While the current implementation shows promising results, we want to discuss several avenues for further improvements and expansion of this approach.

One of the key strengths of our method is its flexibility in incorporating different prior distributions. This study focused on the sparse term introduced by \cite{Ikeda2014} and the TSV for smoothness.
While this combination has been demonstrated in \cite{Morii2019} with X-ray data, and solved by EM algorithm and the proximal gradient method \citep{Beck2009},
our framework allows for straightforward extension to any prior distribution if it is first-order differentiable, e.g., entropy term and total flux density regularizer.
We can explore more sophisticated or physically motivated priors that could further enhance reconstruction quality.
Particularly, we found that applying a smooth prior distribution to the entire sky could blur the structure of the 0.511 MeV emission at the Galactic center.
As the spatial profile of the Galactic center could help to understand the origin of the 0.511 MeV emission, including some dark matter scenarios \citep{Finkbeiner_2007},
it is important to address this issue.

Several approaches are possible to make improvements. One is to adopt spatially varying priors. We can vary the coefficient for the prior depending on the location in the sky.
Since the central region was well-reconstructed without the smooth prior,
weakening the strength of the smooth prior around the Galactic center would mitigate such a problem.
Another approach could be to explore edge-preserving smoothness priors that can maintain sharp transitions, as seen in the boundary of the Galactic central region.
\citet{Allain2006} discussed that the total-variation-type regulator could work well when reconstructing sharp edges with INTEGRAL/SPI simulation data of some artificial spatial distributions.
It is worth investigating a similar prior (or its combination with other priors) using Compton telescope data, assuming science cases like those performed in this paper.
Finally, we want to mention that the gamma distribution introduced in the background can also be used for the source prior in principle.
As discussed in Section~\ref{sec_rl_source}, the sparseness term can be interpreted as a special case of the gamma distribution (see Equation~\ref{eq_gamma_sparse}).
Even if we use a general gamma distribution, the argument in Section~\ref{sec_rl_source} holds after a minor modification, namely replacing $\lambda^{\mathrm{EM}}_j$ with a solution with the general gamma distribution like Equation~\ref{eq_RL_bkg}.
It would allow us to include prior information about the flux and the flux uncertainty of each pixel if we have them, e.g., when we want to incorporate results from previous observations in the data analysis.

While this paper focused on the gamma-ray lines, the proposed method can be applied to other science cases. A promising direction is extending our method to simultaneous spatial-spectral deconvolution. 
This paper focused on applying the TSV prior only in the spatial domain, but it can be naturally extended to include smoothness constraints in the energy dimension as well.
This capability could be particularly effective for analyzing diffuse continuum emission, such as the Galactic diffuse continuum emission (GDCE), where both spatial and spectral distributions are intrinsically smooth.
The origin of the GDCE in the MeV band is not yet fully understood, which is sometimes referred to as ``COMPTEL excess'' or ``MeV hump'' \citep[e.g.,][]{Strong1996,Tsuji2023,Karwin2023}.
The spatial-spectral deconvolution with such smoothness priors could lead to a better determination of the spectral and spatial properties of the GDCE with future COSI data.
Moreover, our method can also be applied to transient events like Gamma-Ray Bursts.
In these cases, the sparseness prior could be particularly effective, as demonstrated in the $^{44}$Ti analysis with relatively limited statistics.
For transient events, where rapid localization is crucial, our algorithm could help to identify point sources by combining the sparseness prior with background constraints.

The basic capability of background optimization was demonstrated within this framework, which is crucial for COSI and MeV gamma-ray observations due to the complex nature of the background.
While our demonstration was a simple case,
we can refine this approach by optimizing several background components separately in future actual data analysis.
For instance, we can assign a single normalization factor
to astronomical backgrounds, e.g., the extragalactic background, for the entire observation period.
Simultaneously, we can prepare independent normalization factors for different time segments to account for time-variable instrumental background \citep{Siegert2019}.
This multi-component approach could further improve the accuracy of source reconstruction.
The main germanium detectors of COSI are surrounded by BGO active shields.
Additionally, the instruments of COSI include the Background Transient Monitor, which is a set of four NaI detectors \citep{Gulick2024}.
The data from these instruments, such as the count rates of saturated events and the light curves from the BGO above 2 MeV, can serve as a tracer of the charged particle flux and provide valuable information for determining the background normalization.
Integrating these data into our algorithm could further improve the accuracy of background estimation in the future.

When applying this method to real COSI observation data,
computational efficiency is also a factor to be considered.
While our current implementation achieves reasonable computation times for the 3.7-degree resolution used in this study,
we may need to use finer spatial resolution in the model and data space to discuss more detailed spatial structure.
Such a finer resolution would be required to analyze the Galactic central region of the 0.511 MeV emission and to conduct point source studies like in the $^{44}$Ti case.
This will significantly increase the computational demands due to the larger data size of the response matrix, which might require handling a large response matrix with a size of O(100) GByte.
In future applications, we can explore parallel computing using supercomputers, acceleration using GPU clusters, and further algorithm optimization. Also, acceleration techniques for the RL algorithm, such as SQUAREM \citep{Du2020}, can improve the computational speed.

Finally, optimizing prior parameters will be crucial when applying this method to real observational data. The L-curve method worked to some extent in our demonstrations, but we needed to rely on visual evaluation in some cases, which caused ambiguity in determining an optimal solution. For longer observation times, data-driven methods such as cross-validation could be an alternative.
For example, \cite{Morii2019,Morii2024} used k-fold cross-validation and showed that it worked successfully for their X-ray image analysis.
Such a technique could potentially be viable for future data analysis.

\section{Conclusions}
In this paper, we have presented a modified RL algorithm tailored for image reconstruction in Compton telescope observations, especially focusing on the upcoming COSI mission. Our method addresses key challenges in MeV gamma-ray astronomy by incorporating Bayesian priors for sparseness and smoothness within a maximum a posteriori framework while simultaneously optimizing background components.

Through the above simulations of $^{44}$Ti, $^{26}$Al, and e$^{+}$/e$^{-}$, we have demonstrated the effectiveness of our approach in reconstructing both point sources and extended emissions. The sparseness prior effectively suppresses artificial structures in point source reconstructions, while the TSV prior significantly improves the reconstruction of extended sources. The combination of these priors allows for accurate reconstruction of complex emission scenarios, such as the 0.511 MeV sky, where both point-like and diffuse components are present. Our demonstration shows that different spatial models of 0.511 MeV emission can be distinguished even with a three-month early phase of COSI operation.

These results suggest that our modified RL algorithm could improve image reconstruction quality across various astrophysical cases. It addresses the unique challenges of MeV gamma-ray astronomy and paves the way for more accurate and detailed studies of gamma-ray sources. Future improvements, such as spatially varying priors and integration of additional background data, could further enhance the capabilities of our method.

As COSI moves toward its launch in 2027, continued refinement and testing of these techniques will be crucial to fully leverage the capabilities of this next-generation Compton telescope, potentially leading to new insights into nucleosynthesis, positron annihilation, and other high-energy phenomena in our Galaxy.

\begin{acknowledgements}
HY acknowledges support by the Bundesministerium f\"{u}r Wirtschaft und Klimaschutz via the Deutsches Zentrum f\"{u}r Luft- und Raumfahrt (DLR) under contract number 50 OO 2219 and JSPS KAKENHI grant number 23K13136.
SM is supported by JSPS KAKENHI Grant Number 23K20232, 20H01895 and JSPS Core-to-Core Program JPJSCCA20200002.
SM and TT are supported by JSPS KAKENHI Grant Number 24H00244 and 20H00153.
YW is supported by JSPS KAKENHI Grant Number 23KJ0470.
SM, TT and YW are supported by the World Premier International Research Center Initiative (WPI), MEXT, Japan (Kavli IPMU).
IM is supported by NASA under award number 80GSFC24M0006.
Computational resources supporting this work were provided by National High Performance Computing (NHR) South-West at Johannes Gutenberg University Mainz, the Discover supercomputer, which is part of the NASA Center for Climate Simulation (NCCS) at NASA Goddard Space Flight Center, and the NASA High-End Computing (HEC) Program through the NASA Advanced Supercomputing (NAS) Division at Ames Research Center.
The Compton Spectrometer and Imager is a NASA Explorer project led by the University of California, Berkeley with funding from NASA under contract 80GSFC21C0059.
\end{acknowledgements}

\bibliographystyle{aa}

\begin{appendix}

\section{Conditions for approximation accuracy in the modified RL algorithm}
\label{sec_condition_approx_Mstep}

Here, we provide a more detailed discussion of the conditions under which the approximation introduced in Section\ref{sec_rl_source} is valid.
We focus on the TSV prior as an example:
\begin{align}
f_{p}\left(\vector{\lambda}\right) &= - c^{\mathrm{TSV}} \sum_j \sum_{j' \in \sigma_{j}} (\lambda_{j} - \lambda_{j'})^2~.
\label{eq_A1}
\end{align}
Its first and second derivatives can be calculated as:
\begin{align}
\dfrac{\partial f_{p}(\vector{\lambda})}{\partial \lambda_{j}} &= - 4 c^{\mathrm{TSV}} \sum_{j' \in \sigma_{j}} (\lambda_{j} - \lambda_{j'})~,\\
\dfrac{\partial^2 f_{p}(\vector{\lambda})}{\partial \lambda_{j} \partial \lambda_{j'}} &= 
\begin{cases}
- 4 c^{\mathrm{TSV}} N(\sigma_{j}) & (j = j')\\
4 c^{\mathrm{TSV}} & (j \neq j', j' \in \sigma_{j})\\
0 & (j \neq j', j' \notin \sigma_{j})~,
\end{cases}
\end{align}
where $N(\sigma_{j})$ is the number of elements in $\sigma_{j}$. Note that for each pair $(j,j' \in \sigma_j)$, the term $(\lambda_j - \lambda_{j'})^2$ appears twice in the sum in Equation~\ref{eq_A1}, resulting in the coefficient 4 in the first derivative.

\begin{figure}
\centering
    \includegraphics[width = 0.9 \linewidth]{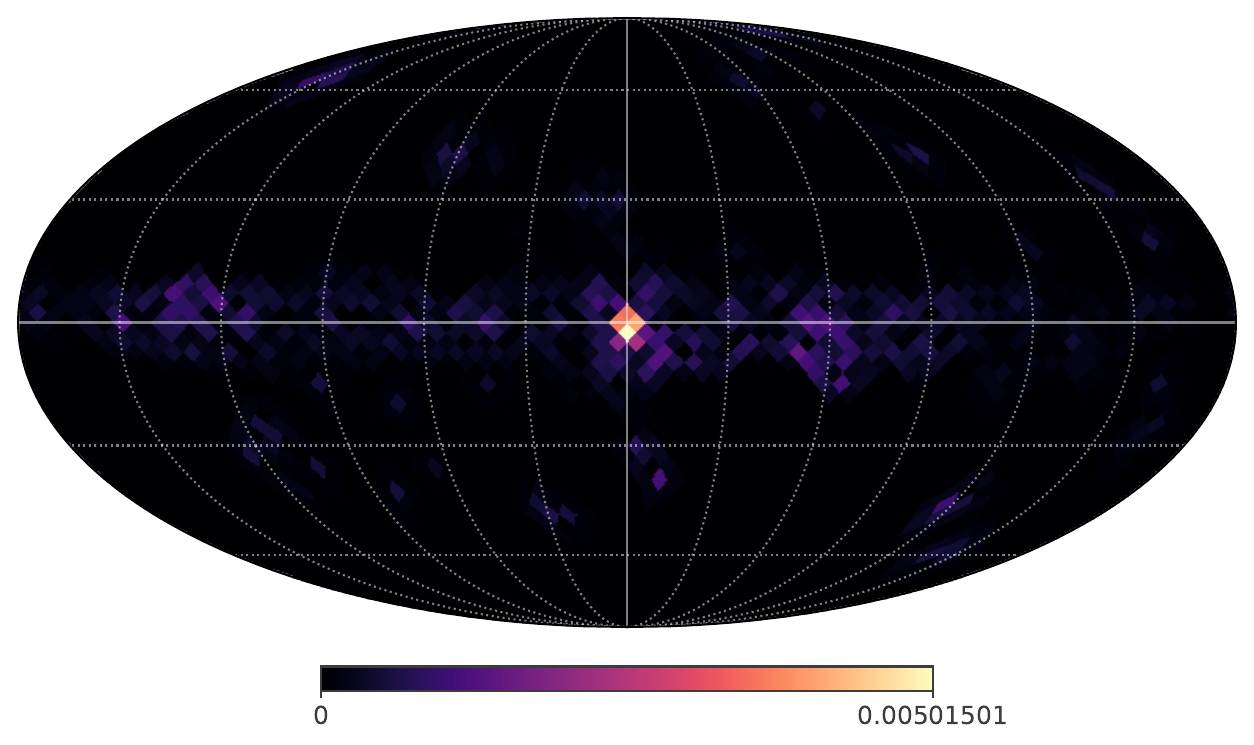}
\caption{Map of the left term in Equation~\ref{eq_approx_map_TSV}. It shows the last iteration of the MAP RL solution in Figure~\ref{fig_511keV_image_comp}.}
\label{fig_511keV_approx_map}
\end{figure}

If we can assume that the first derivative and the sum of the response matrix change smoothly over the image ($\dfrac{\partial f_{p}(\vector{\lambda})}{\partial \lambda_{j}} \approx \dfrac{\partial f_{p}(\vector{\lambda})}{\partial \lambda_{j'}}$,
$\displaystyle \sum_{i} R_{ij} \approx \sum_{i} R_{ij'}$ for $j' \in \sigma_j$), the condition that the updated image should satisfy (Equation~\ref{eq_approx_condition}) can be described as:
\begin{align}
\frac{4 c^{\mathrm{TSV}}}{\displaystyle \sum_{i} R_{ij}} \times \left| \sum_{j' \in \sigma_j} (\lambda_{j} - \lambda_{j'}) \right| \ll 1 ~.
\label{eq_approx_map_TSV}
\end{align}
Figure~\ref{fig_511keV_approx_map} shows the map of the left-hand term over the sky at the last iteration for the result with the 0.511 MeV thin disk model shown in Figure~\ref{fig_511keV_image_comp}.
The maximum value is $\sim 5 \times 10^{-3}$, confirming that the condition for the approximation is satisfied in this case.

\section{Reconstructed 0.511 MeV images with different parameters}
\label{sec_511keV_image_all}

\begin{figure*}
\centering
    \includegraphics[width=0.7\linewidth, keepaspectratio]{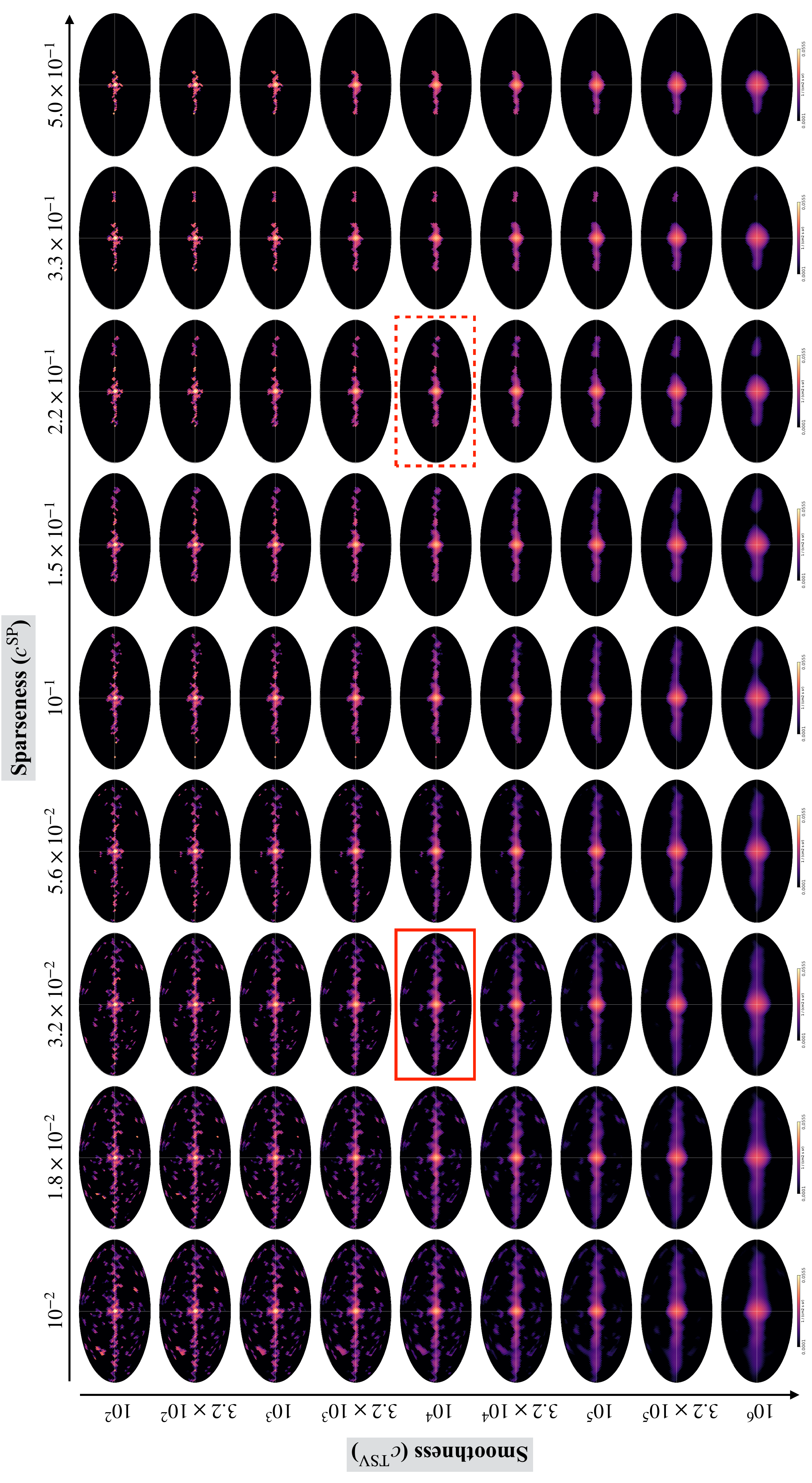}
\caption{Reconstructed images of 0.511 MeV with different values for $c^{\mathrm{TSV}}$ and $c^{\mathrm{SP}}$.}
\label{fig_511keV_images}
\end{figure*}

Figure~\ref{fig_511keV_images} shows the reconstructed 0.511 MeV images using various parameter values in the prior.
The images surrounded by red boxes are candidates for the optimal solution derived from the L-curve method. The solid line surrounds the optimal solution, while the dashed line surrounds the second candidate.

\section{Comparison between the model and the data in the CDS}
\label{sec_comp_model_CDS}

Here, we check how well the reconstructed image (shown in the top right panel of Figure~\ref{fig_511keV_image_comp}) reproduces the observed data in the CDS for the 511 keV thin disk model.
Since the CDS is multidimensional, we project the expected counts onto two axes: the Compton scattering angle ($\phi$) and the scattered gamma-ray direction ($\psi\xi$).

\begin{figure}[H]
\centering
    \includegraphics[width = 0.9 \linewidth]{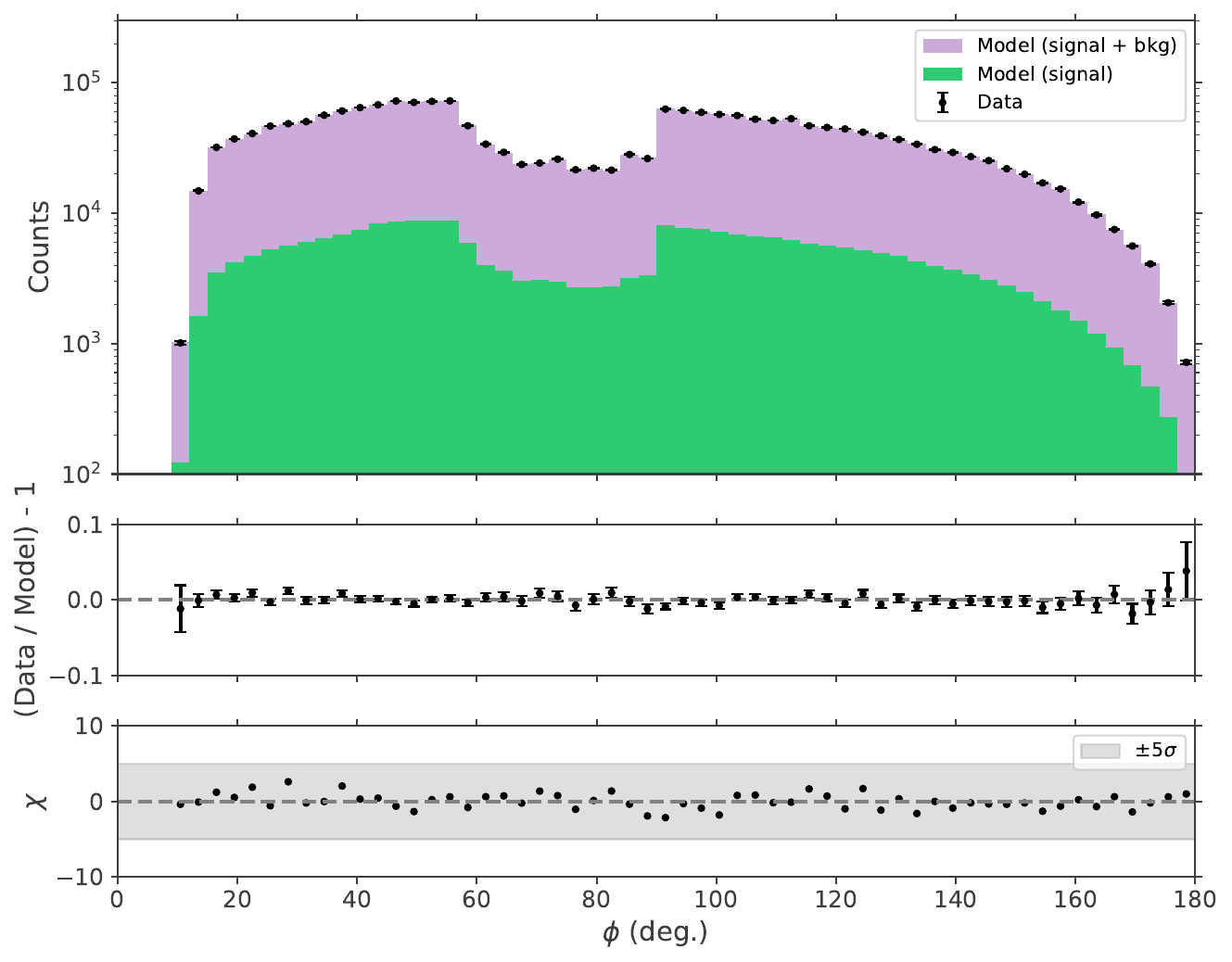}
\caption{Comparison between model predictions and observed data for the 511 keV thin disk model, projected onto the $\phi$ axis.}
\label{fig_comp_model_phi_CDS}
\end{figure}

\begin{figure}[H]
\centering
    \includegraphics[width = 0.9 \linewidth]{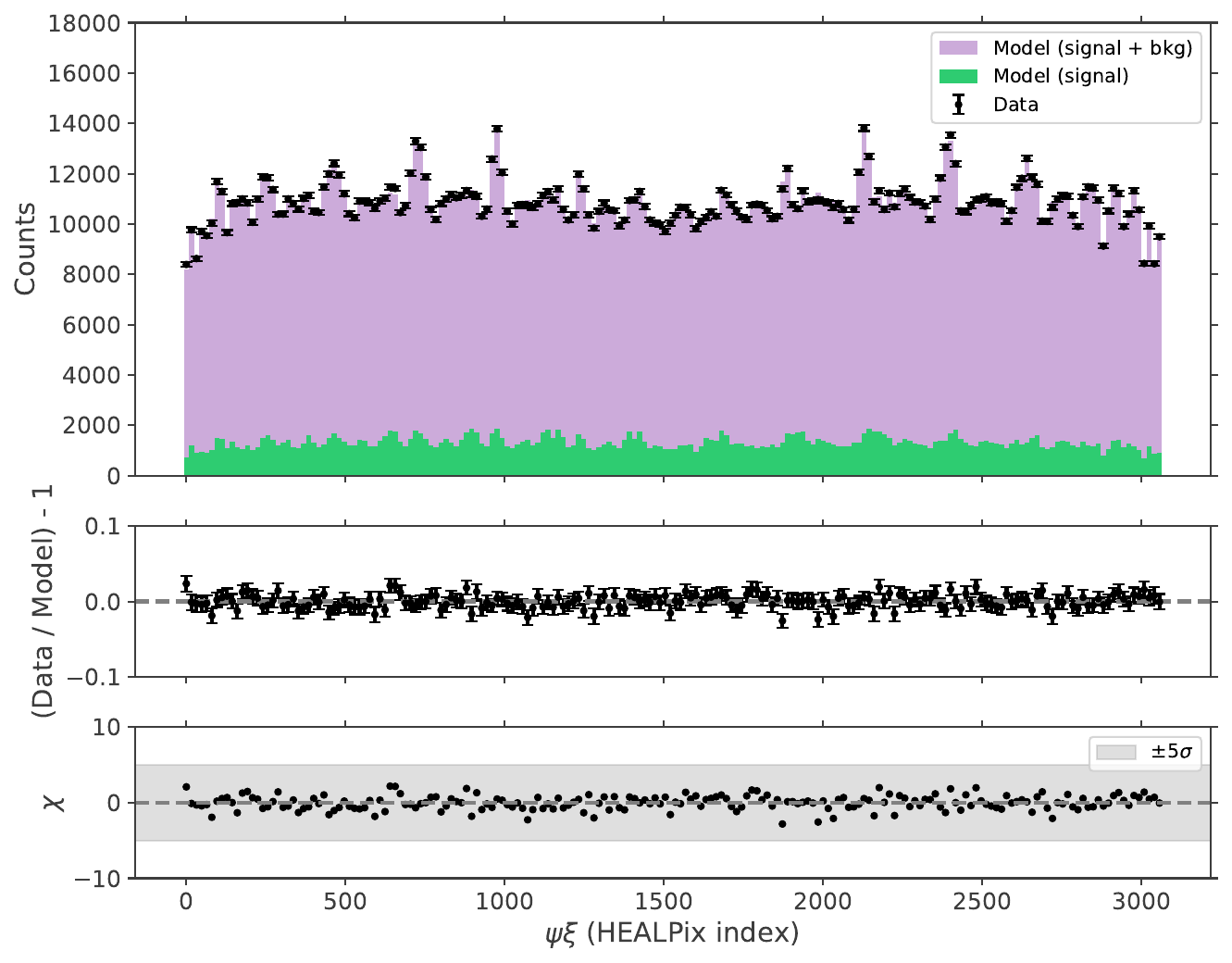}
\caption{Same as Figure~\ref{fig_comp_model_phi_CDS}, but projected onto the $\psi\xi$ axis. For better visualization, the histograms are rebinned every 16 bins.}
\label{fig_comp_model_psichi_CDS}
\end{figure}

Figure~\ref{fig_comp_model_phi_CDS} and \ref{fig_comp_model_psichi_CDS} compare these projected histograms with the actual simulation data. 
In both figures, the top panels show the expected counts from the background and signal (purple) and those from the signal alone (green), overlaid with the observed counts (black).
The middle panels show the residuals between the model prediction and the data. The bottom panels show the difference between expected counts and data divided by the square root of the expected counts, denoted as $\chi$ in the figures.

\end{appendix}
\end{document}